  \documentclass{cambridge6A}
  \usepackage{natbib}

  \usepackage{rotating}
  \usepackage{floatpag}
  \rotfloatpagestyle{empty}
   \usepackage{mdframed}
  \usepackage{amsthm}
  \usepackage{graphicx}
  \usepackage{subcaption}
  \usepackage{nohyperref}

\newcommand{\oversim}[2]{\protect{\mbox{\lower0.5ex\vbox{%
   \baselineskip=0pt\lineskip=0.2ex
   \ialign{$\mathsurround=0pt #1\hfil##\hfil$\crcr#2\crcr\sim\crcr}}}}} 
\newcommand{\simgreat}{\mbox{$\,\mathrel{\mathpalette\oversim>}\,$}} 
\newcommand{\simless} {\mbox{$\,\mathrel{\mathpalette\oversim<}\,$}} 
  \usepackage{multind}
  \makeindex{authors}
  \makeindex{subject}
  \usepackage{color}

  \newcommand\cambridge{cambridge6A}

  \theoremstyle{plain}

  \theoremstyle{definition}

  \theoremstyle{remark}

\begin{document}

  \title[Subtitle, If You Have One]
    {\LaTeXe\ GUIDE FOR AUTHORS USING THE \cambridge\ DESIGN}

  \author{Ali Woollatt\\[3\baselineskip]
    This guide was compiled using \hbox{\cambridge.cls \version}\\[\baselineskip]
    The latest version can be downloaded from:
    https://authornet.cambridge.org/information/productionguide/
      LaTeX\_files/\cambridge.zip}








  \alphafootnotes
  \author[P. Kroupa and T. Jerabkova]
    {Pavel Kroupa\footnotemark\
    and Tereza Jerabkova\footnotemark}

  \chapterauthor{Pavel Kroupa\footnotemark\
    and Tereza Jerabkova\footnotemark}

  \chapter{The initial mass function of stars and the star-formation rates of galaxies}
\footnotetext[1]{Helmholtz-Institut f\"ur Strahlen- und Kernphysik,
    Nussallee 14-16, 53115 Bonn, Germany;\\
    Charles University in Prague, 
Faculty of Mathematics and Physics, 
Astronomical Institute, 
V  Hole\v{s}ovi\v{c}k\'ach 2,
CZ-18000 Praha,
Czech Republic; pavel.kroupa@mff.cuni.cz}
\footnotetext[2]{ESA Fellow; ESTEC/SCI-S,
Keplerlaan 1,
2200 AG Noordwijk,
Netherlands; tereza.jerabkova@esa.int}
\arabicfootnotes

\vspace{-20mm}

{\small(This is Chapter~2 in the book ``Star-formation Rates of Galaxies'', edited byVeroni\-que Buat \& Andreas Zezas, published in April~2021 by Cambridge University Press [ISBN: 9781107184\-169]. This arXiv version contains additional material in Sec.~\ref{sec:terminology}, \ref{sec:ecls}, \ref{sec:iMFmeas}, \ref{sec:shape} and~\ref{sec:mainsequ}.)}

\begin{abstract}

  The measured star-formation rates (SFRs) of galaxies comprise an impor- tant constraint on galaxy evolution and also on their cosmological boundary conditions.  Any available tracer of the SFR depends on the shape of the mass-distribution of formed stars, i.e. on the stellar initial mass function (IMF). The luminous massive stars dominate the observed photon flux while the dim low-mass stars dominate the mass in the freshly formed population. Errors in the number ratio of the massive to low-mass stars lead to errors in SFR measurements and thus to errors concerning the gas-accretion-rates and the gas-consumption time-scales of galaxies. The stellar IMF has tra- ditionally been interpreted to be a scale-invariant probability density distribution function (PDF), but it may instead be an optimal distribution function. In the PDF interpretation, the stellar IMF observed on the stales of individual star clusters is equal to the galaxy-wide IMF (gwIMF) which, by implication, would be invariant. In this Chapter we discuss the fundamental properties of the IMF and of the gwIMF, the nature of both and their systematic variability as indicated by measurements and theoretical expectations, and we discuss the implications for the SFRs of galaxies and on their main sequence. The importance of the putative most-massive-star-mass vs stellar-mass-of-the- hosting-embedded-cluster ($m_{\rm max}-M_{\rm ecl}$) relation and its possible establish- ment during the proto-stellar formation phase, is emphasised.
  
\end{abstract}

\section{Introduction}

\label{sec:introd}

Let $dN$ be the number of stars in the infinitesimal initial-stellar-mass interval $m$ to $m+dm$, then $dN=\xi(m)\,dm$ where $\xi(m)$ is the stellar IMF, that is, the distribution of the initial masses of all stars formed together.  It is usual to represent the IMF with a power-law function, $\xi(m) \propto m^{-\alpha_i}$, where the power-law indices depend on the mass range considered and $\alpha_S = 2.35$ is the Salpeter index valid for the mass range $0.4\,M_\odot-10\,M_\odot$ \citep{Salpeter55}.

It follows that it is in principle straight-forward to construct $\xi(m)$ because the astronomer merely needs to look up and count the stars. The difficulty in the practical execution of this is that stellar masses are only measurable directly when they are in binary orbits. For the vast majority of stars binary-orbits are not available though, and their masses can only be inferred from their luminosities, colours and spectra relying on theoretical stellar models.  Theoretical models assign these radiant properties a mass, provided an age, stellar spin and chemical abundance is known, these properties of the star being also measured from similar data. Massive stars are extremely luminous and shine mostly in the UV with a steep mass--luminosity relation, while low-mass stars are dim and shine largely in the red and infrared with a flat mass--luminosity relation (small uncertainties in the mass translate into large uncertainties in the luminosity). Massive stars are rare and live only for a few Myr while low mass stars dominate the population in numbers and live up to many Hubble times.

The practical construction from observational data of the IMF over all stellar masses is thus subject to very major uncertainties requiring data from different surveys to be put together.  This is particularly true for the Galactic field stellar ensemble nearby to the Sun, where a complete census of the massive stars is reachable to distances of a few kpc but the low-mass stars can only be detected to within a few pc (Sec.~\ref{sec:shape}).

To infer the stellar IMF in the Galactic field the massive-star and low-mass star counts need to be combined appropriately and carefully which requires taking care of the different spatial distribution and the star formation history \citep{Salpeter55, MS79, Scalo86, ES06}.  The surveys of stars in star clusters do not suffer from this complication since the stars are practically at the same distance, clusters typically being further than a few dozen pc away from us while their radii are typically 3~pc, and the stars have nearly the same ages. One problem with star clusters though is that stellar-dynamical processes remove stars from the clusters already starting at the earliest times (a few~$10^5\,$yr) \citep{BM03, Banerjee12b, Haghi15,OKP15, OK16, Kroupa18}, binary-star evolution and dynamically-induced stellar mergers also affect the inferred IMF \citep{deMink+14,Schneider+15,OK18}.  Also, and in all cases, in the Galactic field, star clusters and any other stellar ensemble, the star counts always suffer from unresolved multiple systems such that the observer misses the fainter or unresolved companion. This bias is very significant for late-type stars \citep{KTG91} but is less problematical for more massive stars \citep{Weidner09, KJ18}.

Given its important role in much of astrophysics, the IMF has been the subject of a vast research effort.  Some relevant IMF-related questions that have been being addressed are: Can the IMF be measured? What is the shape of the IMF? What is its mathematical nature (e.g., is it a probability density distribution function)? Does the IMF vary? Does the IMF of a simple stellar population equal that of a composite population?

Reviews which address these questions can be found in \cite{Kroupa02,
  Chabrier03, Bastian10, Offner14, KJ18}. The major effort to assess the IMF
for stars more massive than a few~$M_\odot$ by \cite{Massey03} needs
to be emphasised.  The reviews by \cite{Kroupa13} and particularly
\cite{Hopkins18} cover the significant extra-galactic evidence for a
variable galaxy-wide IMF.  Here we address these five most-important
IMF questions in turn and place them into the context of the
determination of the SFRs of galaxies. Needless to write, the answers
to these questions have a rather major impact on the SFRs which are
calculated given some tracer. This in turn affects our knowledge and
understanding of the matter cycle in the Universe.

\vspace{2mm} \centerline{ \fbox{\parbox{\columnwidth}{ 
     The IMF is the key to understanding the cosmological matter
cycle.
}}}  \vspace{2mm}

In the following two sub-sections we explain some IMF-acronyms and 
discuss some issues concerning how stars emerge from the inter-stellar medium.

\subsection{Used IMF terminology throughout this text}
\label{sec:terminology}

Throughout this text we define several acronyms describing the IMF on different scales. It is useful to introduce these once a variable non-universal IMF is discussed \citep{Hopkins18,Jerabkova+18}.

As in \cite{Jerabkova+18} we refer to the {\it IMF} as being the stellar initial mass function of stars formed during one star formation event in an initially gravitationally-bound region in a molecular cloud (on a 1~pc scale and within about 1~Myr in a spatially and temporarily correlated star-formation event, CSFE, i.e. an embedded cluster; see Sec.~\ref{sec:ecls}).  We refer to the {\it cIMF} as the composite-IMF, i.e.  the sum of the IMFs over larger regions in a galaxy (e.g. stellar associations, cf. \citealt{vanBeveren82, Hopkins18}). The {\it gwIMF} is the initial stellar mass function of {\it newly} formed stars in a whole galaxy, as observed for example with integrated-light indicators.  

\begin{figure}[ht!] \begin{center}
%
\includegraphics[width=0.9\textwidth]{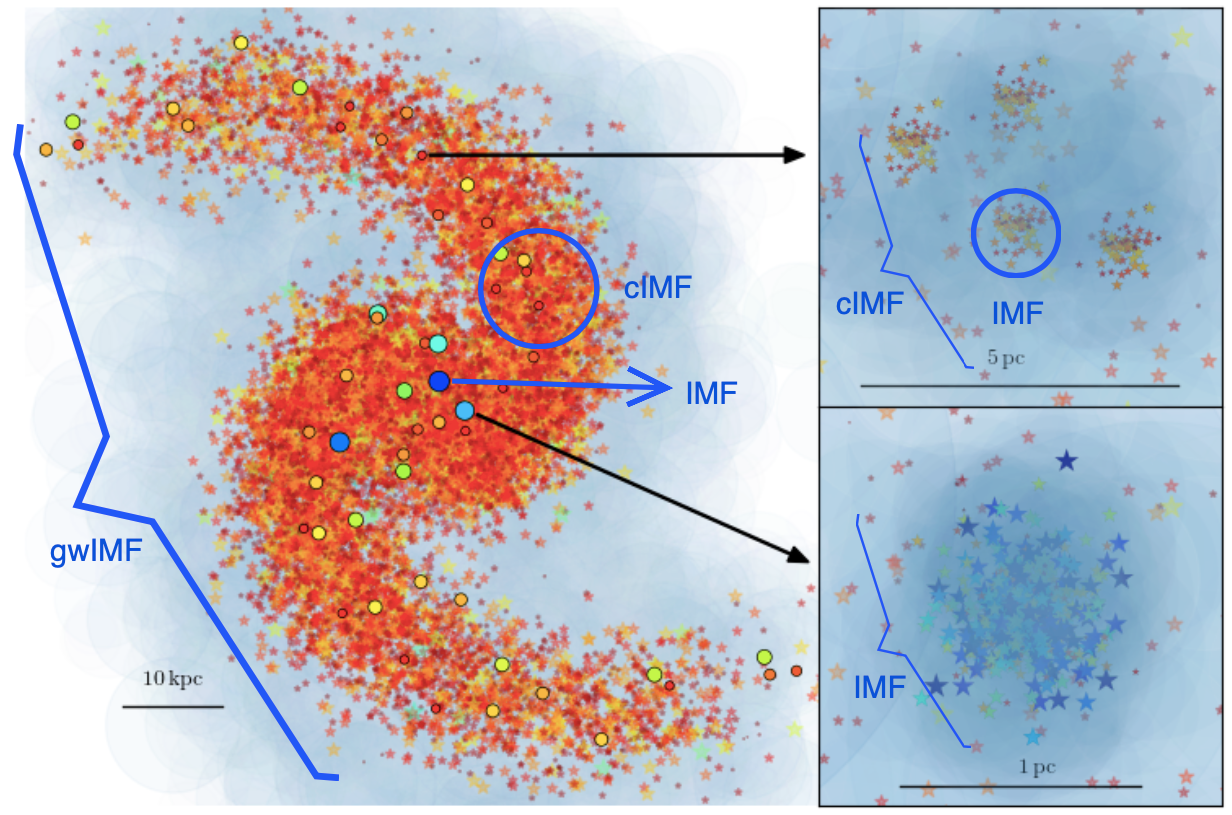}
\end{center}
\vspace{-5mm}
\caption{
Schematic showing a late-type star forming galaxy. Its field population is represented by red and orange stars. The newly formed stellar population is marked by colored circles which represent CSFEs (embedded star clusters/ active star forming units). The colors and sizes of symbols scale with stellar and cluster mass. \textbf{Right bottom panel:} A young massive embedded cluster, which will most likely survive and contribute to the galaxy's open star cluster population \citep{Brinkmann+17}. 
\textbf{Right top panel:} Young embedded cluster complex composed of a number of low-mass embedded clusters (as observed e.g. in the Orion clouds by \citealt{Megeath12, Megeath16}), which will evolve into a T-Tauri association once the embedded clusters expand after loss of their residual gas and disperse into the galaxy field stellar population \citep{KB03,Joncour+18}. Adapted from \cite{Jerabkova+18}.
} \label{fig:gal_sk} 
\end{figure}

These abbreviations are visualised in Fig.~\ref{fig:gal_sk}.

\begin{figure}[ht!] \begin{center}
\includegraphics[width=0.9\textwidth]{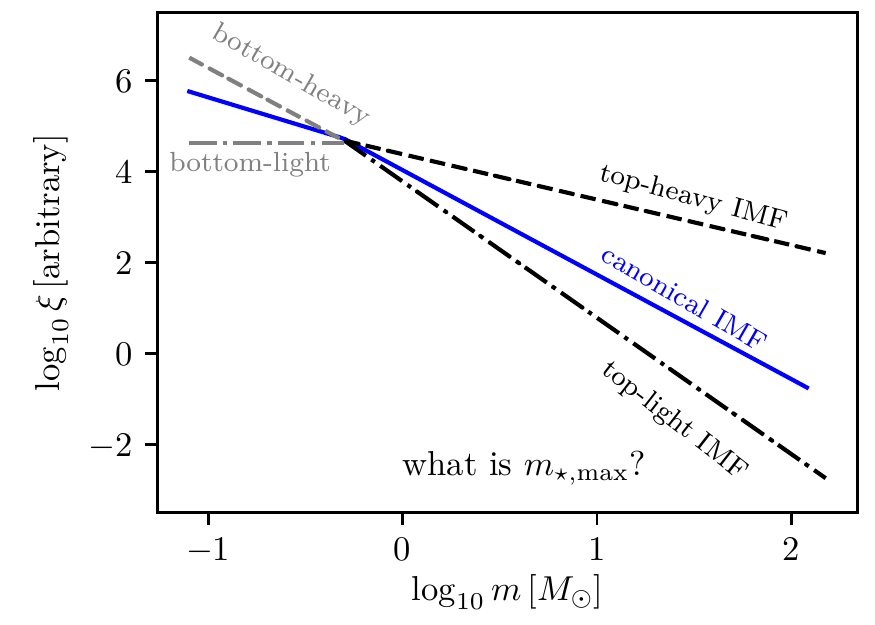}
\end{center}
\vspace{-5mm}
\caption{A sketch to visualise the meaning of a top-heavy, top-light, bottom-heavy and bottom-light IMF, cIMF and gwIMF. The question posed in the panel concerns whether a fundamental maximum physical stellar mass, $m_{\rm max*}$, exists such that all born stars have initial masses $m\le m_{\rm max*}$. 
} \label{fig:toplight} 
\end{figure}

If we define the standard form of the IMF as the {\it canonical IMF} (e.g. Eq.~\ref{eq:imf}), then, relative to this canonical IMF, 
a {\it top-heavy IMF, cIMF or gwIMF} is one which has a surplus of massive ($m\simgreat 10\,M_\odot$) stars,  a {\it top-light} one has a deficit of massive stars and is thus also bottom-heavy, a {\it bottom-heavy} one has a surplus of low-mass ($\simless 0.7\,M_\odot$) stars and a {\it bottom-light} one has a deficit of low-mass stars. But it is clearly the slope of the IMF which sets the terminology.  This is visualised in Fig.~\ref{fig:toplight}

Note that if the IMF would be a universal probability density function (PDF) then IMF$=$cIMF$=$gwIMF (Box The Composite IMF-PDF Theorem on p.~\pageref{box:comp_theorem}). On the 
other hand, if the IMF varies and/or if it is not a PDF, then in general IMF $\ne$ cIMF $\ne$ gwIMF. It is therefore important to introduce this terminology and to discuss various concepts as to what the mathematical nature of the IMF may be and how this may relate to the physics of star formation, as we do in the following text.

Throughout this text we refer to the ECMF as the embedded cluster mass function, i.e. the mass distribution of freshly formed embedded clusters (Eq.~\ref{eq:ECMF} below). The same considerations as for the cIMF and the gwIMF also apply to the ECMF (cECMF and the gwECMF), but we do not discuss these here in further detail apart from mentioning that a Schechter-type form of the galaxy-wide ECMF (the gwECMF) arises naturally if the ECMF is a power-law function \citep{LK17}.

\subsection{Turbulence, filaments, embedded clusters and associations}
\label{sec:ecls}

The problem discussed here is how the galaxy-wide IMF, which is relevant for extragalactic research, is related to the stellar IMF which is observationally deduced in individual star-formation regions within galaxies. One ansatz is to develop a theoretical galaxy-wide description of how a turbulent inter-stellar medium (ISM) generates the gwIMF \citep{Hopkins12}. Another ansatz is to use the IMF deduced from observations of resolved stellar populations as a boundary condition in the problem, which, if successful, has the advantage that this ansatz would be automatically consistent with observations on galaxy-wide scales as well as on molecular-cloud scales. 
 

In Fig.~\ref{fig:gal_sk} the regional IMF (cIMF) and the galaxy-wide IMF (gwIMF) are visualised as arising from the IMF which is established by the star-formation process. An important issue for the discussion below is how this star-formation is arranged. Is a stochastic approach, according to which stellar masses are chosen randomly from the IMF whenever star formation occurs, a reasonable approximation? Or is a more elaborate, physically motivated calculation more appropriate?

The interstellar medium (ISM) is in a state of flow within the potential of the galaxy and comes in different physical, interrelated and constantly time evolving thermally unstable and non-equilibrium phases \citep{Breitschwerdt+12}. The complex structure of the ISM may, to a certain approximation and on a sufficiently large scale, be described as a turbulent dissipative medium (being constantly energised by rotational shear, supernova explosions, stellar winds, \citealt{Breitschwerdt+12,Burkert06,Federrath16}). Where the density becomes sufficiently large the gas can recombine, cool by line emission and form molecules. Observations inform us that a small fraction of the molecular phase achieves a sufficiently dense state to collapse under self-gravity to form stars. When this happens, typically much more mass is available than will end up in a late-type star, and thus small groups of stars to major clusters can form. 
The gas seems to loose its turbulent nature when denser and forms a complex pattern of intertwined filaments which appear to accrete gas from their environment \citep{Andre+10, Andre+14, Federrath16, Shimajiri+17}. 

Protostars (ages $<\,{\rm few}\,10^5\,$yr) are observed to form in these thin ($\approx 0.1\,$pc), phase-space coherent long ($>1\,$pc) molecular cloud filaments with life-times of a few$\,10^5\,$yr (e.g. \citealt{Hacar+13,Andre+14, Konyves+15, Hacar+17,Hacar+17b}). The filaments fragment into protostars, and compact groups of stars (we refer to these as {\it embedded clusters}) appear where such filaments intersect \citep{Schneider+12, Andre+14}. Embedded clusters are those compact ($<1\,$pc) regions which are gravitationally bound (since the gas falls into them along the filaments). Examples of embedded clusters are the compact groups of stars in the Taurus star forming region referred to as Nested Elementary STructures (NESTs) by \cite{Joncour+18}, the many low-mass groups of stars in the southern part of the Orion~A cloud and the Orion Nebula Cluster \citep{Megeath16}. \cite{MK12} derive a half-mass-radius--embedded-stellar-cluster-mass relation (eq.~\ref{eq:radius} below), which leads to a density--mass relation consistent with (a) observed molecular cloud cores and (b) independent estimates of the initial densities of globular clusters. The time when an embedded cluster reaches $r_{\rm h}$ can be identified with $t_{\rm math}$ in Box The Empirical IMF Dependence on p.~\pageref{box:toph_master}. Interesting here is that this radius--mass relation is qualitatively consistent with the dense embedded clusters that form in hydro-dynamical simulations \citep{Bate+14} and that it leads to $r_{\rm h}\approx 0.1\,$pc for $M_{\rm ecl}=1\,M_\odot$, which is the same spatial dimension as that of a filament along which late-type stars form nearly equidistantly \citep{Andre+10,Andre+14}. If the observed open clusters have gone through such a dense state, then they can expand to their present-day radii (about 3~pc) through the expulsion of residual gas assuming a star-formation efficiency of 33~\% (\citealt{BK17}, Eq.~\ref{eq:SFE}).

Numerical simulations of star formation in molecular clouds as well as N-body simulations of realistic (i.e. including a high primordial binary proportion and stellar evolution) embedded clusters prove these to be dynamically active \citep{Bate05, Bate14, Bate+14}, whereby the rate and energy of the activity (ejections of stars and binaries) is small in NESTs \citep{KB03} and dramatic in clusters containing thousands of stars
\citep{OKP15, OK16}. Thus, stars are ejected from the star-forming embedded clusters due to the encounters between stars and due to the high multiplicity fraction. An embedded cluster rapidly (within a few$\,10^5\,$yr) populates the region ($>\,{\rm few} \times 10\,$pc) around it with stars and binaries moving away from it. These may be identified as being part of a distributed population. The discussion of a {\it distributed} versus {\it clustered} mode of star formation \citep{Gieles+12}, based on the observed distribution of very young stars (ages $\simless 1\,$Myr), needs to take this into account. This is an important issue in the consideration of the gwIMF and how it is related to the star-formation process in a galaxy (a purely distributed mode would, for example, be consistent with a PDF interpretation of the IMF).

Given that "clusters" are ill defined in this context, we note that the protostars in the filaments can be mathematically treated as being the low-density outer region of the embedded clusters. Thus, for example, the Plummer model (\citealt{Plummer11}; the simplest solution of the collision-less Boltzmann equation, \citealt{HH03,Kroupa08}, and a good representation of observed clusters) always has individual members beyond a few Plummer radii which may, depending on the chosen definition of cluster membership, be defined to be part of an isolated or distributed population. For practical purposes, and also in good approximation to the observation that the great majority of protostars are found by surveys to be in embedded clusters \citep{LL03, Lada10}, we can assume, as one possible hypothesis, that embedded clusters are the fundamental units of star formation \citep{Kroupa05}. This hypothesis is supported by theoretical studies of the turbulent ISM in galaxies \citep{Hopkins13a} and is useful because it allows us to identify an IMF with an embedded cluster. It is not the sole hypothesis though to be investigated, as we discuss below.

For completeness, in view of the discussion of how OB and T-Tauri associations form, an understanding can be formulated according to which a density wave in a  spiral galaxy sources a minimum potential wave in which the in-falling in-homogeneous ISM condenses as molecular clouds (\citealt{FK10, HD15}; in a sense like bad-weather regions). Stars form in these and move out of the potential minimum along with the molecularised ISM. The molecular clouds thus disperse most likely by themselves just behind the potential minimum wave but also helped by the feedback from the freshly formed stars. 
Each molecular cloud typically spawns a number of embedded clusters which expand once the stars expel the residual gas, whereby their expansion rate depends on the mass of an embedded cluster (slow for low-mass cases; \citealt{Brinkmann+17}). Evidence for a complex kinematical field of young stars in associations has been reported \citep{WM18}. Thus, in this view, a stellar association is always present on the other side of the potential minimum wave as long as the ISM flows relative to the spiral pattern \citep{Egusa+04, Egusa+09, Egusa+17}. This association may be contracting or expanding and may also self-regulate and stimulate further star-formation \citep{Lim+18}, depending on the details of the ISM flow through the potential wave.

Independently of whether embedded clusters may be the fundamental units of star formation, 
the few-pc long, 0.1~pc wide filaments though cannot exist in a turbulent medium, since
it implies that the velocity differences between regions grow randomly with
spatial scale such that the filaments cannot be phase-space coherent
over scales of more than a pc (on scales longer than the smallest turbulence length-scale). 
It is thus the decay of the turbulent-like phase which allows the filaments to form.  

\vspace{2mm} \centerline{ \fbox{\parbox{\columnwidth}{ At the point in
      time when the stars begin to form, the gas therefore cannot be
      described as being a turbulent medium.  }}}  \vspace{2mm}

An important problem thus arises: is the process of star formation (which yields the IMF) given by gravo-thermal-turbulence  according to which the molecular cloud is in a state of turbulence which determines the density peaks which collapse under  self-gravity to form stars \citep{PN02, HB13,Hopkins13,Haugbolle+18}? This theoretical approach requires constant energy input into the cloud on all scales in order to retain the turbulence. But self-consistent computations of turbulent clouds cast doubt on turbulent fragmentation explaining the IMF as low-mass proto-stars tend to be destroyed through shocks before they can form \citep{Bertelli16,Liptai+17}.
On the other hand, the IMF might be related to the fragmentation of filaments instead (e.g. \citealt{Andre+10, Andre+14, Shimajiri+17}). 
That is, the IMF may not be determined by the PDF given by the turbulent molecular cloud, but rather by fragmentation of the dense filaments.  


\section{Can the IMF be measured?}
\label{sec:iMFmeas}


Even if dust obscuration and
reddening were not to matter, the IMF does not exist as a real
distribution function of initial stellar masses because the 
time does not exist when all the stars formed together would be found in a
star-forming region.  The reasons for the non-existence of the IMF in nature
are: (i) while the most massive stars are already evolving off the main
sequence having a substantial mass loss, some of the low mass stars may
not have been formed fully yet in the star-forming event; (ii) stars are
ejected even before the embedded cluster is finished forming
(e.g. \citealt{Kroupa18, WKJ18}), stellar mergers occur \citep{deMink+14, OK18}
and (iii) expulsion of residual gas unbinds
low mass stars from the very young cluster within a crossing time if
clusters form mass segregated \citep{Haghi15}.


Something which does not exist can also not be measured (``The IMF Unmeasurabilty Theorem'' of \citealt{Kroupa13}). But the IMF can be deduced from observations as a {\it mathematical hilfskonstrukt} \citep{KJ18}, to allow calculations and modelling of stellar populations. But in order to do so statistical corrections are necessary to account for the time evolving binary population in the embedded and exposed cluster \citep{Belloni+17,Belloni+18}, loss of stars through ejections \citep{Banerjee12b, OKP15, OK16}, mergers and gas expulsion, as well as the corrections for the changes in stellar mass through stellar and binary evolution \citep{Sana+12, Schneider+15} and energy-equipartition-driven stellar-dynamical evolution \citep{BM03,Baumgardt+08}.

\vspace{2mm} \centerline{ \fbox{\parbox{\columnwidth}{ {\sc
        The existence of the IMF:}\\
      The IMF is a mathematical hilfskonstrukt which does not have a
physical representation, but which is needed for initialising stellar
populations.
}}}  \vspace{2mm}


This is a reasonable description because the early evolutionary processes mentioned above take part largely on short time scales (e.g. the formation of the stellar population in an embedded cluster takes not longer than about one~Myr) such that when studying stellar populations on longer time-scales the non-physical nature of the IMF can be neglected. It is therefore possible to obtain an estimate of the complete sample of stars that formed together and to quantify their birth masses in order to construct the IMF as if this IMF were to represent a complete population of all the stars formed together at exactly the same time, for example in one embedded cluster (Sec.~\ref{sec:ecls}). When doing so for a pre-main sequence population, the flux observed from a young star needs to be converted to a mass, and this calculation needs to be done using pre-main sequence stellar evolutionary tracks which are increasingly uncertain for younger ages \citep{WT03}.

But it should be clear that the concept of
an IMF fails when considering the first Myr of a stellar population. 


\section{What is the shape of the IMF?} 
\label{sec:shape}


The most direct approach to quantify the shape of the IMF is to use resolved stellar populations and these are, by necessity, mostly restricted to the Galaxy.

One important ansatz to assess the shape of the IMF is to measure a representative and complete sample of stars near to the Sun and to calculate their birth masses to construct this Galactic-field average IMF (and to implicitly assume this composite IMF of the Solar neighbourhood to be the IMF). This approach has been followed by many teams and entails two general procedures (the reviews are listed in Section~\ref{sec:introd}) to assess the IMF for stars with $m\simless 1\,M_\odot$: (i) counting stars in a complete volume near to the Sun and (ii) taking pencil beam surveys through the Galactic disk. Approach~(i) has the advantage that the stars are close-by allowing their multiplicity to be better assessed and that trigonometric parallax measurements yield stellar distances directly. But the disadvantage is that the census of very low-mass stars becomes incomplete beyond about 5-10~pc distance from us \citep{Scalo86, Henry+18, Riedel+18}. Approach~(ii) has the advantage that the stellar census is significantly enlarged even at very low masses because telescope-integration times can be chosen to be long per field, but the disadvantage is that distances need to be inferred using photometric parallax and that companions in multiple stellar systems are likely not detected \citep{Kroupa95MF}.  In the course of this research the shape of the late-type stellar luminosity function was discovered to be universal and largely given by the stellar mass-luminosity relation, and unresolved binaries to play a significant role in establishing the difference between star counts under~(i) and~(ii).  A completely independent approach to the above star-count ones is to use microlensing time scales to assess the distribution of lensing masses. This yields constraints on the IMF deducible from main-sequence-field stars in the inner Galaxy which are well-consistent with those obtained from the star-counts under approaches~i and~ii \citep{Wegg+17}. This overall research effort has thus lead to the gwIMF of the Galactic thin disk being now constrained quite well. This Galactic-field average IMF (i.e., the gwIMF as constructed from the local Solar neighbourhood field) can be conveniently written as a three-part power-law form with $\alpha_1 \approx 1.3\pm0.3$ for $0.07\simless m/M_\odot<0.5$, $\alpha_2 \approx 2.3\pm 0.3$ for $0.5<m/M_\odot \le 1$ and $\alpha_3$ for $m>1\,M_\odot$ (Eq.~4.59 in \citealt{Kroupa13}).

The constraints for $m \simgreat 1\,M_\odot$ still stem largely from the analysis by \cite{Scalo86}. This monumental work explains in detail how the $m<1\,M_\odot$ and the $m>1\,M_\odot$ star counts need to be combined since the stellar life-times and thus their distributions in the Galactic thin disk differ greatly and are subject to changes in the SFR \citep{ES06}. 
Assuming the Galactic-field IMF to be continuous across $m\approx 1-10\,M_\odot$, it comes out that it is rather steep for $m\simgreat 1\,M_\odot$, with $\alpha_3=2.7\pm0.4$  being an approximative summary (Eq.~4.59 in \citealt{Kroupa13}). Such a steep Galactic-field IMF deduced from the wider (kpc-scale) Solar-neighbourhood star-counts has been more recently constrained by \cite{Mor+17, Mor+18} using  the Besan\c{c}on Galaxy Model fast approximate simulations technique, which combines photometric data with a spatial and kinematical model of the Galaxy yielding $\alpha_3={2.9}_{-0.2}^{+0.2}$ and $\alpha_3={3.7}_{-0.2}^{+0.2}$ depending on the used extinction map, with an increasing uncertainty for $m>4\,M_\odot$due to the extreme rarity of such stars. This steep index is also found by \cite{RJ15} using their forward modelling method, which also combines kinematical and photometric data with a model of the Milky Way. We note though that a single power-law description for $m>1\,M_\odot$ is only a very rough approximation since the field-IMF, as reconstructed from the star-counts, is curved and this curvature contains information on the competition of the rates with which stars were and are being born and are dying \citep{ES06} as well as on the individual star-formation events which combine to make the Galactic-field IMF (\citealt{Zonoozi+18},
see also Fig.~\ref{fig:IGIMFform} below).

The other major approach to assess the shape of the IMF is to target individual star clusters and/or OB associations in an attempt to infer the IMF within these. The advantage is that the stars have very nearly the same age, metallicity and distance. The problem is that the early and secular dynamical evolution of the clusters leads to a significant time-dependent change in the shape of the stellar mass function through very early stellar ejections, stellar evaporation (which depends on the time-evolution of the tidal field), break-up of binary systems over time as well as the odd stellar merger occurring (see Sec.~\ref{sec:IMFnature} above for more details).


As outlined in the reviews mentioned above, the general finding is that the IMFs reconstructed from the observed populations are consistent with the canonical IMF (Eq.~\ref{eq:imf}). Until recently there has not been a significant evidence for a variation of the IMF within the star-formation events in the Milky Way (but see Sec.~\ref{sec:IMFvar}) and the notion that the IMF equals the composite Galactic-field IMF (from the Solar neighbourhood ensemble of stars) was affirmed for stars with $m\simless 1\,M_\odot$.

\vspace{2mm} \centerline{ \fbox{\parbox{\columnwidth}{ {\sc The Canonical
        IMF:} ($m$ is in units of $M_\odot$, $k$ is the normalisation constant)
\[
\xi_{\rm BD} (m) = {k\over 3}
  \begin{array}{l@{\quad\quad\quad\quad\quad\quad\quad\quad,\quad}l}
   \left({m\over 0.07}\right)^{-0.3\pm0.4} &0.01 < m \simless 0.15,
  \end{array}
\]
\begin{equation}
\xi_\mathrm{star} (m) = k\left\{
  \begin{array}{l@{\quad,\quad}l@{\quad}}
   \left({m\over 0.07}\right)^{-1.3\pm0.3}  &0.07 < m \le 0.5,\\
   \left[\left({0.5\over 0.07}\right)^{-1.3\pm0.3}
       \right] \left({m\over 0.5}\right)^{-2.3\pm0.36} &0.5 < m \le 150.\\
  \end{array}\right.
\label{eq:imf}
\end{equation}
}}}
\vspace{2mm}

We note that the canonical IMF may be either described as a two-part-power-law function (Eq.~\ref{eq:imf} above, Eq.~4.55 in \citealt{Kroupa13}) or a log-normal form (between the hydrogen burning mass limit, $m_{\rm L}\approx 0.07\,M_\odot$, and about $1~M_\odot$) with a power-law extension above $\approx 1\,M_\odot$ (Eq.~4.56 in \citealt{Kroupa13}). The former (Eq.~\ref{eq:imf}) is convenient for use in computations and is adopted throughout this work. It can also be written as a multi-power-law form ($\propto m^{-\alpha_i}$) with $\alpha_1 \approx 1.3 \pm 0.3$ for $0.7 \simless m/M_\odot < 0.5$, $\alpha_2\approx 2.3 \pm 0.3$ for $0.5 \le m/M_\odot \le 1$ and $\alpha_3=\alpha_2$ for $m>1\,M_\odot$ \citep{Jerabkova+18}.  The canonical log-normal$+$power-law form is indistinguishable from the canonical two-part-power-law form.


But as already reported above, for more-massive stars, the field-star-counts consistently yield steeper IMFs than the canonical IMF. This difference in $\alpha_3$ as obtained from individual embedded or young clusters on the one hand side, and from the Galactic-field star counts, which are a result from the combination of many star-formation events on the other hand side, provides an important hint as to how the composite IMF (i.e. cIMF) emerges from the star-formation events in a galaxy (\citealt{KW03,Zonoozi+18}, see Sec.~\ref{sec:optimal}). The bar region of the inner Galaxy has been found to be consistent with the canonical IMF using Made to Measure and chemodynamical modelling \citep{Portail+17}, while there are some indications that the Galactic bulge may have formed with a top-heavy IMF in order to account for the metal-rich stars \citep{Ballero07a}.


Brown dwarfs (BDs) are not discussed here. They add negligible mass and are a by-product of star formation following their own IMF (as planets follow their own IMF, see Eq.~4.55 in \citealt{Kroupa13}) and binary pairing rules different to those of stars and typically they do not pair with stars \citep{Bate+02,Thies15}. The microlensing constraints from the inner Galaxy are consistent with these results \citep{Wegg+17}. Note that in Eq.~\ref{eq:imf} the BD IMF overlaps with the stellar IMF which comes about due to the star-like and BD-like formation processes overlapping \citep{Thies15}.

\section{What is the mathematical nature of the IMF?}
\label{sec:IMFnature}

From a complete co-eval ensemble of stars an IMF can be constructed, and
thus this IMF is a description of the distribution of initial stellar
masses. The physics of star formation determines the form of the IMF, but it also
determines which type of distribution function the IMF is. 

A turbulent-fragmentation process (Sec.~\ref{sec:ecls}) might imply the IMF to be a probability
distribution (PDF) function. A self-regulated growth
process might instead imply the IMF to be an optimal distribution function
(ODF), such that the distribution of stellar masses actually formed does not have
Poisson scatter \citep{Kroupa13}. Nature may well be in-between both extremes. For
example, it may be probabilistic to some degree (the butterfly effect:
two identical gas clouds which differ in some property by an
infinitesimal amount lead to two IMFs, the distance between which is,
in some metric, related to the Poisson uncertainty given by the number of
stars formed). Or it may be deterministic in the sense that the bulk properties
of the cloud (its temperature, density and chemical composition)
determine the shape of the distribution function. But if the butterfly
effect is minor, because the system is highly self-regulated and
insensitive to small differences in initial conditions, or because any physical state of the inter-stellar medium decays to a universal filamentary structure, then the ODF
description may be more appropriate. 

A major aspect of IMF research is 
thus not only to find the form of the IMF and its possible variation,
but also to understand which type of distribution function we are
dealing with. This is important for initialising stellar
populations in galaxies and the mathematical methods how to do so
differ significantly: random vs optimal sampling \citep{Kroupa13,Schulz+15,Applebaum+18}.

\vspace{2mm} \centerline{ \fbox{\parbox{\columnwidth}{ {\sc
        Mathematical nature of the IMF:}\\
      There are thus two main hypotheses as to the mathematical nature
      of the IMF: it is a PDF or an ODF. Each hypothesis can be tested against
      data to infer the valid nature that may lie between these two extreme descriptions.  }}}  \vspace{2mm}

\subsection{The IMF as a PDF}
\label{sec:PDF}

An often-used interpretation is the IMF to be a PDF
(e.g. \citealt{Elm97,Kroupa01,Cervino13a,Cervino13b}): when stars form in 
some system, their masses
appear randomly within the system, subject to being drawn from the
IMF.  This is the stochastic, or probabilistic, description of star
formation within a galaxy.  That is, the probability of finding a star
of mass $m$ in the interval $m_1$ to $m_2$ is
\begin{equation}
X_{12} = \int_{m_1}^{m_2} \xi_{\rm p}(m)\, dm,
\label{eq:PDF}
\end{equation}
where $0 \le X_{12} \le 1$ is uniformly distributed and $X_{12}=1$ for
$m_1 \approx 0.07\,M_\odot$ (the lower-mass limit, $m_{\rm L}$) and
$m_2= m_{\rm max*}$ which is some upper limit which may be infinite or
have a physical limit. Note also that the probability density distribution function is proportional to the IMF,
\begin{equation}
\xi_{\rm p}(m) \propto \xi(m),
\label{eq:propto}
\end{equation}
such that the integral from $m_{\rm L}$ to $m_{\rm max*}$ over the former yields~1 while over the latter it yields the number of stars in the population.

If the IMF is a PDF then any observation of a co-eval stellar
population will be subject to Poisson-differences such that a measured
IMF for population~A will differ from that measured for population~B,
with the difference, $\xi_{\rm A}(m)\,dm -\xi_{\rm B}(m)\,dm$, being
consistent with Poisson scatter for the sizes of the populations
(assuming no observational errors and all individual stars to be
observed). It is thus of much fundamental interest to search if such
variations are observed \citep{Elmegreen99, Kroupa01, Elmegreen04, Dib14, Ashworth+17}.

A prominent test of the hypothesis that the IMF is an unconstrained (scale-free) PDF is
to consider the masses of the most massive stars observed (Box Observational
Constraints I on p.~\pageref{box:obsI}). We would expect to see stars
weighing a few~$1000\,M_\odot$ in populous galaxies \citep{Elm00}.
But the observed most massive stars have $m\approx 300\,M_\odot$
\citep{Crowther10, Schneider+14}.
This suggests that the IMF may be a PDF but with the constraint
$m_{\rm max*}\simless 300\,M_\odot$ (more on this in the Box Observational Constraints I on p.~\pageref{box:obsI}).

It is possible to deduce that the stellar IMF varies without specifying a possible systematic variation with some physical property of the star-forming environment \citep{Dib14, Dib17}. But by taking account of all of the relevant biases, survey differences, observational uncertainties and also additional effects such as arising from patchy reddening in star-forming regions and, importantly, the very large uncertainties in calculating stellar masses from the observed fluxes for pre-main sequence stars \citep{WT03}, the general consensus has been reached that the IMF is invariant, at least for the stellar populations probed in the Milky Way \citep{Kroupa02b, Chabrier03,Bastian10, Kroupa13, Offner14}. More technically, {\it the hypothesis that the IMF is invariant and canonical cannot be discarded with sufficient confidence}, the variation of the IMF shape being found to be too small even compared to the expected Poisson scatter (Box Observational Constraints I on p.~\pageref{box:obsI}).

The notion also emerged in the literature that massive stars can form in isolation. Finding undisputed such cases would constitute a strong argument for the IMF being a PDF and for a stochastic description of star formation (Box Observational Constraints I on p.~\pageref{box:obsI}). Interesting is that the cases of presumed isolated massive star formation in the Milky Way, the Large and Small-Magellanic clouds have successively been shown to be untenable when improved observational data became available, and that meanwhile ever distant dwarf galaxies, such as the Sextans~A dwarf galaxy at about 1.3~Mpc, are now heralded as such cases in-proof \citep{Garcia+19}.  If it were true that star formation is stochastic, then the observationally deduced power-law indices of the IMF, $\alpha_i$, would vary according to the Poisson dispersion (even if the parent IMF were invariant, e.g. \citealt{Kroupa01}), since some O stars may be isolated, others would be in rare clusters with a deficit of low mass stars while other populous clusters would not have any O stars.

\begin{figure}
\centering
\begin{subfigure}[b]{0.58\textwidth}
  \includegraphics[width=1\linewidth]{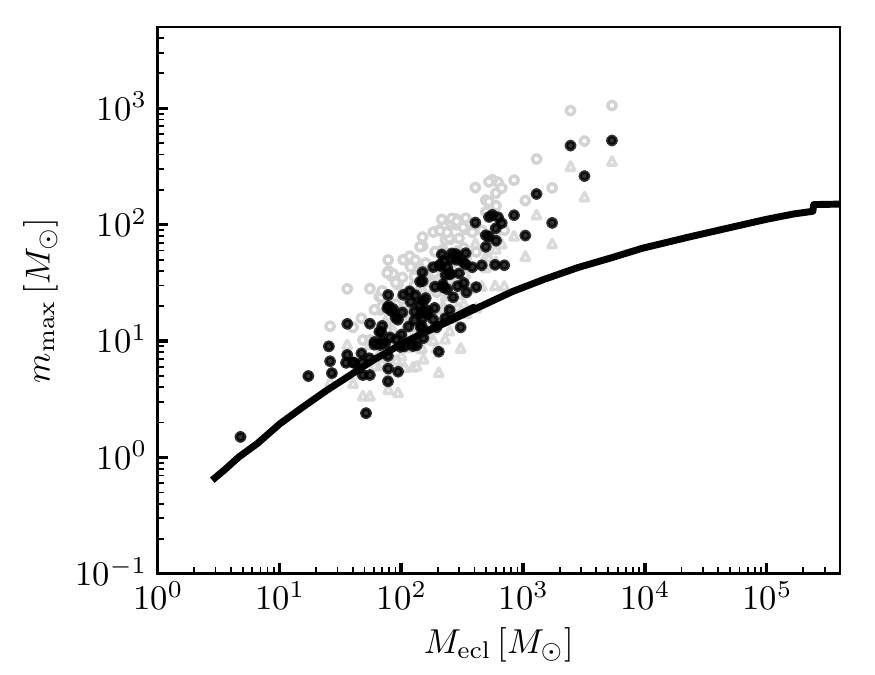}
   \caption{}
\end{subfigure}
\begin{subfigure}[b]{0.55\textwidth}
   \includegraphics[width=1\linewidth]{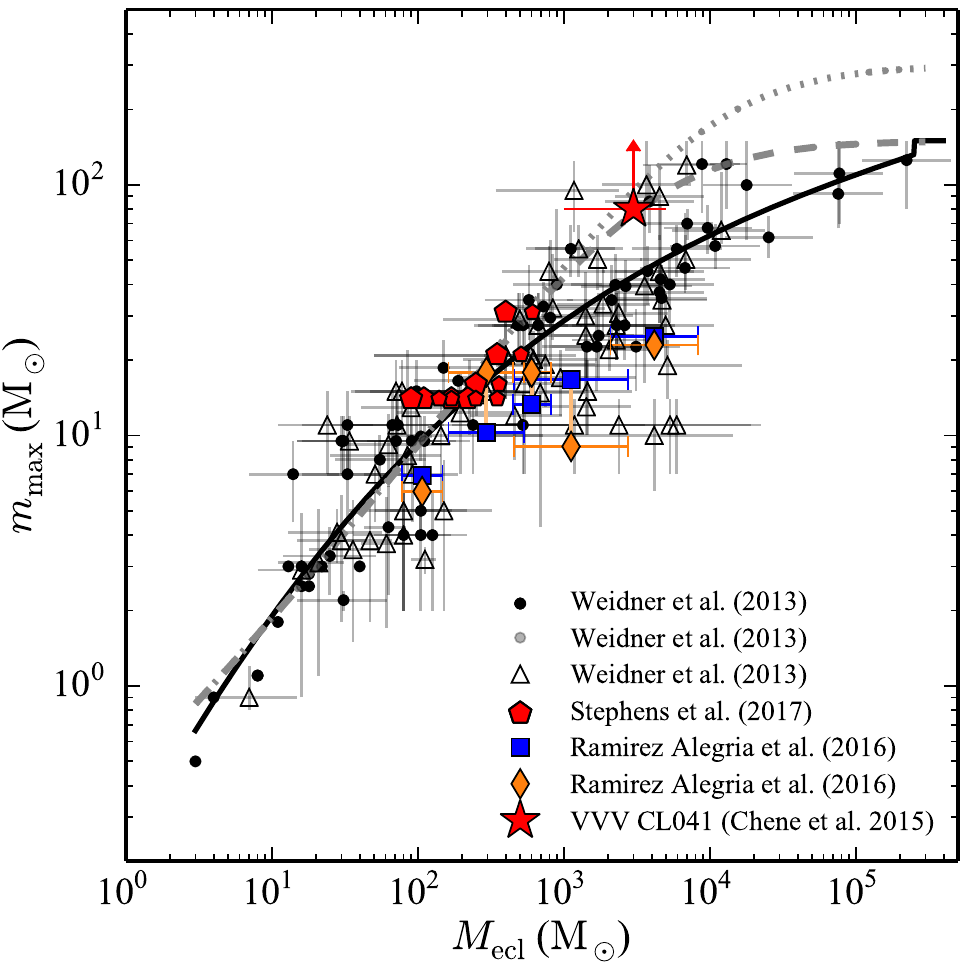}
   \caption{}
\end{subfigure}
\vspace{-2mm}
\caption [The $m_{\rm max}, M_{\rm ecl}$ data]{\small{ {\bf Panel a}: Data from the \cite{Lin+19} APEX/SABOCA survey of star-forming molecular cloud clumps. The measured most-massive sub-clumps with masses $m_{\rm scl} = m_{\rm max} / \epsilon$ are plotted as grey open circles in dependence of their hosting measured proto-cluster of molecular-gas mass $M_{\rm pcl}=M_{\rm ecl} / \epsilon$ assuming each proto cluster and clump has a star formation efficiency of $\epsilon=1/3$. The black circles assume each $m_{\rm max}>10\,M_\odot$ value fragments into two equal mass sub-cores and the open triangles assume fragmentation into three sub-cores. {\bf Panel b}: The $m_{\rm max}, M_{\rm ecl}$ stellar data and the $m_{\rm max}=WK(M_{\rm ecl})$ relation (eq.~1 in \citealt{Weidner13} depicted as the black solid line, shown also in panel~a).  The black circles are the small-error sub-sample data in the $<4\,$Myr old clusters used to obtain this relation.  The open triangles have large uncertainties and are excluded from the fit.  The red star is VVV~CL041 put forward by \cite{Chene+15} as not fitting the $m_{\rm max}=WK(M_{\rm ecl})$ relation. \cite{OK18} explain such deviant cases as being produced through mergers in the young compact embedded host cluster.  The blue squares and orange diamonds are two different $m_{\rm max}$ estimates from the survey of $<10\,$Myr-old clusters by \cite{Ramirez16} and the red pentagons (large for 1~Myr age and small for 2.5~Myr age) are the data from \cite{Stephens+17}. Note how the older objects (blue squares and orange diamonds) lie below the $m_{\rm max}=WK(M_{\rm ecl})$ relation, consistent with stellar evolution having removed the most massive stars.
    The dashed and dotted lines are the semi-analytical relation (Eq.~\ref{eq:mm} and~\ref{eq:Mecl}, from \citealt{Weidner06}) assuming $m_{\rm max*}=150\,M_\odot$ and $300\,M_\odot$, respectively.  Adapted from \cite{OK18}.}}
\label{fig:WK}
\end{figure}

\vspace{2mm} 
\begin{mdframed}
\label{box:obsI}
{\sc Observational Constraints I:}
This box lists some key observations relevant for testing the conjecture
that the IMF is a PDF.
\begin{itemize}
\item The most massive stars weigh $m_{\rm max*}\approx150\,M_\odot$ \citep{WK04,
    Figer05, OC05, Koen06, Jesus07}. \cite{Crowther10, Schneider+14} find evidence for the existence of  $m\approx 300\,M_\odot$ stars. These are consistent with
$m_{\rm max*}\approx 150\,M_\odot$ as a result of mergers in their
dense massive birth clusters \citep{Banerjee12}, but further study is needed to ascertain if the observed number of such super-canonical stars can be obtained through mergers and binary-star evolution in dense star-burst clusters (cf. \citealt{OK18}).
\item The small dispersion of observationally derived IMF power-law
  indices (which contain all the uncertainties and come from different
  teams with different observational procedures),
  $\alpha_3 = 2.36\pm 0.08$, above $2.5\,M_\odot$, compared to models
  which have no observational uncertainties but which include
  binaries, stellar dynamical processes and assume the IMF to be a PDF
  ($\alpha_3 = 2.20\pm 0.63$), suggest the IMF to be remarkably
  invariant, the different populations showing differences in the IMF which are smaller than the expected Poisson scatter if the IMF were a PDF \citep{Kroupa02b}. 
\item Most early-type stars supposed to have formed in isolation have
  been found to be propagating from some young cluster
  \citep{Gvaramadze12}, be in hitherto unknown compact clusters
  \citep{Stephens+17} or be explainable through the two-step ejection
  process \citep{Pflamm10}.
\item \cite{Hsu12} observe in the Orion~A cloud that the very young stellar
  populations formed in low-mass embedded clusters in its southern part
  significantly lack massive stars although statistically the same total number of stars (a few thousand) have
  formed as in the Orion Nebula Cluster in the northern part of the same
  cloud. If the IMF were to be a PDF, then both ensembles ought to have the same number of massive stars within the statistical uncertainties. Thus the IMF seems to differ in the two regions and massive stars appear to correlate with the mass in stars of the embedded clusters. 
  \item \cite{KM11, KM12} find, in their homogeneous survey of very young
  clusters, a pronounced correlation of the mass of the most massive star, $m_{\rm max}$,
  with the stellar mass of the embedded cluster, $M_{\rm ecl}$, for low mass systems (the data lie near the $m_{\rm max}=WK(M_{\rm ecl})$ relation, see the next point).
\item \cite{Stephens+17} performed a high resolution survey of
  previously thought isolated massive stars in the LMC in order to test their isolated nature.
  They found each of
  them to be contained in a compact massive stellar cluster of 1-3~Myr
  age. The $m_{\rm max}$ and $M_{\rm ecl}$ values are correlated (the data lie near the $m_{\rm max}=WK(M_{\rm ecl})$ relation, see the next point). 
  \item \cite{Ramirez16} find a correlation of the most-massive star with the cluster mass in the sample of VVV-survey young ($<10\,$Myr old) clusters.
 \item \cite{Weidner06, Weidner10, Weidner13} collate data on very young clusters from the literature, selecting the clusters to be pre-supernova age ($<4\,$Myr age being the
  only selection criterion) avoiding loss of stars through supernova
  explosions, and significantly limiting dynamical and stellar
  evolution affects. They find $m_{\rm max}$ to significantly
  correlate with $M_{\rm ecl}$, the dispersion of $m_{\rm max}$ values being largely consistent with the observational uncertainties such that intrinsic scatter appears to be small.  They calculate that the hypothesis that the IMF is a PDF is consistent with the data at the 0.1~\% confidence level. The possible existence of a physical $m_{\rm max}=WK(M_{\rm ecl})$ relation is discussed (Fig.~\ref{fig:WK}, see also fig.~1 in \citealt{Yan+17}, which compare all above data by \citealt{KM11, KM12, Weidner13, Ramirez16, Stephens+17}). It may be a consequence of a feedback-self-regulated growth process of stellar masses. Noteworthy is that the APEX/SABOCA survey of star-forming molecular cloud clumps by \cite{Lin+19} shows a similar correlation to exist between the most massive core mass and the proto-cluster clump mass (Fig.~\ref{fig:WK}). The flattening of this to the stellar relation above $m_{\rm max}\approx 10\,M_\odot$ may be due to resolution limiting the identification of less-massive sub-clumps in the former, additional fragmentation in the former and feedback self-regulation during the transformation of the proto-stellar clump to a star \citep{Beuther+07,ZY07}.
 \item Deviations from an assumed $m_{\rm max}=WK(M_{\rm ecl})$ relation, reported by e.g.
 \cite{Chene+15},
 can be understood naturally as arising from stellar mergers and ejections in and from embedded clusters \citep{OK12, OK18}.
  \item Observed very young embedded clusters (not older than a crossing time)
  have been found to be mass segregated \citep{Bontemps+10, Kirk14, Lane+16, Kirk16,
    Plunkett18}. The ALMA data by \cite{Plunkett18} are interesting, as
  they suggest that a proto-cluster is forming perfectly mass segregated \citep{Pavlik+19}. If
  this were to be the case then this suggests a high degree of
  regulation during the star-formation process with the individual
  stellar masses being a sensitive function of the core-density and thus gas reservoir (which varies radially in the embedded cluster),
  which most probably defines the accretion rate \citep{BD98}. Noteworthy is that this consideration appears to be consistent with the possible existence of a physical $m_{\rm max}=WK(M_{\rm ecl})$ relation noted above. Nbody models of the Orion Nebula Cluster ($<2.5\,$Myr old) also indicate it to be primordially mass segregated \citep{BD98}. Observationally, data
  on globular clusters also suggest them to have been born mass
  segregated \citep{Baumgardt+08, Haghi15}. If the IMF were a PDF then the stellar mass drawn randomly from the IMF ought not to correlate with the position of the star. 
\end{itemize}
 \end{mdframed}
 \vspace{2mm}

Given the collation in Box Observational Constraints~I (p.~\pageref{box:obsI}), can the IMF still be a PDF on an embedded cluster scale? One particular observational challenge addresses this question: a physical $m_{\rm max}=WK(M_{\rm ecl})$ relation should not exist and the $m_{\rm max}, M_{\rm ecl}$ data should be consistent with randomly drawing stars from the IMF if it is a PDF. Prior to the more recent observational surveys by \cite{KM11,KM12}, \cite{Ramirez16} and \cite{Stephens+17} mentioned in Box~I (p.~\pageref{box:obsI}), \cite{PG07, MC08, Cervino13a} addressed this problem using their own star-by-star data collations and statistical analysis negating that the relation exists.  \cite{Andrews13, Andrews14} also addressed this problem based on unresolved young star clusters in dwarf galaxies using broad-band photometry, spectral-energy-density (SED) modelling and H$\alpha$ flux measurements to conclude that this relation does not exist, therewith formulating a strong conclusion at $1\,\sigma$ significance.  Being part of this team, this conclusion is echoed in the review by \cite{Krum14}. For completeness, it is noted here that \cite{Weidner14} showed these same extragalactic data to be well consistent (within $1\,\sigma$ uncertainty) with the existence of a physical $m_{\rm max}=WK(M_{\rm ecl})$ relation, largely because the random sampling in connection with $m_{\rm max}$ being assumed to be a truncation limit was applied incorrectly in these and other publications (e.g. \citealt{Fumagalli11}, also part of the Krumholz effort), noting also that field-star contamination in the unresolved observations plays a role in randomising any SED.

Thus, for the time being we conclude that the existence of a physical $m_{\rm max}=WK(M_{\rm ecl})$ relation needs to be studied further but with great care, and we emphasise that if it were to exist then the implications for the gwIMF are major (Sec.~\ref{sec:optimal}).

\subsection{The IMF as an ODF}
\label{sec:ODF}

Independently of the possibility  that a physical $m_{\rm max}=WK(M_{\rm ecl})$ relation may or may not exist (Sec.~\ref{sec:PDF}), the
observed degree of mass segregation in extremely young embedded
clusters may also be seen as an indicator that a star's mass might not be 
random,  but that it instead may be governed by the properties (mostly the density) of
the molecular cloud core and thus gas reservoir within which it forms and from which it accretes.  The IMF may therefore be sensitively dependent on the physical conditions within the embedded cluster. An interpretation of the IMF as an ODF may thus be a useful consideration. 

If the mathematical nature of the IMF is to be understood in terms of an optimal distribution function, then how is such a description to look like? The data (Box Observational Constraints I on p.~\pageref{box:obsI}) suggest that $m_{\rm max}$ is determined by $M_{\rm ecl}$. If the star-formation efficiency of embedded cluster-forming cloud cores, \begin{equation} \epsilon = { M_{\rm ecl} \over M_{\rm ecl} + M_{\rm gas} }, \label{eq:SFE} \end{equation} is typically in the range $0.1-0.3$ \citep{Megeath16} then this dependency may arise if stars form through self-regulated accretion (on a timescale of $10^5\,$yr, \citealt{WT03,Duarte-Cabral+13, KristensenDunham18}) and a link between $m_{\rm max}$ and the total embedded-cluster forming cloud core can be made.  It becomes a density dependency if the birth half-mass radii, $r_{\rm h}$, of the embedded clusters are known\footnote{\label{foot:mkrad} This half-mass-radius--embedded-stellar-cluster-mass relation is derived by \cite{MK12} from the observed binary population in open and very young clusters and assuming the truncation of the binary-binding energy distribution at low values to be a measure of the densest evolution phase of the cluster. It constitutes a theoretical approximation to the initial state of an embedded cluster for Nbody modelling.}  \begin{equation} r_{\rm h}/{\rm pc} = 0.1 \, \left( M_{\rm ecl}/M_\odot \right)^{0.13}.  \label{eq:radius} \end{equation}

Concerning optimal sampling from the IMF, only the mass of the stellar
population formed together in one star-formation event, i.e. an
embedded cluster, is then relevant. A relation between $m_{\rm max}$ and
$M_{\rm ecl}$ is obtained from the following two equations
\begin{equation}
1 = \int_{m_{\rm max}}^{m_{\rm max*}} \xi(m)\,dm,
\label{eq:mm}
\end{equation}
with
\begin{equation}
M_{\rm ecl} - m_{\rm max}(M_{\rm ecl}) = \int_{m_{\rm L}}^{m_{\rm max}}
m\,\xi(m)\,dm,
\label{eq:Mecl}
\end{equation}
which allow solution for the two unknowns, $m_{\rm max}$ and the normalisation constant $k$, for the population which has the physical upper mass limit $m_{\rm max*}$ (Box Observational Constraints~I on p.~\pageref{box:obsI}).  Both equations are valid also if the IMF is a PDF and in this case $m_{\rm max}$ will be the average most massive star in an ensemble of embedded clusters \citep{KW03,OC05}. But the small dispersion of observed $m_{\rm max}$ values does not support this interpretation, as stated above in Box~I.

If, on the other hand, the IMF is an ODF, then Eq.~\ref{eq:mm}--\ref{eq:Mecl}
lead directly to a good representation of the empirical
$m_{\rm max}, M_{\rm ecl}$ data (dotted and dashed lines in Fig.~\ref{fig:WK}, 
\citealt{Weidner13}). In order to ensure
a small or negligible scatter of the $m_{\rm max}$ values, ordered
sampling has been invented \citep{Weidner06}, according to which
$M_{\rm ecl}$ is iteratively solved for with increasing number of
stars to minimise the difference to the observed
$m_{\rm max}={\rm WK}(M_{\rm ecl})$ function. A more elegant method to
efficiently sample stars from the IMF such that this function is
obeyed is given by optimal sampling \citep{Kroupa13, Schulz+15,
  Yan+17}. In optimal sampling (for an improved algorithm see \citealt{Schulz+15}), 
starting with a given $M_{\rm ecl}$
and thus $m_{\rm max}$, a sequence of stellar masses is generated step
by step such that a given $M_{\rm ecl}$ has always exactly the same sequence of
stars. This latter procedure is possible for whichever function
$m_{\rm max} = {\rm WK}(M_{\rm ecl})$ is adopted. One possibility is
given by solving Eqs~\ref{eq:mm} and~\ref{eq:Mecl}, but
physically-motivated functions may also be used
(e.g. \citealt{Weidner10} find evidence for a possible flattening of
WK$(M_{\rm ecl})$ which may be due to feedback regulation in the
formation of $m_{\rm max} \simgreat 25\,M_\odot$ stars).

An optimally-sampled stellar population has no Poisson scatter and thus fulfils two important requirements: it is consistent with the observed small dispersion of $\alpha_3$ values noted above in Box~I (p.~\pageref{box:obsI}) and it is consistent with the small dispersion of observed $m_{\rm max}$ values.\footnote{\label{foot:SLUG}Approaches which adopt the $m_{\rm max}={\rm WK}(M_{\rm ecl})$ relation but sample stars randomly from the IMF (i.e. which interpret it to be a constrained PDF) would lead to, on average, too small $m_{\rm max}$ values for each $M_{\rm ecl}$ such that the respective authors are led to erroneously conclude the WK-relation to be excluded, given unresolved data (e.g. \citealt{Andrews13, Andrews14}). These analysis-shortcomings are elaborated on by \cite{Weidner14}.}  Dynamically stable perfectly mass-segregated embedded clusters can then be constructed readily using the methods introduced in Bonn by \cite{Subr+08, Baumgardt+08}, in order to fulfil the observational constraint that very young embedded clusters are mass segregated. These methods assume the stellar masses to be radially arranged according to their binding energy.

Summarising, the observational  evidence suggests:

\vspace{2mm} \centerline{ \fbox{\parbox{\columnwidth}{
      The IMF may be an ODF rather than a PDF.
}}}  \vspace{2mm}

But more research is needed to further test this assertion. 

\subsection{Emergence of apparent stochasticity}
\label{sec:emergence}

On stating the above, we note the following: Once a perfect (or highly constrained) system of stars is generated (or born), it will evolve towards what appears to be a stochastic version of itself.  This follows from the second law of thermodynamics and comes about naturally because stars are ejected, binaries dissolve and stars merge and age. The Nbody computations of \cite{OKP15, OK16} show a large dynamical activity with the very young binary-rich clusters spurting out stars of all masses within fractions of a Myr.\footnote{\label{foot:oh} Movies
  based on these publications are available on youtube: "Dynamical
  ejection of massive stars from a young star cluster" by Seungkyung
  Oh.}
  
A molecular cloud which forms all its stars in a population of embedded clusters with different masses over a few Myr will thus always lead to a distribution of isolated stars, expanded and already dissolved older clusters and a few deeply embedded still forming embedded clusters with a complex velocity field which also depends on whether the whole cloud was contracting or expanding (Sec.~\ref{sec:ecls}).  Observationally, this activity leaves, in association with observational measurement uncertainties, a population of stars which appears non-optimal and even with an age, spin and chemical spread \citep{Kroupa95,Kroupa95a,Weidner09,MKO11,OK12, OKP15, OK16, OK18} .

The problem, to which degree an apparent probabilistic quality emerges from an
optimally mono (age, metallicity, spin) case may be an interesting
research project worthy of study.

\section{Does the IMF vary?}
\label{sec:IMFvar}

Having discussed above the question of whether the IMF is a PDF or an
ODF, another problem to deal with is whether the shape of the
distribution function varies systematically with the physical
conditions of star formation.

As discussed in Sec.~\ref{sec:shape} the consensus has been that the IMF is an invariant PDF, possibly with an upper stellar mass limit near $150\,M_\odot$ to $300\,M_\odot$. This was taken to be the case in most studies requiring the stellar populations in star clusters and galaxies. But, as discussed at some length in \cite{Kroupa13}, this consensus may deem unnatural, because different physical conditions should lead to different distributions of stellar masses. The most elementary and thus fundamental and solid arguments, based on the Jeans fragmentation scale \citep{Larson98} on the one hand side, and on self regulation on the other \citep{AF96, MatznerMcKee00, Federrath+14, Federrath15}, lead to the same expectation of how the IMF should change with temperature and metallicity of the gas cloud from which the stars form. Additional physical processes, such as the coagulation of contracting pre-stellar cores in very high-density star-forming regions (\citealt{Dib07}, see also \citealt{ZY07} and \citealt{Krum15} for reviews on the formation of massive stars) and the heating of the molecular gas though an elevated flux of supernova-generated cosmic rays in star-burst regions \citep{Papadopoulos13}, are likely to play a role in establishing a systematic variation of the IMF. All these broad physical arguments lead to the following theoretical expectation:

\vspace{2mm} \centerline{ \fbox{\parbox{\columnwidth}{ {\sc The
        theoretically expected variation of the IMF}:\\
      The IMF should become systematically top-heavy with increasing density, increasing temperature and decreasing metallicity. The empirical boundary condition is that, for the star-formation conditions evident in the present-day Milky Way, the IMF should be close to canonical.  This variation may involve $\alpha_3$ decreasing (the IMF becoming more top-heavy) for decreasing metallicity and increasing density,
      and for $\alpha_1$ and $\alpha_2$ to increase (the IMF becoming
      bottom-heavy) for increasing metallicity, but a robust
      prediction as to how the shape of the IMF should vary with
      conditions does not exist to-date.  }}}  \vspace{2mm}

Perhaps the first suggestion, based on direct star counts, as to a
variation of the IMF with metallicity, has been described in
\cite{Kroupa01, Kroupa02}. Here it was pointed out that, taken at face
value, the available resolved stellar populations indicate a
systematic flattening of $\alpha_1$ and of $\alpha_2$ with decreasing
metallicity valid for [Fe/H]$\simgreat -2$.  If such a
systematic effect is present, then for $m < 1\,M_\odot$ ($i=1,2$) and
to first order,
\begin{equation}
\alpha_i \approx 1.3 + \Delta\alpha\, {\rm log}_{10}(Z/Z_\odot),
\label{eq:lowmassvar}
\end{equation}
with $\Delta\alpha \approx0.5$ \citep{Kroupa01} (here writing the dependent on metallicity, $Z$, rather than the iron abundance).  The evidence was limited to late-type stars because most of the populations at disposal were old. Concerning massive stars, in a great observational effort, \cite{Massey03} demonstrated a high degree of invariance of $\alpha_3$ on density and metallicity for young populations of stars in the Milky Way, the Large and Small Magellanic Clouds.  More recently, due to improved data reaching into more extreme star-forming environments, explicit evidence for a systematic variation of $\alpha_3$ with physical conditions may have begun to emerge on star-cluster scales (Box Observational Constraints II).

\vspace{2mm} 
\begin{mdframed}
{\sc Observational Constraints II}:
\label{box:obsII}
\begin{itemize}
\item The high dynamical mass to light ratios of many ultra compact
  dwarf (UCD) galaxies, which have present-day stellar masses larger
  than $10^6\,M_\odot$, can be understood if they formed under high
  density, $\rho$, with $\alpha_3$ becoming smaller (more
  top-heavy) with increasing UCD mass (thus increasing $\rho$), such that a
  larger fraction of their initial mass ends up in stellar 
  remnants \citep{Dab09, Jerabkova+17}.
\item The high X-ray luminosity of many UCDs can be understood if they
  have a larger fraction of low-mass X-ray binaries which need a
  larger fraction of neutron stars and stellar black holes. The data
  suggest $\alpha_3$ to become smaller with increasing UCD mass and
  thus $\rho$ \citep{Dab12}.
\item The above two points lead to similar constraints on
  how $\alpha_3$ varies with $\rho$. This is an important
  consistency check. The systematic variation of $\alpha_3$ with
  star-formation density implies significant expansion of the UCDs
  with time due to mass loss from expelled residual gas and from
  stellar evolution \citep{Dab10}. This work demonstrates that the
  solutions for $\alpha_3$ also lead to stellar-dynamical solutions to
  the observed UCDs (they do not dissolve due to too much mass loss).  
\item Combining the information on an observed
  systematic deficit of low mass stars in present-day
  stellar mass functions in observed globular clusters (GCs) with their
  present-day radii and metallicities, with stellar-dynamical models
  which are initially mass segregated and allow for expulsion of
  residual gas and thus expansion of the young GCs and with the above constraints
  on $\alpha_3(\rho)$, yields the description of how
  $\alpha_3$ depends on $\rho$ and $Z$ given by
  Eq.~\ref{eq:toph_master} below from \cite{Marks12} (see also \citealt{Kroupa13}).
\item
The observed star clusters in M31 are consistent with the dependency
formulated as Eq.~\ref{eq:toph_master} \citep{Zonoozi16, Haghi17}.
\item The young ($\approx1\,$Myr), compact (half-mass radius
  $r_h\approx 1\,$pc) $M_{\rm st}\approx10^5\,M_\odot$-heavy star-burst cluster
  R136 in the 30~Dor star-forming region in the Large Magellanic Cloud
  has been ejecting its massive stars efficiently. Adding these back
  statistically into the star-counts in the cluster yields a top-heavy
  IMF \citep{Banerjee12b}.
\item
The stellar census in the massive star-forming region 30~Dor in the
Large Magellanic Cloud has been found to have a top-heavy IMF
\citep{Schneider18}, confirming the \cite{Banerjee12b} prediction.
\item Deep adaptive optics imaging of the Magellanic Bridge Cluster
  NGC 796 reveals a top-heavy IMF \citep{Kalari18}.
\item Using multi-epoch Hubble Space Telescope observations of the
  Arches cluster ($2-4\,$Myr old,
  $M_{\rm ecl} \approx 4-6 \times 10^4\,M_\odot$) near the Galactic Centre,
  \cite{Hosek18} find the IMF to be top-heavy.
\end{itemize}
\end{mdframed}
\vspace{2mm}

\vspace{2mm} 
\begin{mdframed}
{\sc The
        Empirical IMF Dependence on Density and Metallicity -- the
        initial conditions of stellar populations}: Resolved stellar
      populations largely show an invariant IMF (Eq.~\ref{eq:imf}), but for
      star-formation rate densities
      $SFRD \simgreat 0.1\,M_\odot$/(yr~pc$^3$) in embedded clusters,
      the IMF appears to become top-heavy \citep{Marks12}. The dependence of
      $\alpha_3$ on cluster-forming cloud density, $\rho$, (stars plus
      gas) and $Z$ can be parametrised as
\begin{equation}
          \begin{array}{l@{\;\;}ll@{\;\;}l}
\alpha_3 = \alpha_2,  
      &m> 1\,M_\odot  \quad \quad       &x' < -0.87, \\
\alpha_3 = -0.41 \times x' + 1.94,
      &m> 1\,M_\odot \quad \quad         &x'\ge -0.87,\\[3mm]
 \end{array}
         \label{eq:toph_master}
\end{equation}
\begin{equation}
x' = -0.14 \,{\rm log}_{10}(Z/Z_\odot) + 0.99\, {\rm log}_{10}\left( \rho/ \left(10^6
    \, M_\odot \, {\rm pc}^{-3}\right) \right). 
\end{equation}
A possible variation of the IMF for late-type stars with $Z$ is
suggested by Eq.~\ref{eq:lowmassvar}.  In Eq.~\ref{eq:toph_master},
the density (stars plus gas at a mathematical time, $t_{\rm math}$, 
when all the stars
have been born and the remaining gas has not yet been expelled) of the
star-forming molecular cloud clump is
\begin{equation}
\rho = {3 \, \left(M_{\rm ecl}/2\right) \over 4\, \pi \epsilon \, r_h^3},
\label{eq:embeddedcluster}
\end{equation}
where the star-formation efficiency, $\epsilon$, is Eq.~\ref{eq:SFE} and $r_{\rm h}$ is Eq.~\ref{eq:radius} being the half-mass radius of the clump. Note that this clump, or rather embedded cluster, is a mathematical hilfskonstrukt which in reality does not exist because the alluded to mathematical time does not exist. In reality the stars of the embedded cluster form over about 1~Myr and the gas is blown out over a time which may be approximated by the sound speed of ionised gas ($\simgreat 10\,$km/s, e.g. the extremely young Treasure Chest cluster, \citealt{Smith+05}, and the massive star-burst clusters in the Antennae galaxies, \citealt{Whitmore+99, Zhang+01}) and $r_h$.  The stars form in filaments rather than spherical Plummer models.  Nevertheless, by dynamical equivalence, any two dynamical structures (e.g. the Plummer model and a filament) lead to comparable dynamical processing of the stars formed within them if the two structures are dynamically equivalent \citep{Kroupa95,
  Belloni+18}.  By conservation of angular momentum, stars need to
form in multiple systems, whereby most form as binaries \citep{GK05}
with well defined orbital parameter distributions and internal (eigen)
evolutionary processes \citep{Kroupa95a, Belloni+17}.  But at the
extreme density needed for the top-heavy IMF regime, our knowledge of
the processes within the forming cluster remain poor. Formally, in
this regime, the crossing time is shorter than the formation time of a
single proto star ($\approx 10^5\,$yr).
\label{box:toph_master}
\end{mdframed} 
\vspace{2mm}

\begin{figure}
\begin{center}
\scalebox{1.0}{\includegraphics{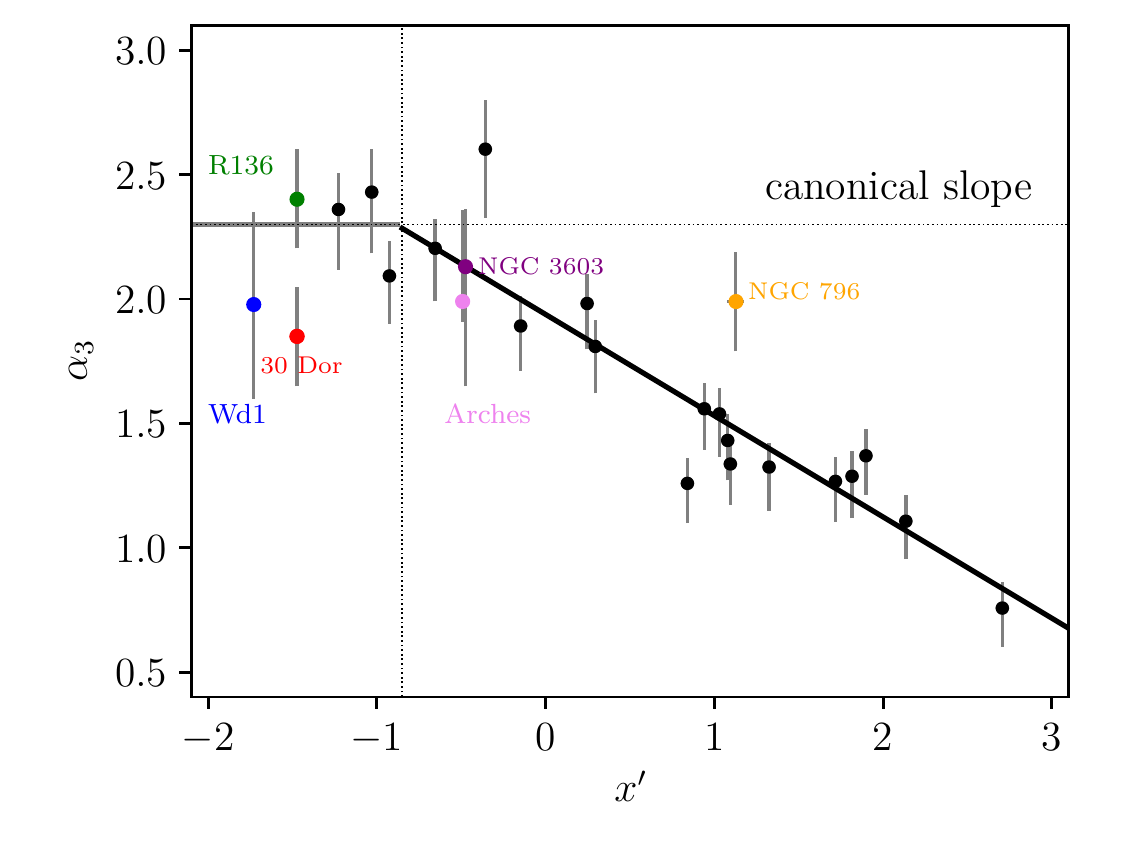}}
\vspace{-6mm}
\caption[$\alpha_3$ as a function of density and metallicity]{\small{
    The variation of the upper end of the stellar IMF with metallicity and cloud
    density (solid line, Eq.~\ref{eq:toph_master}) as
    deduced from deep observations of MW GCs using a
    principal-component-type analysis \citep{Marks12}. The recent observational determination of top-heavy IMFs in two metal-poor very young clusters (30Dor: \citealt{Schneider18}, NGC796: \citealt{Kalari18}) and the recent Arches data \citep{Hosek18} are included with the data shown here.
        Adapted from \cite{Marks12}. 
 }}
\label{fig:toph_master}
\end{center}
\end{figure}

In conclusion, high-quality star-count data appear to 
suggest a systematic variation of the IMF. Eq~\ref{eq:toph_master}
and~\ref{eq:lowmassvar} are possible quantifications of this
variation with density and metallicity of the star-forming gas
cloud. The above observational findings are compared in
Fig.~\ref{fig:toph_master}.  This variation is consistent with the
qualitative expectations from star-formation theory noted at the
beginning of this section. Extending the analysis to unresolved young clusters 
in other galaxies will prove to be an important consistency test whether this 
formulation of the variation is a good approximation (e.g. \citealt{Ashworth+17}).

If true, this IMF variation has important
implications for galaxy-wide IMFs, and also impinges on the
interpretation of the IMF as a PDF. If the IMF varies as suggested
here and if it is to be a PDF, then it can only be a constrained PDF,
because, before beginning the stochastic drawing process, the
functional form of the IMF needs to be calculated based on the
properties of the embedded cluster, which however itself is supposedly
drawn stochastically from an embedded cluster mass function. It
remains to be seen, perhaps more from a philosophical point of view,
if and how a PDF interpretation of the IMF is satisfyingly compatible
with a systematic variation of the IMF given by the physical
conditions at star formation, since one would, on the one hand side, be
resorting to a stochastic process, but on the other hand side, also to
to a process driven by physical conditions. 

\section{Is the IMF of a simple stellar population equal to that
  of a composite population?}

Consider a single star-forming region and a region containing many
star-forming regions, such as a whole galaxy. We can then define the
IMF of the larger region to be the composite IMF, cIMF.  Considering a
whole galaxy, the cIMF becomes equal to the galaxy-wide IMF
(gwIMF, \citealt{KW03}). 

\vspace{2mm} \centerline{ \fbox{\parbox{\columnwidth}{ {\sc The
        composite IMF Statement}:\\
    Logically, the cIMF is the sum of all IMFs assembled
throughout the region within some time $\delta t$.
\label{box:compositeIMF} }}}  \vspace{2mm}

From Eq.~\ref{eq:PDF} and~\ref{eq:propto}: for any pair $m_1, m_2$, $X_{12}$ 
(calculated assuming $\xi_{\rm p}(m)$ is the PDF corresponding to the IMF)
will be equal to another $X_{12}$ (calculated assuming $\xi_{\rm p}(m)$ is the PDF
corresponding to the cIMF) if and only if both PDFs are equal because
$m_1$ and $m_2$ are arbitrary. This is the fundamental composite IMF-PDF theorem (see also \citealt{KJ18}):

\vspace{2mm} \centerline{ \fbox{\parbox{\columnwidth}{ {\sc The
        composite IMF-PDF theorem}:\\
     If the IMF is a PDF then the IMF of all stars formed in a galaxy
within a short time interval $\delta t$, the galaxy-wide IMF (gwIMF),
is equal to the IMF. 
\label{box:comp_theorem}
 }}}  \vspace{2mm}

This theorem has been applied in most studies of star-formation in
galaxies. In particular, the Kennicutt SFR--H$\alpha$ tracer rests on
assuming this theorem holds \citep{Kennicutt98, Jerabkova+18}. It
assumes that any sum of IMFs will yield the same PDF which is true if
all individual IMFs are fully sampled over all possible stellar
masses.  But is this formulation applicable to real galaxies?

Solving this problem is of paramount importance for measuring SFRs. If the IMF is invariant and equal to the gwIMF then the tracer can be calibrated simply and becomes invariant to galaxy type and mass. For example, the number of recombination photons per unit time, i. e. the H$\alpha$ flux, which is proportional to the number per unit time of ionising photons emitted by a fresh stellar population, depends on the number of the ionising (i.e. massive, $m\simgreat 10\,M_\odot$) stars in the young population. If the gwIMF is invariant and equal to the IMF and both are fully sampled up to $m_{\rm max*}$, then the H$\alpha$ flux is a direct measure of the total mass of stars born per unit time \citep{Kennicutt98}. For this calibration to remain valid, the IMF needs to be an invariant PDF.  In star-by-star models of galaxies this interpretation is equivalent to a stochastic description of star formation without constraints.

\subsection{Clustered star formation, but star formation is stochastic}
\label{sec:stoch}

If stars form in embedded clusters (e.g. \citealt{Hopkins13a}) which are randomly distributed throughout a galaxy and which follow a distribution of masses, i.e. an ECMF (Eq.~\ref{eq:ECMF} below), then two possibilities arise:

\begin{enumerate}

\item \label{point:purePDF} The cluster mass in stars, $M_{\rm ecl}$, plays no role and
  stars appear randomly filling a cluster to a pre-specified maximum
  number of stars, $N_{\rm ecl}$. This yields a distribution of
  clusters with different numbers of stars, and thus to an
  ECMF. Within each cluster the IMF is sampled randomly without
  constraints, apart from the condition that stars have masses
  $m\le m_{\rm max*}$.  In
  this case the composite IMF-PDF theorem (p.~\pageref{box:comp_theorem}) holds, making gwIMF equal in
  form to the IMF and any tracer of star formation rate (e.g. the
  Kennicutt SFR--H$\alpha$ tracer, \citealt{Kennicutt98}) will yield a
  correct value of the SFR upon measurement of the tracer (e.g. the
  H$\alpha$ flux) subject to stochastic variations.  
  This interpretation implies that a ``cluster'' may
  consist of one massive star only. Thus, isolated massive stars 
  would pose an important argument for this purely
  stochastic approach to galaxy evolution. This possibility implies
  that a $m_{\rm max}-N_{\rm ecl}$ relation does not exist. A
  $m_{\rm max} - M_{\rm ecl}$ relation emerges but has a large
  scatter consistent with random drawing from the IMF \citep{MC08}.  An issue with
  this possibility is that the primary variable, $N_{\rm ecl}$, needs to be interpreted as a physical 
  parameter.

  If nature were to follow this mathematical recipe, then the measured SFRs, using the Kennicutt SFR-H$\alpha$ tracer for example, 
  will appropriately assess the true SFRs. Galaxies with very small
  SFRs will show a dispersion of SFRs which increases with decreasing
  SFR as a result of stochasticity.

\item \label{point:constrPDF} Considering $M_{\rm ecl}$ to be the primary variable, and choosing it randomly from the ECMF, the embedded cluster is filled with stars randomly from the IMF until
  the mass of the stellar population matches $M_{\rm ecl}$. This is
  constrained random sampling \citep{Weidner06, Weidner10, Weidner13}.
  If the ECMF contains clusters with $M_{\rm ecl}<m_{\rm max*}$ then
  in these clusters, the stellar population will lack the
  most massive stars. Such clusters are undersampling the IMF at large
  stellar masses. The deficit arises because a cluster of mass
  $M_{\rm ecl} < m_{\rm max*}$ cannot contain a star weighing more
  than the cluster.  The whole population of all embedded clusters
  will thus have a systematic deficit of the most massive stars,
  i.e. the gwIMF will be under-sampled at high stellar masses and will
  not equal the IMF. This possibility implies a
  $m_{\rm max}-M_{\rm ecl}$ relation which has a large scatter, since
  there may exist a cluster of mass $M_{\rm ecl}$ being composed
  off one star of mass $m=M_{\rm ecl}$. The existence of
  isolated massive stars is if paramount importance for this
  possibility to be valid. 

This implies that any tracer of ionising stars, calibrated assuming
the first possibility above, will systematically underestimate the
true SFR of the region or galaxy in cases when under-sampled clusters
are involved. This case has been extensively studied in the 
SLUG approach \citep{daSilva12, daSilva14}. 

If nature were to follow this mathematical recipe, then the measured SFRs, using the Kennicutt SFR-H$\alpha$ tracer for example, will deviate systematically towards smaller-than-$SFR_{\rm true}$-values at low SFRs. Galaxies with very small SFRs will show a dispersion of SFRs which increases with decreasing SFR as a result of stochasticity (fig.~3 in \citealt{daSilva14}).

\end{enumerate}

\subsection{Clustered star formation, but star formation is
  optimal}
\label{sec:optimal}

The above two stochastic approaches are consistent with the notion that galaxies have a stochastic history which comes about from the need of a very large number of mergers to build-up Milky-Way-class disk galaxies in the standard LCDM cosmological model. The observed simplicity of galaxies \citep{Disney08} and lack of evidence for an intrinsic dispersion in the baryonic-Tully-Fischer relation \citep{McGaugh05, McGaugh12}, in the mass-discrepancy relation \citep{McGaugh04} and in the radial-acceleration relation \citep{McGaugh+2016, Lelli+17} as well as the existence of a tight and well defined main sequence of galaxies over different red-shifts (\citealt{Speagle14}), indicate that the dynamical structure and the star-formation behaviour of galaxies may follow precise rules and that the expected stochasticity may be absent \citep{Disney08}.

Observations have shown that stars form in molecular clouds (Sec.~\ref{sec:ecls}). This
suggests that if the IMF were to be a PDF, then this PDF should be
subject to constraints: stars do not form at arbitrary positions
within a galaxy, they do so only where the conditions allow their
formation. Since the conditions and properties of molecular clouds
change with position in a galaxy (notably, inner region versus the far
outer region, \citealt{FK10, HD15}) the possibility might be given that the
constraints subjecting the PDF also depend on position within the
galaxy and within molecular clouds (see Box Observational Constraints III on p.~\pageref{box:obsIII}).

\vspace{2mm} 
\begin{mdframed}
 {\sc Box Observational Constraints III}:
\label{box:obsIII}
\begin{itemize}
\item Stars do not form randomly throughout molecular clouds but in embedded clusters (Sec.~\ref{sec:ecls}).
The observed distributed population is explainable through the dynamical activity of the forming embedded
clusters (Sec.~\ref{sec:emergence}).
\item The mixture of embedded
clusters determines the cIMF of a region in the molecular cloud
\citep{Hsu12}. According to the discussion in Sec.~\ref{sec:PDF}, stars
form in filaments which combine to embedded clusters which are
mass-segregated (Box Observational Constraints I on p.~\pageref{box:obsI}).
\item The distribution of young star-cluster masses shows a radial
  gradient with the most-massive cluster within a radial annulus being
  smaller at larger galactocentric distance in the disk galaxy M33 for
  example \citep{Pflamm13}. This is a result of the exponentially
  declining surface mass-density of gas in a disk galaxy.
\item A decreasing cluster-mass with galactocentric-distance relation is also found in the interacting interacting LIRG Arp 299 system as a result of the gas density distribution \citep{Rand18}.
  \item The formation of the embedded most massive stellar sources shows a Galactocentric radially decreasing trend in the Milky Way
    \citep{Urquhart+14} reminiscent of the above M33 result.
  \item Late-type galaxies show a pronounced correlation between their
  SFR and their most massive very young cluster, the $M_{\rm
    ecl,max}-SFR$ relation \citep{Weidner04}. The dispersion of the data is
  consistent with observational uncertainty, and \cite{Rand13} point
  out that their own data imply that the dispersion is not consistent
  with stochastic scatter by being too small.
\end{itemize}
 \end{mdframed}
 \vspace{2mm}

Is there thus an alternative to the above two stochastic approaches
(Sec. \ref{sec:stoch}) for describing stellar populations in galaxies?
Is it possible to derive a theory which has no intrinsic scatter, taking the observed 
simplicity of galaxies as a motivation? Even in such a theory, 
an observable dispersion would arise naturally  (Sec.~\ref{sec:emergence}), 
and some intrinsic dispersion can always
be added if needed. It might be educational to develop a theory which
has no intrinsic scatter to test how applicable this perhaps extreme
description may be.

\vspace{2mm} \centerline{ \fbox{\parbox{\columnwidth}{
If successful, we will have uncovered the laws of
nature which describe how the interstellar medium in a galaxy
transforms into a new stellar population and at which rate it does
so.  }}}  \vspace{2mm}

One possibility is to begin with the rules (we may also refer to them
as axioms, cf. \citealt{Recchi15, Yan+17}) deduced from observations of star formation within nearby
molecular clouds and from very young stellar populations in embedded
and older star clusters (e.g. Eqs~\ref{eq:lowmassvar}
and~\ref{eq:toph_master}, Observational Constrains I--III), assume these
rules are valid independently of cosmological epoch and calculate how
they imply what galaxies with different properties ought to look like
in their star formation behaviour. This approach has the advantage
that it (by construction) fulfils all observational constraints (as
outlined in Observational Constraints I--III), and that it is, in particular,
consistent with the stellar mass functions in the observed star
clusters and the Galaxy. These observations suggest that star
formation may be significantly regulated, possibly following precise
and clear rules.

\vspace{2mm} \centerline{ \fbox{\parbox{\columnwidth}{ 
     In addition to the two above stochastic approaches
(Sec.~\ref{sec:stoch}), we may therefore entertain a model in which
star-formation is entirely optimal on every scale. 
 }}}  \vspace{2mm}

Once we know how to compute such a model, it is clear that it becomes completely predictive. Some scatter in observable properties then enters via physical processes (e.g. galaxy--galaxy encounters which change the mass distribution and SFR over some time), other physical processes (Sec.~\ref{sec:emergence}) and measurement errors. The question is now how to construct such a deterministic, optimal, model?

We thus begin with ``The composite IMF statement'' (p.~\pageref{box:compositeIMF}): If each
embedded cluster\footnote{An embedded cluster can be constructed to be
  optimal by it being mass segregated initially and by it following
  the $m_{\rm max}-M_{\rm ecl}$ relation \citep{Pavlik+19}. Initially mass segregated
  models in dynamical equilibrium, such that the mass segregation
  would persist in collision-less systems, can be constructed using the
  methods developed in Bonn, namely by energy-stratifying the binary
  systems in the clusters  \citep{Subr+08, Baumgardt+08}. The embedded clusters can be
  optimally distributed throughout the galaxy by associating the most
  massive cluster at any galactocentric radius with the local gas
  surface density \citep{PAK08}.}  spawns a population of stars (this takes $\simless 1\,$Myr) with combined mass $M_{\rm ecl}$ per cluster, then the gwIMF becomes an integral over all such embedded clusters within the galaxy, yielding the integrated galaxy-wide initial mass function (IGIMF, Eq.~\ref{eq:igimf_t}; \citealt{KW03, Weidner05, Yan+17, Jerabkova+18}). A clue that this may be an interesting avenue to investigate is provided by the galactic-field gwIMF being steeper than the canonical IMF for $m>1\,M_\odot$ (\citealt{Scalo86, KTG93, RJ15, Mor+17, Mor+18}; Sec.~\ref{sec:shape}): this difference might be related to the gwIMF being a sum of the IMFs in the star-forming units \citep{KW03, Zonoozi+18}.  The IGIMF is a particular mathematical formulation of the gwIMF. The embedded clusters need to be sampled (e.g. optimally, \citealt{Schulz+15}) from the ECMF. This means that over some time interval, $\delta t$, the ECMF is optimally assembled, whereby this time interval needs to be investigated with care\footnote{\label{foot:SLUG2}Note that
  here the optimal model is being formulated. It is possible to relax
  the inherent non-dispersion nature and to retain the IGIMF
  formulation (Eq.~\ref{eq:igimf_t}) but to treat the sampling of
  embedded star cluster masses and of stars within the embedded
  clusters stochastically. This was the original notion followed by
  \cite{KW03} and forms the basis of the SLUG approach
  \citep{daSilva14}.}.

\vspace{2mm}
\noindent\fbox{\parbox{\columnwidth}{
{\sc Definition}: The IGIMF is an integral over all
star-formation events (embedded clusters, Sec.~\ref{sec:ecls}) 
in a given star-formation ``epoch'' $t, t+\delta t$,

\begin{equation} 
\xi_{\rm IGIMF}(m;t) = 
\int_{M_{\rm ecl,min}}^{M_{\rm
ecl,max}(SFR(t))} \xi\left(m\le m_{\rm max}\left(M_{\rm
ecl}\right)\right)~\xi_{\rm ecl}(M_{\rm ecl})~dM_{\rm ecl},
\label{eq:igimf_t}
\end{equation}
with the normalisation conditions Eqs.~\ref{eq:maxMecl}
and~\ref{eq:SFR} below and also with the conditions
Eqs.~\ref{eq:mm} and~\ref{eq:Mecl}
which together yield the $m_{\rm max}-M_{\rm ecl}$ relation
 }}

\vspace{2mm}

\noindent Here
$\xi(m\le m_{\rm max})~\xi_{\rm ecl}(M_{\rm ecl})~dM_{\rm ecl}$ is the
composite stellar IMF (i.e. the cIMF) contributed by $\xi_{\rm ecl}~dM_{\rm ecl}$
embedded clusters with stellar mass in the interval
$M_{\rm ecl}, M_{\rm ecl}+dM_{\rm ecl}$.  The ECMF is often taken to
be a power-law,
\begin{equation}
\xi_{\rm ecl}(M_{\rm ecl}) \propto M_{\rm ecl}^{-\beta},
\label{eq:ECMF}
\end{equation}
with $\beta \approx 2$ (\citealt{LL03, Schulz+15} and references
herein). For the Milky Way and extra-galactic data on non-star-bursting galaxies, 
$\beta\approx 2.3$ may be favoured \citep{Weidner04, Mor+18}, and $\beta$ may vary with the metallicity and SFR \citep{Yan+17}.

The most-massive embedded cluster forming in a galaxy,
$M_{\rm ecl,max}(SFR)$, can be assumed to come from the empirical
maximum star-cluster-mass {\it vs}
global-star-formation-rate-of-the-galaxy relation,
\begin{equation}
M_{\rm ecl,max}=8.5\times 10^4\;\left({ {\rm SFR} \over M_\odot/{\rm
      yr} }\right)^{0.75},
\label{eq:maxcl_sfr}
\end{equation}
(eq.~1 in \citealt{Weidner05}, as derived by \citealt{Weidner04} using
observed maximum star cluster masses). A relation between
$M_{\rm ecl,max}$ and $SFR$, which is a good description of the
empirical data, can also be arrived at by resorting to optimal
sampling (Sec.~\ref{sec:ODF}). Thus, when a galaxy has, at a time $t$, a $SFR(t)$ which is
approximately constant over a time span $\delta t$ over which an
optimally sampled embedded star cluster distribution builds up with
total mass $M_{\rm tot}(t)$, then there is one most massive embedded
cluster with mass $M_{\rm ecl, max}$,
\begin{equation}
1 = \int_{M_{\rm ecl, max}(t)}^{M_{\rm U}} \xi_{\rm ecl}(M_{\rm
  ecl})\,dM_{\rm ecl},
\label{eq:maxMecl}
\end{equation}
with $M_{\rm U}$ being the physical maximum star cluster than can form
(for practical purposes $M_{\rm U}>10^8\,M_\odot$), and
\begin{equation}
SFR(t) = {M_{\rm tot}(t) \over \delta t} = {1\over \delta t}
\int_{M_{\rm ecl, min}}^{M_{\rm ecl,max}(t)} M_{\rm ecl}\,\xi_{\rm
  ecl}(M_{\rm ecl})\,dM_{\rm ecl}.
\label{eq:SFR}
\end{equation}
$M_{\rm ecl,min}\,=\,5\,M_{\odot}$ is adopted in the standard
modelling and corresponds to the smallest ``star-cluster'' units
observed (the few$\,M_\odot$-heavy embedded clusters in fig.~1 in
\citealt{Yan+17} such as are observed in Taurus-Auriga,
e.g. \citealt{KB03,Joncour+18}). Note the similarity of these equations with Eqs.~\ref{eq:mm} and~\ref{eq:Mecl}. Perhaps the physical meaning of this is that within a galaxy the formation of embedded clusters may be a self-regulated growth process from the ISM, like stars in a molecular cloud core which spawns an embedded cluster.

But what is $\delta t$?
\citet{Weidner04} define $\delta t$ to be a ``star-formation epoch'',
within which the ECMF is sampled optimally, given a SFR. This
formulation leads naturally to the observed $M_{\rm ecl,max}(SFR)$
correlation if the ECMF is invariant, $\beta\approx2.35$ and if the
``epoch'' lasts about $\delta t=10$~Myr.  Under these conditions, a
galaxy forms embedded clusters with stellar masses ranging 
from $M_{\rm ecl, min}=5\,M_\odot$ to
$M_{\rm ecl, max}$, the value of which increases with the SFR of the
galaxy in accordance with the observed young most-massive cluster vs
SFR data (Box Observational Constraint III on p.~\pageref{box:obsIII}).  Thus, the embedded
cluster mass function is optimally sampled in about 10~Myr intervals,
independently of the SFR. 

It is interesting to note now that 
this time-scale is consistent with
the star-formation time-scale in normal galactic disks measured by
\citet{Egusa+04, Egusa+09, Egusa+17} using 
the offset of HII regions from the molecular clouds in spiral-wave
patterns. In this view, the ISM takes about 10~Myr to transform via
molecular cloud formation to a new population of young stars which
optimally sample the embedded-cluster mass function (see also Sec.~\ref{sec:ecls}).  
\cite{Schulz+15}
discuss (in their Sec.~3) the meaning and various observational
indications for the time-scale $\delta t\approx10\,$Myr. This
agreement between the independent methods to yield
$\delta t\approx 10\,$Myr is encouraging, since the \cite{Weidner04}
argument is independent of the arguments based on molecular cloud life
times and spiral arm phase-velocities.

\begin{figure}
\begin{center}
\rotatebox{0}{\resizebox{0.85 \textwidth}{!}{\includegraphics{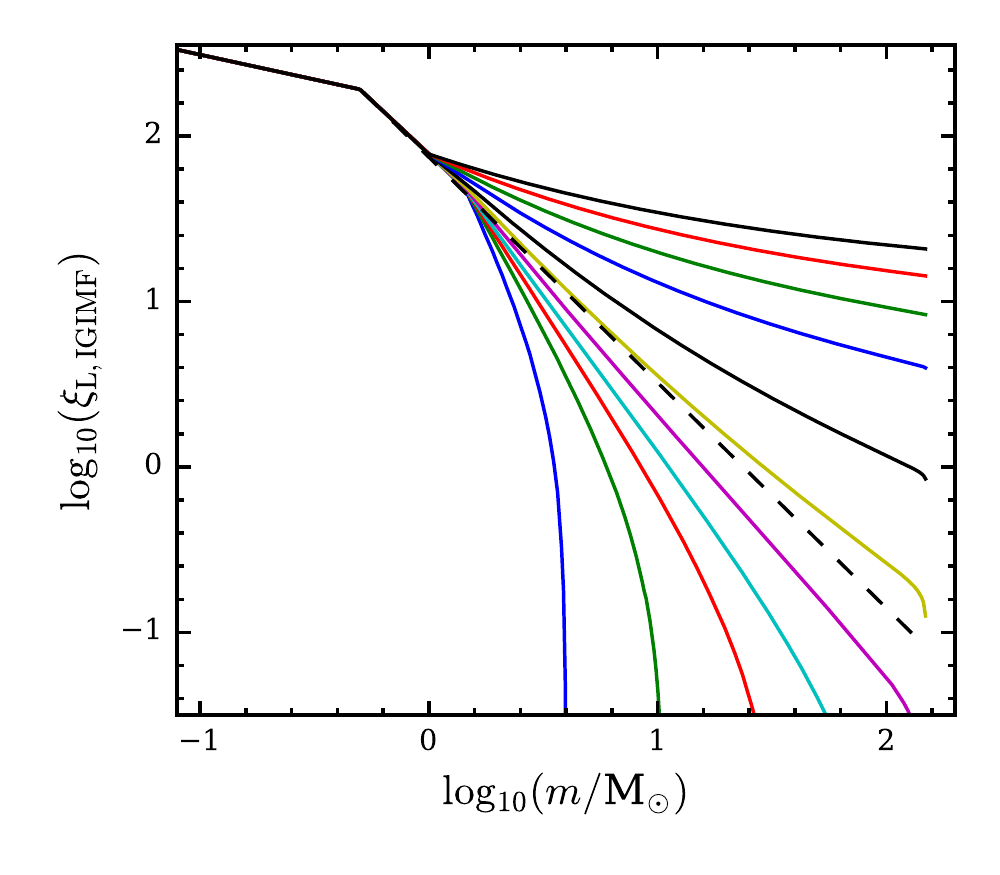}}}
\vspace{-7mm}
\caption[IGIMF in dependence of the SFR]{\small{
Logarithmic integrated gwIMFs (in number of stars per log-mass interval, using the transformation $\xi_{\rm L}=m\, {\rm ln}(10)\, \xi(m)$, eq.~14 in \citealt{Yan+17}), for different SFRs and formed over a $\delta t=10\,$Myr epoch. Each line is normalised to the same values at $m < 1\,M_\odot$. Solid curves are IGIMFs for $SFR = 10^{-5}, 10^{-4} ... \, 10^5\,M_\odot$/yr from bottom left to top right. The dashed line is the canonical IMF (Eq.~\ref{eq:imf}). 
For further details see \cite{Yan+17}. Note that the explicit metal dependency of the IGIMF is not included in these calculations and that the $\alpha_3^{\rm gal}$ values plotted in Fig.~\ref{fig:gwIMFvar} are the slopes of the here shown IGIMFs in different stellar mass ranges. See \cite{Jerabkova+18} for a full grid of IGIMF models in dependency of metallicity and SFR. }}
\label{fig:IGIMFform}
\end{center}
\end{figure}

\begin{figure}
\begin{center}
\rotatebox{0}{\resizebox{0.95 \textwidth}{!}{\includegraphics{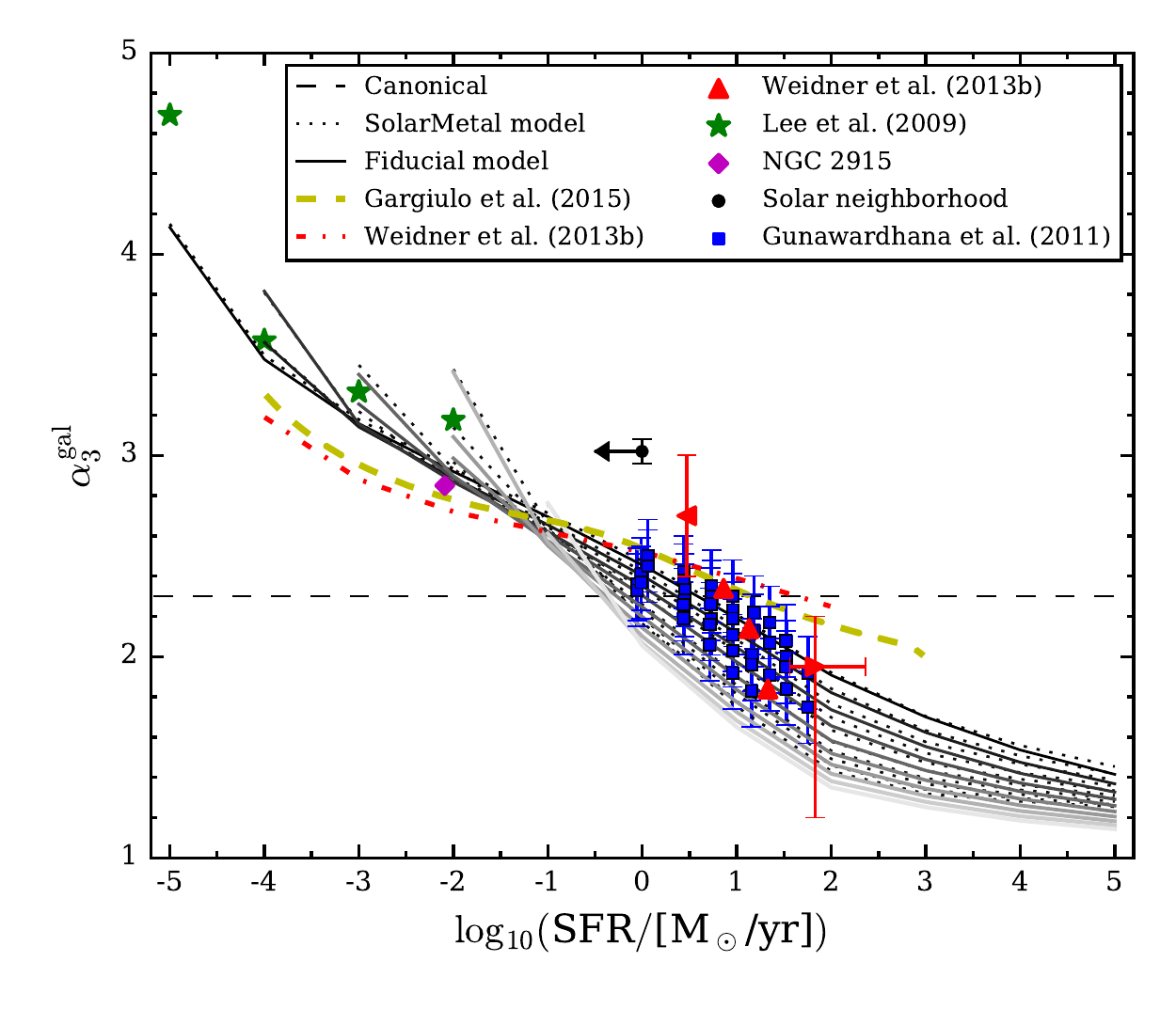}}}
\vspace{-5mm}
\caption[$\alpha_3$ as a function of the SFR]{\small{
    The variation of the IGIMF power-law index $\alpha_3^{\rm gal}$ (for stars with $m>1\,M_\odot$) with the galaxy-wide SFR is shown as the lines.  Each line is an evaluation of the index at a particular stellar mass, showing that the IGIMF is curved since $\alpha_3^{\rm gal}$ changes with $m$ (Fig.~\ref{fig:IGIMFform}). The invariant canonical IMF is shown as the horizontal dashed line, values above it are top-light gwIMFs and below it are top-heavy gwIMFs.  The solid and dotted lines constitute, respectively, the fiducial model (which includes an effective metallicity dependence in the density dependence of the IMF) and the Solar-abundance model (for details see \citealt{Yan+17}). The dot-dashed and dashed coloured lines are IGIMF models from \cite{Gargiulo+15} and \cite{Weidner13b}. The symbols are various observational constraints as indicated in the key.
    From \cite{Yan+17}.  }}
\label{fig:gwIMFvar}
\end{center}
\end{figure}

The IGIMF (Eq.~\ref{eq:igimf_t}) can be calculated under various assumptions on how the IMF, $\xi(m)$, varies with physical conditions. Ignoring the explicit metallicity dependence but taking into account the effective density dependence, which includes an intrinsic metallicity dependence, in Eq.~\ref{eq:toph_master} above, \cite{Yan+17} studies the prediction of the IGIMF theory for the variation of the shape of the gwIMF in comparison with observational constraints (Fig.~\ref{fig:gwIMFvar}). The change of the IGIMF as a function of the SFR is shown in Fig.~\ref{fig:IGIMFform}. The observational constraints, which indicate the gwIMF to change from a top-light form (i.e. with a deficit of massive stars) in dwarf galaxies (which have low SFRs, \citealt{Lee09}) to top-heavy gwIMFs in massive late-type galaxies (which have high SFRs, \citealt{Gun11}), are well covered by the IGIMF theory. It is noted here for completeness that the top-light gwIMF variation for dwarf galaxies was predicted on the basis of the IGIMF theory \citep{Pflamm07,Pflamm09b} before the data were available, with observational support from star-counts being found by \cite{Watts+18} in DDO154.  The calculations of the IGIMF for galaxies with high SFRs became physically relevant once the variation of the IMF (Eq.~\ref{eq:lowmassvar} and~\ref{eq:toph_master}) in extreme star-burst clusters became quantified based on observational data \citep{Marks12}.

A full-gird of IGIMF models is provided by \cite{Jerabkova+18} in which three cases
are considered: (1) The IMF is invariant, (2), the IMF varies only at
the massive end (Eq.~\ref{eq:toph_master}), and (3) the IMF varies
at low (Eq.~\ref{eq:lowmassvar}) and at high
(Eq.~\ref{eq:toph_master}) masses. This latter case, IGIMF3, is
considered to be the most realistic, and appears to be able to account
for the complicated time-evolving gwIMF variation deduced to have
occurred when elliptical galaxies formed, as well as for the gwIMF
variations deduced for late-type galaxies. 

\subsection{Some observational constraints}

Any of the three approaches, namely pure stochastic sampling (Sec.~\ref{sec:stoch}), constrained stochastic sampling (Sec.~\ref{sec:stoch}) and the IGIMF formulation in terms of optimal systems (Sec.~\ref{sec:optimal}), need to be consistent with observational constraints. Some relevant ones can be found in Box Observational Constraints IV (p.~\pageref{box:obsIV}).

\vspace{2mm} 
\begin{mdframed}
\label{box:obsIV}
{\sc Box Observational Constraints IV}: 
\begin{itemize} 
\item Dwarf
        late-type galaxies, with typically small
        ($\approx 0.0001-0.1\,M_\odot$/yr) to extremely small
        ($\simless 10^{-4}\,M_\odot$/yr) SFRs, have been found to have
        a systematically increasing deficit of H$\alpha$ emission
        relative to their UV emission with decreasing SFR. This can be
        readily understood as a result of an increasingly top-light
        gwIMF \citep{Lee09}.
\item Direct (star-count) evidence supporting this comes from the
        nearby dwarf galaxy DDO154 \citep{Watts+18} and from Leo~P 
       \citep{Jerabkova+18}. Both galaxies have a deficit of massive stars, 
       i.e. appear to have a top-light gwIMF. 
 \item The chemical properties of the low-mass satellite galaxies of the Milky Way, for
  which sufficient data exist, suggest they had gwIMFs with a
  deficit of massive stars \citep{Tsujimoto11}. This is consistent with the above two points. 
\item Massive late-type galaxies, with typically large
        ($\simgreat 1\,M_\odot$/yr) SFRs, have been found to have a
        systematically more top-heavy gwIMF with increasing SFR, as
        deduced from their H$\alpha$ flux and broad-band optical
        colours. This can be understood as a result of an
        increasingly top-heavy gwIMF \citep{Gun11}. Evidence for this
        has been found before also \citep{HG08,Meurer09}.
\item Massive elliptical galaxies have been found to have formed
  rapidly (within a Gyr) with $SFR \simgreat 10^3\,M_\odot/$yr, as deduced
  from their high alpha-element abundances \citep{Thomas05, Recchi09}. Their high 
  metallicity required them to have had top-heavy gwIMFs to generate
  the large mass in metals within the short time
  \citep{Matteucci94,  Vazdekis97, Weidner13b, Weidner13c,
    Ferreras+15, MartinNavarro16}. This fits
  well into the above three points which suggest a general shift of
  the gwIMF towards producing more massive stars relative to the
  low-mass stellar content with increasing SFR.  
\item The isotopes, $^{13}$C and $^{18}$O, are, respectively, released mainly by low- and intermediate-mass stars ($m < 8\,M_\odot$) and massive stars ($m > 8\,M_\odot$).  Using the ALMA facility, \cite{Zhang+18} measured rotational transitions of the $^{13}$CO and C$^{18}$O isotopologues in a number of star-bursting galaxies ($SFR\simgreat 10^3\,M_\odot$/yr) finding strong evidence for the galaxy-wide IMF to be top-heavy.
\item Massive elliptical galaxies also show evidence that they had a
  significantly bottom-heavy IMF (\citealt{vanDokkum10, CvD12, MartinNavarro16} 
  and references therein).
\item Early-type galaxies have a trend in metallicity and
  alpha-element abundances explainable with an increasingly
  top-light gwIMF with decreasing galaxy mass and thus decreasing
  SFR during their formation \citep{Koeppen07, Recchi09, Recchi15}. Alternatively, element-selective outflows from star-forming galaxies may account for the observed mass--metallicity relation among galaxies. Observational surveys to establish this have failed to detect evidence for outflows
  \citep{Lelli+14, Concas+17}, although \cite{McQuinn+18} report observation of hot gas leaving star-bursting dwarf galaxies.
\item Disk galaxies have an H$\alpha$ cutoff radius beyond which
  H$\alpha$ emission is significantly reduced or absent compared to
  the UV emission \citep{Boissier07}. These UV-extended disks can be
  understood in terms of a galactocentric-radial dependency of the ECMF 
  (see Box Observational Constraints III on p.~\pageref{box:obsIII}) in combination
  with the $m_{\rm max}-M_{\rm ecl}$ relation (Fig.~\ref{fig:WK}) such that the typically
  low-mass embedded clusters forming in the outer regions come along
  with a stellar population which is deficient in ionising radiation
  \citep{PAK08}.
\end{itemize}
\end{mdframed}
\vspace{2mm}

Is it possible that instead of the optimal star formation theory developed above (Sec.~\ref{sec:optimal}), the gwIMF remains universal and invariant and that star formation is stochastic, perhaps in a constrained manner such as in the SLUG approach (Sec.~\ref{sec:stoch}), and that the observed correlations in and among galaxies (Box Observational Constraints IV on p.~\pageref{box:obsIV}) are due to photon leakage, dust obscuration and redenning and other physical effects \citep{Calzetti08,Calzetti13}?  It is likely that these are relevant, but the authors of the original research papers reporting the mentioned effects \citep{HG08,Meurer09,Lee09, Gun11} discuss these biases at great length disfavouring them. It is nevertheless probably useful to study how invariant but stochastic models might be able to lead to the observed correlations.  For example, a systematic deficit of massive stars in dwarf galaxies could come about if the dwarf galaxies are, as a population, going through a current lull in star-formation activity \citep{KE12}.  But this appears contrived and extremely unlikely \citep{Lee09}, especially since one would need to resort to an in-step height in the SFR of nearby massive disk galaxies in order to explain the general observed shift of a large H$\alpha$-deficiency becoming smaller through to an H$\alpha$-emission overabundance when increasing the stellar mass from dwarf to major disk galaxies (Fig.~\ref{fig:gwIMFvar}).

The predictability of the IGIMF theory is an advantage though.  It allows testing of the models against data (but the calculations need to be done correctly before drawing conclusions, see the end of Sec.~\ref{sec:PDF}). Also, if found to be a good representation of the observational data, we can use it to learn about star-formation at high redshift and use galaxy-wide star-formation behaviour to constrain star-formation physics on the scales of embedded clusters \citep{Jerabkova+18}.

\section{Implications for the SFRs of galaxies}
\label{sec:SFRimpl}

Most measures of the SFR of galaxies rely on the photon output from massive stars ($m \simgreat 10\,M_\odot$) which are the most luminous objects. These are a good probe of the current SFR due to their short life-times ($\simless 50\,$Myr before exploding as a core-collapse supernovae, SNII events).  For an empirical comparison of the various SFR tracers the reader is referred to \cite{Mahajan+19}.

For the purpose of this discussion of how different treatments of the gwIMF may affect SFR measurements we assume the observer detects all relevant photons (i.e. that dust obscuration, photon leakage and other effects -- see \citealt{Calzetti08, Calzetti13} for a discussion -- have been corrected for) and we concentrate on the H$\alpha$-based SFR measure \citep{Kennicutt89, Pflamm09b}. It assumes that a fraction of ionising photons ionises hydrogen atoms in the nearby ISM and that these recombine. A fraction of the recombination photons is emitted as H$\alpha$ photons and it has been shown that the flux of these is proportional to the ionising flux such that the H$\alpha$ luminosity becomes a direct measure of the current population of massive stars. Given their short life-times ($\simless 50\,$Myr), we thus obtain an estimate of the SFR if we know which mass in all young stars is associated with these ionising stars, i.e if we know the shape of the gwIMF. The H$\alpha$ luminosity is the most sensitive measure of the population of ionising stars (unless direct star-counts can be performed in nearby dwarf galaxies, e.g. \citealt{Watts+18}). For completeness, we note that the GALEX far-UV (FUV) flux is a measure of the stellar population which includes late-B stars and thus summarises the SFR activity over the recent $400\,$Myr time window (see \citealt{Pflamm09b} who used the PEGASE code, and chapter~7 [in the present book] for a detailed discussion of the timescales of each SFR indicator but assuming an invariant gwIMF).  We do not address this measure here except to note that in the limit where only a few ionising stars form, the FUV-flux derived SFRs are more robust and these are indeed consistent with the higher SFRs as calculated using the IGIMF1 formulation \citep{Jerabkova+18} as shown explicitly in fig. 8 of \cite{Lee09}, who compare FUV and H$\alpha$-based SFR indicators for dwarf galaxies. The FUV flux is thus a more robust measure of $SFR_{\rm true}$ than the H$\alpha$ flux because it assesses a much more populous stellar ensemble therewith being less susceptible to Poisson noise, but it is more sensitive to the gwIMF of intermediate-mass stars and also only offers a poorer time resolution (maximally $400$ vs maximally $50\,$Myr, respectively, \citealt{Pflamm09b}).  The rate of SNII also provides a measure, but we do not know which fraction of massive stars implode into a black hole without producing an explosion and how this depends on metallicity and thus redshift. SNII events are too rare on a human life-time to provide reliable global measurements except in profusely star-bursting systems (e.g. as in Arp~220, \citealt{Dab12} and references therein, \citealt{Jerabkova+17}).

We refer to $SFR_{{\rm H}\alpha}$ as being the SFR measure using the H$\alpha$ flux, and with $SFR_{\rm K}$ we mean $SFR_{{\rm H}\alpha}$ in the specific case of using the \cite{Kennicutt98} calibration which assumes a fully sampled and invariant standard IMF (the Kennicutt IMF, which is very similar to the canonical IMF, \citealt{Pflamm09b}). Therefore, if the gwIMF differs from the invariant canonical one, then $SFR_{\rm K} \ne SFR_{\rm true}$.  If the gwIMF were to be invariant and a PDF then the average of $SFR_{\rm K}$ over a sufficiently large number of galaxies of similar baryonic mass \citep{Speagle14} would provide the correct measure of the SFR with increasing dispersion with decreasing SFR$_{\rm true}$ (Sec.~\ref{sec:PDF}), $SFR_{\rm true}=\overline{SFR}_{\rm K}$.  We also consider $SFR_{\rm SLUG}$ which is the H$\alpha$-based SFR computed within the SLUG approach (Sec.~\ref{sec:stoch}, footnote~\ref{foot:SLUG} on p.~\pageref{foot:SLUG}, footnote~\ref{foot:SLUG2} on p.~\pageref{foot:SLUG2}).  If reality were to correspond to the assumptions underlying the SLUG approach (see text below) then $SFR_{\rm SLUG}=SFR_{\rm true}$, since in any particular model galaxy, the SLUG methodology knows what the model gwIMF is, such that the SFR is calculated properly. However, the SLUG approach does not allow to infer the true SFR, $SFR_{\rm true}$, for an observed galaxy, since, by the stochastic aspect inherent to SLUG, the observer does not know the actual momentary gwIMF. This comes about because for a SFR$_{\rm true}$ in the observed galaxy, the observer does not know whether the gwIMF is more or less top-heavy for example (due to stochastic fluctuations) as long as this gwIMF is consistent with the SFR-tracer (for example, the same H$\alpha$ flux can be obtained by differently shaped gwIMFs).  The SFR measure, $SFR_{\rm IGIMF}$, is likewise based on the H$\alpha$ flux but is calculated taking into account the number of ionising stars in the IGIMF theory (Sec.~\ref{sec:ODF}) such that $SFR_{\rm IGIMF}=SFR_{\rm true}$ without scatter (apart from variations at the very small $SFR_{\rm true}$ level where a single ionising star may be born or die, see \citealt{Pflamm07, Jerabkova+18} for a discussion of this limit). Details of the IGIMF calculations can be found in \cite{Jerabkova+18}.  We define the correction factor (eq.~17 in \citealt{Jerabkova+18}), 
\begin{equation}
    \Psi = { SFR_{\rm K \; or\; SLUG \; or \, IGIMF}  \over SFR_{\rm K}}.
\label{eq:corrfact}
\end{equation}

Assuming the IMF is a PDF (Sec.~\ref{sec:stoch}, point~\ref{point:purePDF}), the average over an ensemble of galaxies with the same baryonic mass, $\overline{\Psi}=1$, for all $SFR_{\rm K}$ with increasing scatter as $SFR_{\rm K}$ decreases. The scatter in $SFR_{\rm K}$ will constitute a few orders of magnitude for $SFR_{\rm true}\simless 0.1\,M_\odot$/yr as a result of randomly sampling stars from the IMF without constraints. For example, there can be galaxies which contain no ionising stars (such that $SFR_{\rm K}\approx 0\,M_\odot$/yr) despite having $SFR_{\rm true}=1\,M_\odot$/yr. Alternatively, there can be galaxies consisting only of ionising stars and with the same SFR$_{\rm true}$.

In the SLUG approach (Sec.~\ref{sec:stoch}, point~\ref{point:constrPDF}), $\overline{\Psi}>1$ for $SFR_{\rm true}\simless 1\,M_\odot$/yr, increasing with decreasing $SFR_{\rm K}$. For decreasing $SFR_{\rm true}$, the scatter in $SFR_{\rm K}$ increases up to a few orders of magnitude. For $SFR_{\rm true}\simgreat 1\,M_\odot$/yr, $\overline{\Psi}=1$ (fig.~2 in \citealt{daSilva14}).  The reason for this comes about because at low $SFR_{\rm true}$, the galaxy is populated typically with less-massive embedded clusters such that the gwIMF is not fully sampled to the highest allowed stellar masses. As a consequence, the true SFR, $SFR_{\rm true}$, is larger than that measured from the H$\alpha$ flux assuming the gwIMF is invariant and fully sampled (in this case $SFR_{\rm K}$) because the gwIMF contains fewer ionising stars per low-mass star.  The scatter comes about because the IMF and ECMF are assumed to be PDFs (the IMF is a constrained PDF, the constraint being that the random drawing of stars from the IMF must add up to the mass of the pre-determined embedded cluster, $M_{\rm ecl}$). Thus, in the SLUG approach it is possible to generate a galaxy which has no ionising stars despite having $SFR_{\rm true}=1\,M_\odot$/yr. In the SLUG approach it is possible that a galaxy with $SFR_{\rm true}=1\times 10^{-4}\,M_\odot$/yr forms only a single $10\,M_\odot$ star which needs this SFR ($10\,M_\odot$ being assembled in $10^5\,$yr). In this case an observer would however conclude that $SFR_{\rm K}\gg 10^{-4}\,M_\odot$/yr since the H$\alpha$ flux would be associated with a full gwIMF (modelled via the Kennicutt IMF).  This is, by the way, also true for the pure stochastic approach above, such that there $\Psi$ varies both sides of the value of one.

In the IGIMF approach (Sec.~\ref{sec:ODF}) galaxies always form populations of stars and these do not contain O~stars when $SFR_{\rm true}<10^{-4}\,M_\odot$/yr (table~3 in \citealt{Weidner13b}, fig.~7 in \citealt{Yan+17}). This has led, in the literature, to confusion, since claims were made that observed galaxies with $SFR_{\rm K}<10^{-4}\,M_\odot$/yr are producing O~stars therewith ruling out the IGIMF theory. But these claims forgot that $\Psi \gg 1$ in these cases according to the IGIMF theory (see discussion of this in \citealt{Jerabkova+18}).
It is evident from Figs.~\ref{fig:IGIMFform} and~\ref{fig:gwIMFvar} that the number of ionising stars relative to the number of low-mass stars decreases systematically with decreasing $SFR_{\rm true}$. Consequently, $\Psi>1$ for $SFR_{\rm K}\simless 1\,M_\odot$/yr, increasing with decreasing $SFR_{\rm K}$. This is a similar but stronger effect than in the SLUG approach. If the IMF is invariant and canonical, then $\Psi=1$ for $SFR_{\rm K}\simgreat1\,M_\odot$/yr, as in the SLUG approach, which also assumes (in the current published version and in the spirit of the IMF being a PDF, i.e. not being subject to physical limits apart from the constraint given by $M_{\rm ecl}$ when drawing stars) that the IMF is invariant. In the IGIMF theory, the IMF is however assumed to systematically vary with the physical conditions of the molecular cloud (Eqs~\ref{eq:lowmassvar} and \ref{eq:toph_master}) such that $\Psi<1$ for $SFR_{\rm K}\simgreat 1\,M_\odot$/yr.

\cite{KE12} discuss the differences of these approaches and conclude that the systematic deviations observed between $SFR_{{\rm H}\alpha}$ and $SFR_{\rm UV}$ (which is a reasonably good approximation of $SFR_{\rm true}$, \citealt{Pflamm09b}) at small values of the SFR can be produced instead through temporal variations in SFRs, without having to resort to modifying the IMF. The galaxies would, however, need to be in an unexplained synchronised lull of their SFRs. Even if this were the correct conclusion, it would not be able to accommodate the systematic variation observed at high SFRs, which appear to merely be a natural continuation of the overall trend of the observationally-constrained gwIMF becoming increasingly top-heavy with increasing galaxy-wide SFR, beginning with a significantly top-light gwIMF in the least-massive dwarf galaxies (Fig.~\ref{fig:gwIMFvar}).

\begin{figure}
\begin{center}
\rotatebox{0}{\resizebox{0.8 \textwidth}{!}{\includegraphics{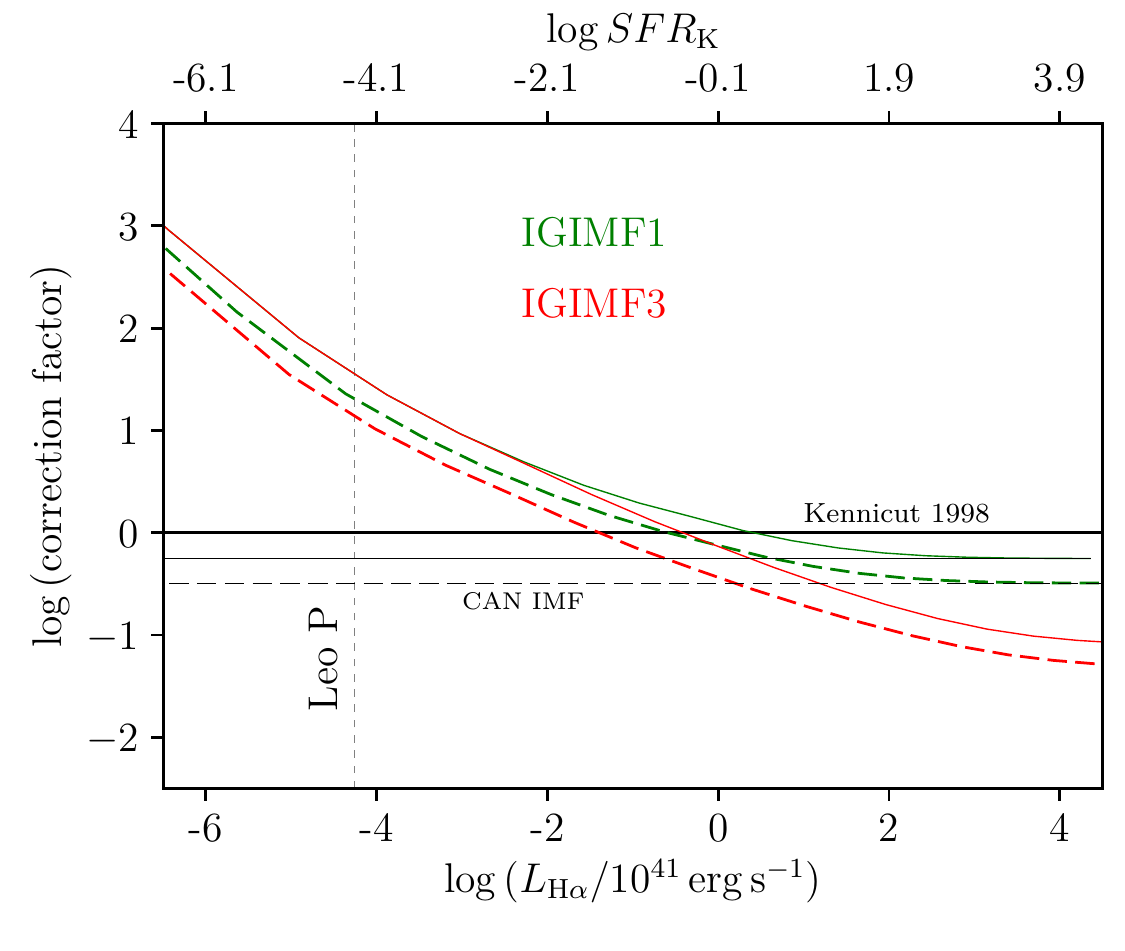}}}
\vskip -5mm
\caption[The IGIMF correction factor to $SFR_{{\rm H}\alpha}$]{\small{
The correction factor $\Psi$ (Eq.~\ref{eq:corrfact}, ${\rm log}x \equiv {\rm log}_{10}x$ for any $x$) is plotted for various cases: the horizontal solid line is $\Psi=1$ whereby the gwIMF used by \cite{Kennicutt98} is assumed, the thin solid and dashed horizontal lines are for an invariant canonical IMF (Eq.~\ref{eq:imf}) for [Fe/H]$=0$ and $-2$, respectively. The solid green and red curves are IGIMF models for [Fe/H]$=0$ assuming, respectively, the IMF varies only through $\alpha_3$ (Eq.~\ref{eq:toph_master}, "IGIMF1") or ("IGIMF3" which is the most realistic case) at the low-mass end (Eq.~\ref{eq:lowmassvar}) and at the high-mass end (Eq.~\ref{eq:toph_master}). The corresponding dashed lines are for [Fe/H]$=-2$. Note that for [Fe/H]$=0$ IGIMF1 and IGIMF3 bifurcate for $SFR\simgreat 10^{-1.5}\,M_\odot$/yr because the IMFs are identical to the canonical IMF at the low mass end at this metallicity, while at [Fe/H]$=-2$ the IGIMF1 and IGIMF3 models remain separated at all SFRs because the IMFs differ at the low- and at the high-mass end at this metallicity. 
 Thus, for example taking the case of a dwarf galaxy with [Fe/H]$=-2$ and measured $SFR_{\rm K}\approx 10^{-4.2}\,M_\odot$/yr, it would have, according to the IGIMF3 model $SFR_{\rm true}\approx \Psi\times 10^{-4.2}\,M_\odot$/yr with $\Psi\approx 10^{1.2}$. A massive disk galaxy with [Fe/H]$=0$ and $SFR_{\rm K}\approx 10^{1.9}\,M_\odot$/yr would have a $\Psi\approx 10^{-0.7}$ times larger $SFR_{\rm true}$. If the IGIMF theory is applicable, then the Leo~P dwarf galaxy has a $SFR_{\rm true} \approx \Psi  \times 10^{-4.2}\,M_\odot$/yr with $\Psi\approx 10^{1.2}$ explaining why it has one or two O stars. The presence of these stars has been leading to confusion in the literature if the $SFR_{\rm K}$ value for its SFR is assumed \citep{McQuinn+15}.  Adapted from \cite{Jerabkova+18}. }}
\label{fig:corrfact}
\end{center}
\end{figure}

The above discussion is quantified in Fig.~\ref{fig:corrfact} which plots $\Psi$ for the IGIMF case (the pure PDF case, point~\ref{point:purePDF} in Sec.~\ref{sec:PDF}, has $\overline{\Psi}=1$ independently of $SFR_{\rm K}$ with a scatter which is comparable to that evident in the SLUG case) and in fig.~3 in \cite{daSilva14} for the SLUG approach.

\subsection{The main sequence of galaxies}
\label{sec:mainsequ}

\begin{figure}
\begin{center}
\rotatebox{0}{\resizebox{0.9 \textwidth}{!}{\includegraphics{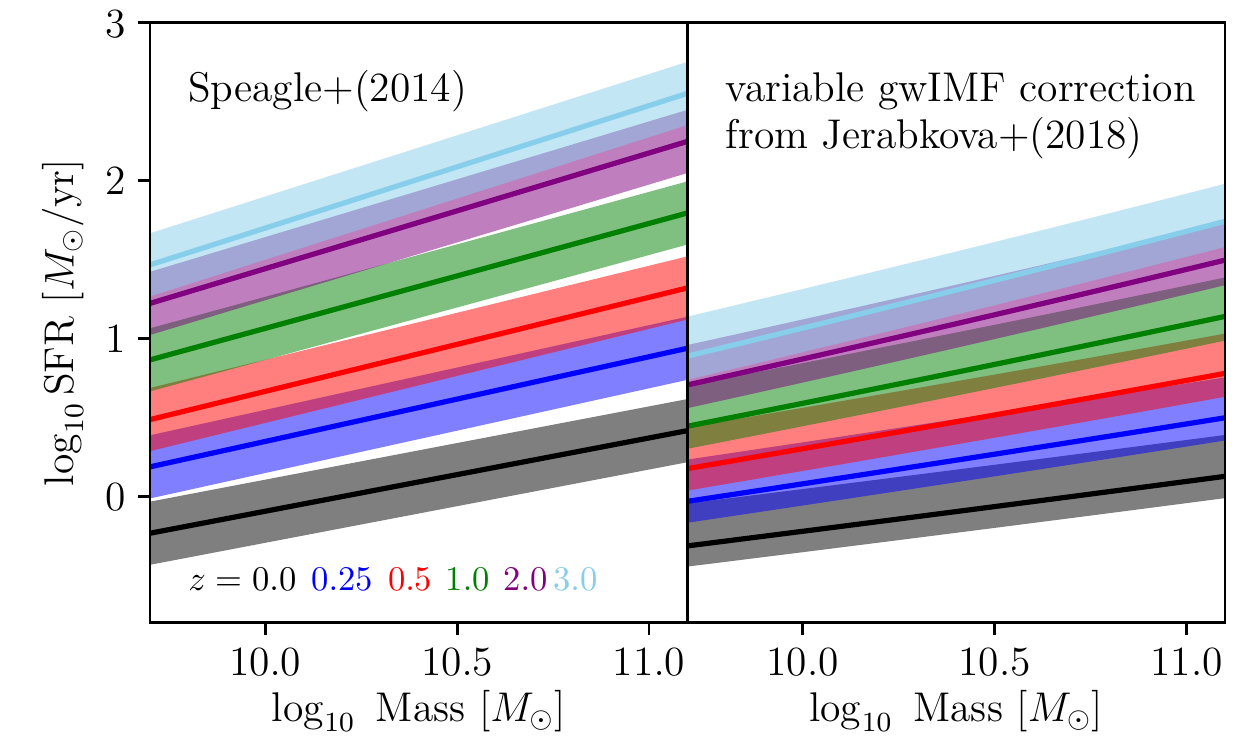}}}
\vskip 0mm
\caption[The main sequence of galaxies corrected for gwIMF variation]{\small{
       The galaxy main sequence according to which the true SFR of a galaxy (on the y-axis) depends on the stellar mass, $M_*$ (on the x axis) of the galaxy. The {\bf left panel} shows the original quantification by \cite{Speagle14} with the colour coding according to redshift, $z$, as indicated at the bottom of the panel. The {\bf right panel} shows the new main sequence after applying the correction factor $\Psi$ (Eq.~\ref{eq:corrfact}, Fig.~\ref{fig:corrfact}) assuming the IGIMF3 model. Assuming the correction factors to be correct, the new main sequence would have a shallower slope, the SFRs would be lower and the evolution with redshift would be reduced. This is a consequence of the gwIMF depending on the metallicity and SFR, which is driven by the IMF becoming top-heavy in star-burst star clusters. 
       }}
\label{fig:galmainseq}
\end{center}
\end{figure}

The vast majority ($\approx 97$~per cent) of galaxies with a baryonic mass larger than about $10^{10}\,M_\odot$ are disk galaxies today and $t\approx 6\,$Gyr ago \citep{Delgado10}. These lie on a main sequence according to which the galaxy-wide stellar mass, $M_*$, correlates strongly with the SFR \citep{Speagle14}.
The SFR tracers employed are mostly sensitive to the luminous massive (ionising) stellar population in the galaxies and it was assumed that the gwIMF is invariant. 
Given the above correction factors (Eq.~\ref{eq:corrfact}, Fig.~\ref{fig:corrfact}), we can investigate how these affect the galaxy main sequence. We correct each $SFR_{\rm K}$ to obtain a new main sequence,
under the assumption that the IGIMF3 model is valid and on assuming $SFR_{\rm Speagle} = SFR_{\rm K}$. This new main sequence and its dependence with redshift (assuming, as in \citealt{Speagle14}, that the standard LCDM cosmological model applies) is shown in Fig~\ref{fig:galmainseq}. 
It transpires that the change of the SFR$_{\rm true}$ values, at a given baryonic galaxy mass, from $z=3$ until today is reduced to an order of magnitude in comparison to two orders of magnitude according to \cite{Speagle14}. The slope is also lessened. Further work on how the gwIMF variation and a possibly different cosmological model (giving different luminosity and angular diameter distances) affect the true physical main sequence at different redshifts is required before firm conclusions can be reached on the true cosmological evolution of the main sequence.  Since on a time-scale $\simgreat 2.5\,$Gyr \citep{Pflamm09} the SFR needs to be up-kept by gas accretion, it follows that the gas-accretion rate is proportional to the stellar mass of the galaxy.

\section{Conclusion}
\label{sec:conc}

The relation of the IMF to the composite and the galaxy-wide IMF has been discussed. The mathematical treatment of the IMF (as a probability distribution function or an optimal distribution function) is closely associated with the physical processes according to which the distribution of stellar masses emerges from the interstellar medium. Three types of possibilities are being discussed and investigated in the community, and these lead to different predictions on the relation between the tracer used to asses the star formation rate of a galaxy, e.g. $SFR_{\rm H\alpha}$, and the true star formation rate, $SFR_{\rm true}$.  The H$\alpha$ and FUV flux is often employed to assess the $SFR$ (over the past $\approx 50$ and $400\,$Myr, respectively) and we used the former as an example to show the differences in the predictions. Assuming in all cases that dust obscuration, photon leakage and other effects \citep{Calzetti08, Calzetti13} do not play a role, and that the SFR does not vary over time-scales of less than about 50~Myr:
\begin{enumerate}
\item  \label{point:st} If the IMF is a pure PDF without constraints apart from the possible existence of an upper mass limit to stars ($m_{\rm max*}$, Sec.~\ref{sec:stoch}) then $SFR_{{\rm H}\alpha}$ is an unbiased tracer of $SFR$ ($SFR_{{\rm H}\alpha} \propto SFR_{\rm true}$) and the dispersion of the SFRs, $\sigma_{SFR({\rm H}\alpha)}$, increases with decreasing $SFR_{\rm true}$ due to stochastic variations. This dispersion can be orders of magnitude, since galaxies can be sampled which have no ionising stars or which consist only of ionising stars. This model cannot reproduce the observed systematic trend of the gwIMF becoming increasingly top-light with decreasing $SFR_{\rm true}\simless 1\,M_\odot$/yr, nor 
the observed systematic trend of the gwIMF becoming increasingly top-heavy with increasing $SFR_{\rm true}\simgreat 1\,M_\odot$/yr (Fig. \ref{fig:gwIMFvar}), nor the $m_{\rm max}-M_{\rm ecl}$ data (Fig.~\ref{fig:WK}). If a galaxy has an observed H$\alpha$ flux, then this approach does not allow the determination of $SFR_{\rm true}$ unless one is in the regime where $\sigma_{SFR({\rm H}\alpha)}$ is sufficiently small.
\item \label{point:cst} If it is assumed that stars from in embedded clusters which are drawn stochastically from the ECMF and that the IMF is sampled stochastically within each embedded cluster with the constraint that $M_{\rm ecl}$ be fulfilled (the SLUG approach, Sec.~\ref{sec:stoch}), then 
$SFR_{{\rm H}\alpha}$ increasingly underestimates $SFR_{\rm true}$ with decreasing $SFR_{\rm true}\simless 1\,M_\odot$/yr. The dispersion, $\sigma_{SFR({\rm H}\alpha)}$, also increases with decreasing $SFR_{\rm true}$ due to stochastic variations, but to a lesser extend than under point~\ref{point:st}, since the constraint $M_{\rm ecl}$ limits the possible variations. This model allows for galaxies to contain no ionising stars and also for galaxies to consist only of ionising stars and so $\sigma_{SFR({\rm H}\alpha)}$ reaches orders of magnitude at low $SFR_{\rm true}$ \citep{daSilva14}.
This model can approximately reproduce the observed systematic trend of the gwIMF becoming increasingly top-light with decreasing $SFR_{\rm true}$ (\citealt{daSilva14}, Fig.~\ref{fig:gwIMFvar}, Eq.~\ref{eq:corrfact}, Fig.~\ref{fig:corrfact}), but not the $m_{\rm max}-M_{\rm ecl}$ data (Fig.~\ref{fig:WK}). It can also not reproduce the observed systematic trend of the gwIMF becoming increasingly top-heavy with increasing $SFR_{\rm true}\simgreat 1\,M_\odot$/yr (Fig.~\ref{fig:gwIMFvar}). If a galaxy has an observed H$\alpha$ flux, then this approach does not allow the determination of $SFR_{\rm true}$ unless one is in the regime where $\sigma_{SFR({\rm H}\alpha)}$ is sufficiently small.
\item \label{point:igimf} If the IGIMF theory is assumed, then $SFR_{{\rm H}\alpha}$ increasingly underestimates $SFR_{\rm true}$ with decreasing $SFR_{\rm true}\simless 1\,M_\odot$/yr with a somewhat larger systematic change than under point~\ref{point:cst}. A small dispersion, $\sigma_{SFR({\rm H}\alpha)}$, at small $SFR$ is present due to the birth and death of individual most-massive stars in the galaxy. Since galaxies are interacting and have internal dynamical instabilities which affect the assembly of molecular clouds (e.g., the bar instability), a natural $\sigma_{SFR({\rm H}\alpha)}$ will be observable, but this has not yet been calculated.  This model can reproduce the observed systematic trend of the gwIMF becoming increasingly top-light with decreasing $SFR_{\rm true}$ (\citealt{Lee09}, Fig.~\ref{fig:gwIMFvar}, Eq.~\ref{eq:corrfact}, Fig.~\ref{fig:corrfact}) and the $m_{\rm max}-M_{\rm ecl}$ data (Fig.~\ref{fig:WK}). It can also reproduce the observed systematic trend of the gwIMF becoming increasingly top-heavy with increasing $SFR_{\rm true}\simgreat 1\,M_\odot$/yr (Fig.~\ref{fig:gwIMFvar}) due to the incorporation of the metallicity and density dependent IMF (Eq.~\ref{eq:lowmassvar} and~\ref{eq:toph_master}). If a galaxy has an observed H$\alpha$ flux, then this approach provides a unique determination of $SFR_{\rm true}$, since the expected variation, $\sigma_{SFR({\rm H}\alpha)}$, for galaxies of the same baryonic mass can (we expect) be related to the distribution of morphologies of the galaxies and there is no stochasticity.  \end{enumerate}

Efficient IMF-sampling algorithms are available as downloads: for stochastic sampling see \cite{PAK06} and \cite{Kroupa13} and for optimal sampling see \cite{Yan+17} and \cite{Kroupa13}. For the SLUG approach (relevant for points~\ref{point:st} and~\ref{point:cst} in Sec.~\ref{sec:stoch}) programs are available \citep{daSilva12, daSilva14, Ashworth+17}, while for the IGIMF approach also \citep{Yan+17, Jerabkova+18}.

The IGIMF theory has been computed with and without a variable IMF, and the most realistic case is IGIMF3 \citep{Jerabkova+18} which incorporates the full IMF dependency on density and metallicity (Eq.~\ref{eq:lowmassvar} and~\ref{eq:toph_master}) currently known from observational data.  The two stochastic approaches above (points~\ref{point:st} and~\ref{point:cst} in Sec.~\ref{sec:stoch}) may be used with a variable IMF also. But an IMF which varies systematically with physical conditions may be in tension with an interpretation of the IMF as a PDF since it implies that conditions need to be applied on the PDF. If the IMF becomes top-heavy at high SFR density (e.g. as formulated in Eq.~\ref{eq:toph_master}) then galaxies with a $SFR_{\rm true}\simgreat 1\,M_\odot$/yr will have $SFR_{{\rm H}\alpha}$ overestimating $SFR_{\rm true}$, just as in the IGIMF theory.

The observational constraints (Boxes I--IV, p.~\pageref{box:obsI}, \pageref{box:obsII},\pageref{box:obsIII}, \pageref{box:obsIV}) are useful for assessing which of the above possibilities are relevant for nature: The existence or not existence of isolated massive stars and of a physical $m_{\rm max}-M_{\rm ecl}$ relation (Fig.~\ref{fig:WK}) and its dispersion are thus central issues in understanding SFR measurements of galaxies. Important for the interpretation of the IMF and thus on how to compute gwIMFs and galaxy-wide SFRs given some tracer is also whether embedded clusters can be viewed as being fundamental building blocks of galaxies \citep{Kroupa05} and if stars form mass segregated in these embedded clusters. If this is the case, then it is also important if the ECMF varies systematically within a galaxy, e.g. radially.  To further clarify these points, which all play a role in our interpretation of the IMF and its relation to the cIMF and gwIMF, further observational work is required. But that these issues are being discussed shows how rich and informative this topic is. An important constraint on any formulation of the gwIMF is that this formulation must be consistent with the IMFs deduced from resolved stellar populations (star clusters, stellar associations). The implications for our understanding of how galaxies evolve are potentially major, as discussed in this contribution. Certainly, as an example, the quantification of the main sequence of galaxies depends on how we understand the gwIMF (Sec.~\ref{sec:mainsequ}).

Returning to elliptical galaxies (Box Observational Constraint IV, page~\pageref{box:obsIV}), the need for a bottom-heavy and a top-heavy gwIMF, possibly in terms of the evolution of the gwIMF, is a challenge for any theory of how elliptical galaxies formed and evolved \citep{Weidner13c, Ferreras+15, Jerabkova+18}. It is clear that elliptical galaxies are unusual, making only a few per cent of the population of galaxies heavier than about $10^{10}\,M_\odot$ \citep{Delgado10}, and that elliptical galaxies formed rapidly and thus with $SFR_{\rm true}>1000\,M_\odot$/yr. Any such theory needs to be consistent with the data on star-forming galaxies. Important in this context are the constraints on the allowed dark matter (faint M~dwarfs, stellar remnants) from lensing observations \citep{SL13, Smith14} which appear to be inconsistent with the bottom-heavy gwIMF to provide a significant amount of mass. This problem is discussed also in the review on elliptical galaxies by \cite{Cappellari16} and significantly advanced by \cite{Yan+21}.






\bibliography{/Users/pavel/PAPERS/BIBL_REFERENCES/kroupa_ref}

\begin{thebibliography}{225}
\expandafter\ifx\csname natexlab\endcsname\relax\def\natexlab#1{#1}\fi
\expandafter\ifx\csname selectlanguage\endcsname\relax
  \def\selectlanguage#1{\relax}\fi

\bibitem[\protect\citename{{Adams} and {Fatuzzo}, }1996]{AF96}
{Adams}, F.~C., and {Fatuzzo}, M. 1996.
\newblock {A Theory of the Initial Mass Function for Star Formation in
  Molecular Clouds}.
\newblock {\em \apj}, {\bf 464}(June), 256.

\bibitem[\protect\citename{{Andr{\'e}} {et~al.}, }2010]{Andre+10}
{Andr{\'e}}, P., {Men'shchikov}, A., {Bontemps}, S., {K{\"o}nyves}, V.,
  {Motte}, F., {Schneider}, N., {Didelon}, P., {Minier}, V., {Saraceno}, P.,
  {Ward-Thompson}, D., {di Francesco}, J., {White}, G., {Molinari}, S.,
  {Testi}, L., {Abergel}, A., {Griffin}, M., {Henning}, T., {Royer}, P.,
  {Mer{\'{\i}}n}, B., {Vavrek}, R., {Attard}, M., {Arzoumanian}, D., {Wilson},
  C.~D., {Ade}, P., {Aussel}, H., {Baluteau}, J.-P., {Benedettini}, M.,
  {Bernard}, J.-P., {Blommaert}, J.~A.~D.~L., {Cambr{\'e}sy}, L., {Cox}, P.,
  {di Giorgio}, A., {Hargrave}, P., {Hennemann}, M., {Huang}, M., {Kirk}, J.,
  {Krause}, O., {Launhardt}, R., {Leeks}, S., {Le Pennec}, J., {Li}, J.~Z.,
  {Martin}, P.~G., {Maury}, A., {Olofsson}, G., {Omont}, A., {Peretto}, N.,
  {Pezzuto}, S., {Prusti}, T., {Roussel}, H., {Russeil}, D., {Sauvage}, M.,
  {Sibthorpe}, B., {Sicilia-Aguilar}, A., {Spinoglio}, L., {Waelkens}, C.,
  {Woodcraft}, A., and {Zavagno}, A. 2010.
\newblock {From filamentary clouds to prestellar cores to the stellar IMF:
  Initial highlights from the Herschel Gould Belt Survey}.
\newblock {\em \aap}, {\bf 518}(July), L102.

\bibitem[\protect\citename{{Andr{\'e}} {et~al.}, }2014]{Andre+14}
{Andr{\'e}}, P., {Di Francesco}, J., {Ward-Thompson}, D., {Inutsuka}, S.-I.,
  {Pudritz}, R.~E., and {Pineda}, J.~E. 2014.
\newblock {From Filamentary Networks to Dense Cores in Molecular Clouds: Toward
  a New Paradigm for Star Formation}.
\newblock {\em Protostars and Planets VI},  27--51.

\bibitem[\protect\citename{{Andrews} {et~al.}, }2013]{Andrews13}
{Andrews}, J.~E., {Calzetti}, D., {Chandar}, R., {Lee}, J.~C., {Elmegreen},
  B.~G., {Kennicutt}, R.~C., {Whitmore}, B., {Kissel}, J.~S., {da Silva},
  R.~L., {Krumholz}, M.~R., {O'Connell}, R.~W., {Dopita}, M.~A., {Frogel},
  J.~A., and {Kim}, H. 2013.
\newblock {An Initial Mass Function Study of the Dwarf Starburst Galaxy NGC
  4214}.
\newblock {\em \apj}, {\bf 767}(Apr.), 51.

\bibitem[\protect\citename{{Andrews} {et~al.}, }2014]{Andrews14}
{Andrews}, J.~E., {Calzetti}, D., {Chandar}, R., {Elmegreen}, B.~G.,
  {Kennicutt}, R.~C., {Kim}, H., {Krumholz}, M.~R., {Lee}, J.~C., {McElwee},
  S., {O'Connell}, R.~W., and {Whitmore}, B. 2014.
\newblock {Big Fish in Small Ponds: Massive Stars in the Low-mass Clusters of
  M83}.
\newblock {\em \apj}, {\bf 793}(Sept.), 4.

\bibitem[\protect\citename{{Applebaum} {et~al.}, }2020]{Applebaum+18}
{Applebaum}, Elaad, {Brooks}, Alyson~M., {Quinn}, Thomas~R., and {Christensen},
  Charlotte~R. 2020.
\newblock {A stochastically sampled IMF alters the stellar content of simulated
  dwarf galaxies}.
\newblock {\em \mnras}, {\bf 492}(1), 8--21.

\bibitem[\protect\citename{{Ashworth} {et~al.}, }2017]{Ashworth+17}
{Ashworth}, G., {Fumagalli}, M., {Krumholz}, M.~R., {Adamo}, A., {Calzetti},
  D., {Chandar}, R., {Cignoni}, M., {Dale}, D., {Elmegreen}, B.~G.,
  {Gallagher}, III, J.~S., {Gouliermis}, D.~A., {Grasha}, K., {Grebel}, E.~K.,
  {Johnson}, K.~E., {Lee}, J., {Tosi}, M., and {Wofford}, A. 2017.
\newblock {Exploring the IMF of star clusters: a joint SLUG and LEGUS effort}.
\newblock {\em \mnras}, {\bf 469}(Aug.), 2464--2480.

\bibitem[\protect\citename{{Ballero} {et~al.}, }2007]{Ballero07a}
{Ballero}, S.~K., {Kroupa}, P., and {Matteucci}, F. 2007.
\newblock {Testing the universal stellar IMF on the metallicity distribution in
  the bulges of the Milky Way and M 31}.
\newblock {\em \aap}, {\bf 467}(May), 117--121.

\bibitem[\protect\citename{{Banerjee} and {Kroupa}, }2012]{Banerjee12b}
{Banerjee}, S., and {Kroupa}, P. 2012.
\newblock {On the true shape of the upper end of the stellar initial mass
  function. The case of R136}.
\newblock {\em \aap}, {\bf 547}(Nov.), A23.

\bibitem[\protect\citename{{Banerjee} and {Kroupa}, }2017]{BK17}
{Banerjee}, S., and {Kroupa}, P. 2017.
\newblock {How can young massive clusters reach their present-day sizes?}
\newblock {\em \aap}, {\bf 597}(Jan.), A28.

\bibitem[\protect\citename{{Banerjee} {et~al.}, }2012]{Banerjee12}
{Banerjee}, S., {Kroupa}, P., and {Oh}, S. 2012.
\newblock {The emergence of super-canonical stars in R136-type starburst
  clusters}.
\newblock {\em \mnras}, {\bf 426}(Oct.), 1416--1426.

\bibitem[\protect\citename{{Bastian} {et~al.}, }2010]{Bastian10}
{Bastian}, N., {Covey}, K.~R., and {Meyer}, M.~R. 2010.
\newblock {A Universal Stellar Initial Mass Function? A Critical Look at
  Variations}.
\newblock {\em \araa}, {\bf 48}(Sept.), 339--389.

\bibitem[\protect\citename{{Bate}, }2005]{Bate05}
{Bate}, M.~R. 2005.
\newblock {The dependence of the initial mass function on metallicity and the
  opacity limit for fragmentation}.
\newblock {\em \mnras}, {\bf 363}(Oct.), 363--378.

\bibitem[\protect\citename{{Bate}, }2014]{Bate14}
{Bate}, M.~R. 2014.
\newblock {The statistical properties of stars and their dependence on
  metallicity: the effects of opacity}.
\newblock {\em \mnras}, {\bf 442}(July), 285--313.

\bibitem[\protect\citename{{Bate} {et~al.}, }2002]{Bate+02}
{Bate}, M.~R., {Bonnell}, I.~A., and {Bromm}, V. 2002.
\newblock {The formation mechanism of brown dwarfs}.
\newblock {\em \mnras}, {\bf 332}(May), L65--L68.

\bibitem[\protect\citename{{Bate} {et~al.}, }2014]{Bate+14}
{Bate}, M.~R., {Tricco}, T.~S., and {Price}, D.~J. 2014.
\newblock {Collapse of a molecular cloud core to stellar densities:
  stellar-core and outflow formation in radiation magnetohydrodynamic
  simulations}.
\newblock {\em \mnras}, {\bf 437}(Jan.), 77--95.

\bibitem[\protect\citename{{Baumgardt} and {Makino}, }2003]{BM03}
{Baumgardt}, H., and {Makino}, J. 2003.
\newblock {Dynamical evolution of star clusters in tidal fields}.
\newblock {\em \mnras}, {\bf 340}(Mar.), 227--246.

\bibitem[\protect\citename{{Baumgardt} {et~al.}, }2008]{Baumgardt+08}
{Baumgardt}, H., {De Marchi}, G., and {Kroupa}, P. 2008.
\newblock {Evidence for Primordial Mass Segregation in Globular Clusters}.
\newblock {\em \apj}, {\bf 685}(Sept.), 247--253.

\bibitem[\protect\citename{{Belloni} {et~al.}, }2017]{Belloni+17}
{Belloni}, D., {Askar}, A., {Giersz}, M., {Kroupa}, P., and {Rocha-Pinto},
  H.~J. 2017.
\newblock {On the initial binary population for star cluster simulations}.
\newblock {\em \mnras}, {\bf 471}(Nov.), 2812--2828.

\bibitem[\protect\citename{{Belloni} {et~al.}, }2018]{Belloni+18}
{Belloni}, D., {Kroupa}, P., {Rocha-Pinto}, H.~J., and {Giersz}, M. 2018.
\newblock {Dynamical equivalence, the origin of the Galactic field stellar and
  binary population, and the initial radius-mass relation of embedded
  clusters}.
\newblock {\em \mnras}, {\bf 474}(Mar.), 3740--3745.

\bibitem[\protect\citename{{Bertelli Motta} {et~al.}, }2016]{Bertelli16}
{Bertelli Motta}, C., {Clark}, P.~C., {Glover}, S.~C.~O., {Klessen}, R.~S., and
  {Pasquali}, A. 2016.
\newblock {The IMF as a function of supersonic turbulence}.
\newblock {\em \mnras}, {\bf 462}(Nov.), 4171--4182.

\bibitem[\protect\citename{{Beuther} {et~al.}, }2007]{Beuther+07}
{Beuther}, H., {Churchwell}, E.~B., {McKee}, C.~F., and {Tan}, J.~C. 2007
  (Jan.).
\newblock {The Formation of Massive Stars}.
\newblock {Page  165 of:} {Reipurth}, Bo, {Jewitt}, David, and {Keil}, Klaus
  (eds), {\em Protostars and Planets V}.

\bibitem[\protect\citename{{Boissier} {et~al.}, }2007]{Boissier07}
{Boissier}, S., {Gil de Paz}, A., {Boselli}, A., {Madore}, B.~F., {Buat}, V.,
  {Cortese}, L., {Burgarella}, D., {Mu{\~n}oz-Mateos}, J.~C., {Barlow}, T.~A.,
  {Forster}, K., {Friedman}, P.~G., {Martin}, D.~C., {Morrissey}, P., {Neff},
  S.~G., {Schiminovich}, D., {Seibert}, M., {Small}, T., {Wyder}, T.~K.,
  {Bianchi}, L., {Donas}, J., {Heckman}, T.~M., {Lee}, Y.-W., {Milliard}, B.,
  {Rich}, R.~M., {Szalay}, A.~S., {Welsh}, B.~Y., and {Yi}, S.~K. 2007.
\newblock {Radial Variation of Attenuation and Star Formation in the Largest
  Late-Type Disks Observed with GALEX}.
\newblock {\em \apjs}, {\bf 173}(Dec.), 524--537.

\bibitem[\protect\citename{{Bonnell} and {Davies}, }1998]{BD98}
{Bonnell}, I.~A., and {Davies}, M.~B. 1998.
\newblock {Mass segregation in young stellar clusters}.
\newblock {\em \mnras}, {\bf 295}(Apr.), 691.

\bibitem[\protect\citename{{Bontemps} {et~al.}, }2010]{Bontemps+10}
{Bontemps}, S., {Motte}, F., {Csengeri}, T., and {Schneider}, N. 2010.
\newblock {Fragmentation and mass segregation in the massive dense cores of
  Cygnus X}.
\newblock {\em \aap}, {\bf 524}(Dec.), A18.

\bibitem[\protect\citename{{Breitschwerdt} {et~al.}, }2012]{Breitschwerdt+12}
{Breitschwerdt}, D., {de Avillez}, M.~A., {Feige}, J., and {Dettbarn}, C. 2012.
\newblock {Interstellar medium simulations}.
\newblock {\em Astronomische Nachrichten}, {\bf 333}(June), 486.

\bibitem[\protect\citename{{Brinkmann} {et~al.}, }2017]{Brinkmann+17}
{Brinkmann}, N., {Banerjee}, S., {Motwani}, B., and {Kroupa}, P. 2017.
\newblock {The bound fraction of young star clusters}.
\newblock {\em \aap}, {\bf 600}(Apr.), A49.

\bibitem[\protect\citename{{Burkert}, }2006]{Burkert06}
{Burkert}, A. 2006.
\newblock {The turbulent interstellar medium}.
\newblock {\em Comptes Rendus Physique}, {\bf 7}(Apr.), 433--441.

\bibitem[\protect\citename{{Calzetti}, }2008]{Calzetti08}
{Calzetti}, D. 2008 (June).
\newblock {Star Formation Rate Determinations}.
\newblock {Page  121 of:} {Knapen}, J.~H., {Mahoney}, T.~J., and {Vazdekis}, A.
  (eds), {\em Pathways Through an Eclectic Universe}.
\newblock Astronomical Society of the Pacific Conference Series, vol. 390.

\bibitem[\protect\citename{{Calzetti}, }2013]{Calzetti13}
{Calzetti}, D. 2013.
\newblock {\em {Star Formation Rate Indicators}}.
\newblock Page  419.

\bibitem[\protect\citename{{Cappellari}, }2016]{Cappellari16}
{Cappellari}, M. 2016.
\newblock {Structure and Kinematics of Early-Type Galaxies from Integral Field
  Spectroscopy}.
\newblock {\em \araa}, {\bf 54}(Sept.), 597--665.

\bibitem[\protect\citename{{Cervi{\~n}o} {et~al.}, }2013a]{Cervino13a}
{Cervi{\~n}o}, M., {Rom{\'a}n-Z{\'u}{\~n}iga}, C., {Luridiana}, V., {Bayo}, A.,
  {S{\'a}nchez}, N., and {P{\'e}rez}, E. 2013a.
\newblock {Crucial aspects of the initial mass function. I. The statistical
  correlation between the total mass of an ensemble of stars and its most
  massive star}.
\newblock {\em \aap}, {\bf 553}(May), A31.

\bibitem[\protect\citename{{Cervi{\~n}o} {et~al.}, }2013b]{Cervino13b}
{Cervi{\~n}o}, M., {Rom{\'a}n-Z{\'u}{\~n}iga}, C., {Bayo}, A., {Luridiana}, V.,
  {S{\'a}nchez}, N., and {P{\'e}rez}, E. 2013b.
\newblock {Crucial aspects of the initial mass function. II. The inference of
  total quantities from partial information on a cluster}.
\newblock {\em \aap}, {\bf 553}(May), A32.

\bibitem[\protect\citename{{Chabrier}, }2003]{Chabrier03}
{Chabrier}, G. 2003.
\newblock {Galactic Stellar and Substellar Initial Mass Function}.
\newblock {\em \pasp}, {\bf 115}(July), 763--795.

\bibitem[\protect\citename{{Chen{\'e}} {et~al.}, }2015]{Chene+15}
{Chen{\'e}}, A.-N., {Ram{\'{\i}}rez Alegr{\'{\i}}a}, S., {Borissova}, J.,
  {O'Leary}, E., {Martins}, F., {Herv{\'e}}, A., {Kuhn}, M., {Kurtev}, R.,
  {Consuelo Amigo Fuentes}, P., {Bonatto}, C., and {Minniti}, D. 2015.
\newblock {Massive open star clusters using the VVV survey. IV. WR 62-2, a new
  very massive star in the core of the VVV CL041 cluster}.
\newblock {\em \aap}, {\bf 584}(Dec.), A31.

\bibitem[\protect\citename{{Concas} {et~al.}, }2017]{Concas+17}
{Concas}, A., {Popesso}, P., {Brusa}, M., {Mainieri}, V., {Erfanianfar}, G.,
  and {Morselli}, L. 2017.
\newblock {Light breeze in the local Universe}.
\newblock {\em \aap}, {\bf 606}(Oct.), A36.

\bibitem[\protect\citename{{Conroy} and {van Dokkum}, }2012]{CvD12}
{Conroy}, C., and {van Dokkum}, P.~G. 2012.
\newblock {The Stellar Initial Mass Function in Early-type Galaxies From
  Absorption Line Spectroscopy. II. Results}.
\newblock {\em \apj}, {\bf 760}(Nov.), 71.

\bibitem[\protect\citename{{Crowther} {et~al.}, }2010]{Crowther10}
{Crowther}, P.~A., {Schnurr}, O., {Hirschi}, R., {Yusof}, N., {Parker}, R.~J.,
  {Goodwin}, S.~P., and {Kassim}, H.~A. 2010.
\newblock {The R136 star cluster hosts several stars whose individual masses
  greatly exceed the accepted 150M$_{solar}$ stellar mass limit}.
\newblock {\em \mnras}, {\bf 408}(Oct.), 731--751.

\bibitem[\protect\citename{{da Silva} {et~al.}, }2012]{daSilva12}
{da Silva}, R.~L., {Fumagalli}, M., and {Krumholz}, M. 2012.
\newblock {SLUG: Stochastically Lighting Up Galaxies. I. Methods and Validating
  Tests}.
\newblock {\em \apj}, {\bf 745}(Feb.), 145.

\bibitem[\protect\citename{{da Silva} {et~al.}, }2014]{daSilva14}
{da Silva}, R.~L., {Fumagalli}, M., and {Krumholz}, M.~R. 2014.
\newblock {SLUG - Stochastically Lighting Up Galaxies - II. Quantifying the
  effects of stochasticity on star formation rate indicators}.
\newblock {\em \mnras}, {\bf 444}(Nov.), 3275--3287.

\bibitem[\protect\citename{{Dabringhausen} {et~al.}, }2009]{Dab09}
{Dabringhausen}, J., {Kroupa}, P., and {Baumgardt}, H. 2009.
\newblock {A top-heavy stellar initial mass function in starbursts as an
  explanation for the high mass-to-light ratios of ultra-compact dwarf
  galaxies}.
\newblock {\em \mnras}, {\bf 394}(Apr.), 1529--1543.

\bibitem[\protect\citename{{Dabringhausen} {et~al.}, }2010]{Dab10}
{Dabringhausen}, J., {Fellhauer}, M., and {Kroupa}, P. 2010.
\newblock {Mass loss and expansion of ultra compact dwarf galaxies through gas
  expulsion and stellar evolution for top-heavy stellar initial mass
  functions}.
\newblock {\em \mnras}, {\bf 403}(Apr.), 1054--1071.

\bibitem[\protect\citename{{Dabringhausen} {et~al.}, }2012]{Dab12}
{Dabringhausen}, J., {Kroupa}, P., {Pflamm-Altenburg}, J., and {Mieske}, S.
  2012.
\newblock {Low-mass X-Ray Binaries Indicate a Top-heavy Stellar Initial Mass
  Function in Ultracompact Dwarf Galaxies}.
\newblock {\em \apj}, {\bf 747}(Mar.), 72.

\bibitem[\protect\citename{{de Mink} {et~al.}, }2014]{deMink+14}
{de Mink}, S.~E., {Sana}, H., {Langer}, N., {Izzard}, R.~G., and {Schneider},
  F.~R.~N. 2014.
\newblock {The Incidence of Stellar Mergers and Mass Gainers among Massive
  Stars}.
\newblock {\em \apj}, {\bf 782}(Feb.), 7.

\bibitem[\protect\citename{{Delgado-Serrano} {et~al.}, }2010]{Delgado10}
{Delgado-Serrano}, R., {Hammer}, F., {Yang}, Y.~B., {Puech}, M., {Flores}, H.,
  and {Rodrigues}, M. 2010.
\newblock {How was the Hubble sequence 6 Gyr ago?}
\newblock {\em \aap}, {\bf 509}(Jan.), A78.

\bibitem[\protect\citename{{Dib}, }2014]{Dib14}
{Dib}, S. 2014.
\newblock {Testing the universality of the IMF with Bayesian statistics: young
  clusters}.
\newblock {\em \mnras}, {\bf 444}(Oct.), 1957--1981.

\bibitem[\protect\citename{{Dib} {et~al.}, }2007]{Dib07}
{Dib}, S., {Kim}, J., and {Shadmehri}, M. 2007.
\newblock {The origin of the Arches stellar cluster mass function}.
\newblock {\em \mnras}, {\bf 381}(Oct.), L40--L44.

\bibitem[\protect\citename{{Dib} {et~al.}, }2017]{Dib17}
{Dib}, S., {Schmeja}, S., and {Hony}, S. 2017.
\newblock {Massive stars reveal variations of the stellar initial mass function
  in the Milky Way stellar clusters}.
\newblock {\em \mnras}, {\bf 464}(Jan.), 1738--1752.

\bibitem[\protect\citename{{Disney} {et~al.}, }2008]{Disney08}
{Disney}, M.~J., {Romano}, J.~D., {Garcia-Appadoo}, D.~A., {West}, A.~A.,
  {Dalcanton}, J.~J., and {Cortese}, L. 2008.
\newblock {Galaxies appear simpler than expected}.
\newblock {\em \nat}, {\bf 455}(Oct.), 1082--1084.

\bibitem[\protect\citename{{Duarte-Cabral} {et~al.}, }2013]{Duarte-Cabral+13}
{Duarte-Cabral}, A., {Bontemps}, S., {Motte}, F., {Hennemann}, M., {Schneider},
  N., and {Andr{\'e}}, P. 2013.
\newblock {CO outflows from high-mass Class 0 protostars in Cygnus-X}.
\newblock {\em \aap}, {\bf 558}(Oct.), A125.

\bibitem[\protect\citename{{Egusa} {et~al.}, }2004]{Egusa+04}
{Egusa}, F., {Sofue}, Y., and {Nakanishi}, H. 2004.
\newblock {Offsets between H{$\alpha$} and CO Arms of a Spiral Galaxy, NGC
  4254: A New Method for Determining the Pattern Speed of Spiral Galaxies}.
\newblock {\em \pasj}, {\bf 56}(Dec.), L45--L48.

\bibitem[\protect\citename{{Egusa} {et~al.}, }2009]{Egusa+09}
{Egusa}, F., {Kohno}, K., {Sofue}, Y., {Nakanishi}, H., and {Komugi}, S. 2009.
\newblock {Determining Star Formation Timescale and Pattern Speed in Nearby
  Spiral Galaxies}.
\newblock {\em \apj}, {\bf 697}(June), 1870--1891.

\bibitem[\protect\citename{{Egusa} {et~al.}, }2017]{Egusa+17}
{Egusa}, F., {Mentuch Cooper}, E., {Koda}, J., and {Baba}, J. 2017.
\newblock {Gas and stellar spiral arms and their offsets in the grand-design
  spiral galaxy M51}.
\newblock {\em \mnras}, {\bf 465}(Feb.), 460--471.

\bibitem[\protect\citename{{Elmegreen}, }1997]{Elm97}
{Elmegreen}, B.~G. 1997.
\newblock {The Initial Stellar Mass Function from Random Sampling in a
  Turbulent Fractal Cloud}.
\newblock {\em \apj}, {\bf 486}(Sept.), 944--954.

\bibitem[\protect\citename{{Elmegreen}, }1999]{Elmegreen99}
{Elmegreen}, B.~G. 1999.
\newblock {The Stellar Initial Mass Function from Random Sampling in
  Hierarchical Clouds. II. Statistical Fluctuations and a Mass Dependence for
  Starbirth Positions and Times}.
\newblock {\em \apj}, {\bf 515}(Apr.), 323--336.

\bibitem[\protect\citename{{Elmegreen}, }2000]{Elm00}
{Elmegreen}, B.~G. 2000.
\newblock {Modeling a High-Mass Turn-Down in the Stellar Initial Mass
  Function}.
\newblock {\em \apj}, {\bf 539}(Aug.), 342--351.

\bibitem[\protect\citename{{Elmegreen}, }2004]{Elmegreen04}
{Elmegreen}, B.~G. 2004.
\newblock {Variability in the stellar initial mass function at low and high
  mass: three-component IMF models}.
\newblock {\em \mnras}, {\bf 354}(Oct.), 367--374.

\bibitem[\protect\citename{{Elmegreen} and {Scalo}, }2006]{ES06}
{Elmegreen}, B.~G., and {Scalo}, J. 2006.
\newblock {The Effect of Star Formation History on the Inferred Stellar Initial
  Mass Function}.
\newblock {\em \apj}, {\bf 636}(Jan.), 149--157.

\bibitem[\protect\citename{{Federrath}, }2015]{Federrath15}
{Federrath}, C. 2015.
\newblock {Inefficient star formation through turbulence, magnetic fields and
  feedback}.
\newblock {\em \mnras}, {\bf 450}(July), 4035--4042.

\bibitem[\protect\citename{{Federrath}, }2016]{Federrath16}
{Federrath}, C. 2016.
\newblock {On the universality of interstellar filaments: theory meets
  simulations and observations}.
\newblock {\em \mnras}, {\bf 457}(Mar.), 375--388.

\bibitem[\protect\citename{{Federrath} {et~al.}, }2014]{Federrath+14}
{Federrath}, C., {Schr{\"o}n}, M., {Banerjee}, R., and {Klessen}, R.~S. 2014.
\newblock {Modeling Jet and Outflow Feedback during Star Cluster Formation}.
\newblock {\em \apj}, {\bf 790}(Aug.), 128.

\bibitem[\protect\citename{{Ferreras} {et~al.}, }2015]{Ferreras+15}
{Ferreras}, I., {Weidner}, C., {Vazdekis}, A., and {La Barbera}, F. 2015.
\newblock {Further evidence for a time-dependent initial mass function in
  massive early-type galaxies}.
\newblock {\em \mnras}, {\bf 448}(Mar.), L82--L86.

\bibitem[\protect\citename{{Figer}, }2005]{Figer05}
{Figer}, D.~F. 2005.
\newblock {An upper limit to the masses of stars}.
\newblock {\em \nat}, {\bf 434}(Mar.), 192--194.

\bibitem[\protect\citename{{Fukui} and {Kawamura}, }2010]{FK10}
{Fukui}, Y., and {Kawamura}, A. 2010.
\newblock {Molecular Clouds in Nearby Galaxies}.
\newblock {\em \araa}, {\bf 48}(Sept.), 547--580.

\bibitem[\protect\citename{{Fumagalli} {et~al.}, }2011]{Fumagalli11}
{Fumagalli}, M., {da Silva}, R.~L., and {Krumholz}, M.~R. 2011.
\newblock {Stochastic Star Formation and a (Nearly) Uniform Stellar Initial
  Mass Function}.
\newblock {\em \apjl}, {\bf 741}(Nov.), L26.

\bibitem[\protect\citename{{Garcia} {et~al.}, }2019]{Garcia+19}
{Garcia}, Miriam, {Herrero}, Artemio, {Najarro}, Francisco, {Camacho},
  In{\'e}s, and {Lorenzo}, Marta. 2019.
\newblock {Ongoing star formation at the outskirts of Sextans A: spectroscopic
  detection of early O-type stars}.
\newblock {\em \mnras}, {\bf 484}(1), 422--430.

\bibitem[\protect\citename{{Gargiulo} {et~al.}, }2015]{Gargiulo+15}
{Gargiulo}, I.~D., {Cora}, S.~A., {Padilla}, N.~D., {Mu{\~n}oz Arancibia},
  A.~M., {Ruiz}, A.~N., {Orsi}, A.~A., {Tecce}, T.~E., {Weidner}, C., and
  {Bruzual}, G. 2015.
\newblock {Chemoarchaeological downsizing in a hierarchical universe: impact of
  a top-heavy IGIMF}.
\newblock {\em \mnras}, {\bf 446}(Feb.), 3820--3841.

\bibitem[\protect\citename{{Gieles} {et~al.}, }2012]{Gieles+12}
{Gieles}, M., {Moeckel}, N., and {Clarke}, C.~J. 2012.
\newblock {Do all stars in the solar neighbourhood form in clusters? A
  cautionary note on the use of the distribution of surface densities}.
\newblock {\em \mnras}, {\bf 426}(Oct.), L11--L15.

\bibitem[\protect\citename{{Goodwin} and {Kroupa}, }2005]{GK05}
{Goodwin}, S.~P., and {Kroupa}, P. 2005.
\newblock {Limits on the primordial stellar multiplicity}.
\newblock {\em \aap}, {\bf 439}(Aug.), 565--569.

\bibitem[\protect\citename{{Gunawardhana} {et~al.}, }2011]{Gun11}
{Gunawardhana}, M.~L.~P., {Hopkins}, A.~M., {Sharp}, R.~G., {Brough}, S.,
  {Taylor}, E., {Bland-Hawthorn}, J., {Maraston}, C., {Tuffs}, R.~J.,
  {Popescu}, C.~C., {Wijesinghe}, D., {Jones}, D.~H., {Croom}, S., {Sadler},
  E., {Wilkins}, S., {Driver}, S.~P., {Liske}, J., {Norberg}, P., {Baldry},
  I.~K., {Bamford}, S.~P., {Loveday}, J., {Peacock}, J.~A., {Robotham},
  A.~S.~G., {Zucker}, D.~B., {Parker}, Q.~A., {Conselice}, C.~J., {Cameron},
  E., {Frenk}, C.~S., {Hill}, D.~T., {Kelvin}, L.~S., {Kuijken}, K., {Madore},
  B.~F., {Nichol}, B., {Parkinson}, H.~R., {Pimbblet}, K.~A., {Prescott}, M.,
  {Sutherland}, W.~J., {Thomas}, D., and {van Kampen}, E. 2011.
\newblock {Galaxy and Mass Assembly (GAMA): the star formation rate dependence
  of the stellar initial mass function}.
\newblock {\em \mnras}, {\bf 415}(Aug.), 1647--1662.

\bibitem[\protect\citename{{Gvaramadze} {et~al.}, }2012]{Gvaramadze12}
{Gvaramadze}, V.~V., {Weidner}, C., {Kroupa}, P., and {Pflamm-Altenburg}, J.
  2012.
\newblock {Field O stars: formed in situ or as runaways?}
\newblock {\em \mnras}, {\bf 424}(Aug.), 3037--3049.

\bibitem[\protect\citename{{Hacar} {et~al.}, }2013]{Hacar+13}
{Hacar}, A., {Tafalla}, M., {Kauffmann}, J., and {Kov{\'a}cs}, A. 2013.
\newblock {Cores, filaments, and bundles: hierarchical core formation in the
  L1495/B213 Taurus region}.
\newblock {\em \aap}, {\bf 554}(June), A55.

\bibitem[\protect\citename{{Hacar} {et~al.}, }2017a]{Hacar+17b}
{Hacar}, A., {Tafalla}, M., and {Alves}, J. 2017a.
\newblock {Fibers in the NGC 1333 proto-cluster}.
\newblock {\em \aap}, {\bf 606}(Oct.), A123.

\bibitem[\protect\citename{{Hacar} {et~al.}, }2017b]{Hacar+17}
{Hacar}, A., {Alves}, J., {Tafalla}, M., and {Goicoechea}, J.~R. 2017b.
\newblock {Gravitational collapse of the OMC-1 region}.
\newblock {\em \aap}, {\bf 602}(June), L2.

\bibitem[\protect\citename{{Haghi} {et~al.}, }2015]{Haghi15}
{Haghi}, H., {Zonoozi}, A.~H., {Kroupa}, P., {Banerjee}, S., and {Baumgardt},
  H. 2015.
\newblock {Possible smoking-gun evidence for initial mass segregation in
  re-virialized post-gas expulsion globular clusters}.
\newblock {\em \mnras}, {\bf 454}(Dec.), 3872--3885.

\bibitem[\protect\citename{{Haghi} {et~al.}, }2017]{Haghi17}
{Haghi}, H., {Khalaj}, P., {Hasani Zonoozi}, A., and {Kroupa}, P. 2017.
\newblock {A Possible Solution for the M/L-[Fe/H] Relation of Globular Clusters
  in M31. II. The Age-Metallicity Relation}.
\newblock {\em \apj}, {\bf 839}(Apr.), 60.

\bibitem[\protect\citename{{Haugb{\o}lle} {et~al.}, }2018]{Haugbolle+18}
{Haugb{\o}lle}, T., {Padoan}, P., and {Nordlund}, {\AA}. 2018.
\newblock {The Stellar IMF from Isothermal MHD Turbulence}.
\newblock {\em \apj}, {\bf 854}(Feb.), 35.

\bibitem[\protect\citename{{Heggie} and {Hut}, }2003]{HH03}
{Heggie}, D., and {Hut}, P. 2003.
\newblock {\em {The Gravitational Million-Body Problem: A Multidisciplinary
  Approach to Star Cluster Dynamics}}.

\bibitem[\protect\citename{{Hennebelle} and {Chabrier}, }2013]{HB13}
{Hennebelle}, P., and {Chabrier}, G. 2013.
\newblock {Analytical Theory for the Initial Mass Function. III. Time
  Dependence and Star Formation Rate}.
\newblock {\em \apj}, {\bf 770}(June), 150.

\bibitem[\protect\citename{{Henry} {et~al.}, }2018]{Henry+18}
{Henry}, T.~J., {Jao}, W.-C., {Winters}, J.~G., {Dieterich}, S.~B., {Finch},
  C.~T., {Ianna}, P.~A., {Riedel}, A.~R., {Silverstein}, M.~L., {Subasavage},
  J.~P., and {Vrijmoet}, E.~H. 2018.
\newblock {The Solar Neighborhood XLIV: RECONS Discoveries within 10 parsecs}.
\newblock {\em \aj}, {\bf 155}(June), 265.

\bibitem[\protect\citename{{Heyer} and {Dame}, }2015]{HD15}
{Heyer}, M., and {Dame}, T.~M. 2015.
\newblock {Molecular Clouds in the Milky Way}.
\newblock {\em \araa}, {\bf 53}(Aug.), 583--629.

\bibitem[\protect\citename{{Hopkins}, }2018]{Hopkins18}
{Hopkins}, A.~M. 2018.
\newblock {The Dawes Review 8: Measuring the Stellar Initial Mass Function}.
\newblock {\em \pasa}, {\bf 35}(Nov.), e039.

\bibitem[\protect\citename{{Hopkins}, }2012]{Hopkins12}
{Hopkins}, P.~F. 2012.
\newblock {The stellar initial mass function, core mass function and the
  last-crossing distribution}.
\newblock {\em \mnras}, {\bf 423}(July), 2037--2044.

\bibitem[\protect\citename{{Hopkins}, }2013a]{Hopkins13}
{Hopkins}, P.~F. 2013a.
\newblock {Variations in the stellar CMF and IMF: from bottom to top}.
\newblock {\em \mnras}, {\bf 433}(July), 170--177.

\bibitem[\protect\citename{{Hopkins}, }2013b]{Hopkins13a}
{Hopkins}, P.~F. 2013b.
\newblock {Why do stars form in clusters? An analytic model for stellar
  correlation functions}.
\newblock {\em \mnras}, {\bf 428}(Jan.), 1950--1957.

\bibitem[\protect\citename{{Hosek} {et~al.}, }2019]{Hosek18}
{Hosek}, Matthew~W., Jr., {Lu}, Jessica~R., {Anderson}, Jay, {Najarro},
  Francisco, {Ghez}, Andrea~M., {Morris}, Mark~R., {Clarkson}, William~I., and
  {Albers}, Saundra~M. 2019.
\newblock {The Unusual Initial Mass Function of the Arches Cluster}.
\newblock {\em \apj}, {\bf 870}(1), 44.

\bibitem[\protect\citename{{Hoversten} and {Glazebrook}, }2008]{HG08}
{Hoversten}, E.~A., and {Glazebrook}, K. 2008.
\newblock {Evidence for a Nonuniversal Stellar Initial Mass Function from the
  Integrated Properties of SDSS Galaxies}.
\newblock {\em \apj}, {\bf 675}(Mar.), 163--187.

\bibitem[\protect\citename{{Hsu} {et~al.}, }2012]{Hsu12}
{Hsu}, W.-H., {Hartmann}, L., {Allen}, L., {Hern{\'a}ndez}, J., {Megeath},
  S.~T., {Mosby}, G., {Tobin}, J.~J., and {Espaillat}, C. 2012.
\newblock {The Low-mass Stellar Population in L1641: Evidence for Environmental
  Dependence of the Stellar Initial Mass Function}.
\newblock {\em \apj}, {\bf 752}(June), 59.

\bibitem[\protect\citename{{Je{\v r}{\'a}bkov{\'a}} {et~al.},
  }2017]{Jerabkova+17}
{Je{\v r}{\'a}bkov{\'a}}, T., {Kroupa}, P., {Dabringhausen}, J., {Hilker}, M.,
  and {Bekki}, K. 2017.
\newblock {The formation of ultra compact dwarf galaxies and massive globular
  clusters. Quasar-like objects to test for a variable stellar initial mass
  function}.
\newblock {\em \aap}, {\bf 608}(Dec.), A53.

\bibitem[\protect\citename{{Je{\v{r}}{\'a}bkov{\'a}} {et~al.},
  }2018]{Jerabkova+18}
{Je{\v{r}}{\'a}bkov{\'a}}, T., {Hasani Zonoozi}, A., {Kroupa}, P., {Beccari},
  G., {Yan}, Z., {Vazdekis}, A., and {Zhang}, Z.~Y. 2018.
\newblock {Impact of metallicity and star formation rate on the time-dependent,
  galaxy-wide stellar initial mass function}.
\newblock {\em \aap}, {\bf 620}(Nov), A39.

\bibitem[\protect\citename{{Joncour} {et~al.}, }2018]{Joncour+18}
{Joncour}, Isabelle, {Duch{\^e}ne}, Gaspard, {Moraux}, Estelle, and {Motte},
  Fr{\'e}d{\'e}rique. 2018.
\newblock {Multiplicity and clustering in Taurus star forming region. II. From
  ultra-wide pairs to dense NESTs}.
\newblock {\em \aap}, {\bf 620}(Nov.), A27.

\bibitem[\protect\citename{{Kalari} {et~al.}, }2018]{Kalari18}
{Kalari}, V.~M., {Carraro}, G., {Evans}, C.~J., and {Rubio}, M. 2018.
\newblock {The Magellanic Bridge Cluster NGC 796: Deep Optical AO Imaging
  Reveals the Stellar Content and Initial Mass Function of a Massive Open
  Cluster}.
\newblock {\em \apj}, {\bf 857}(Apr.), 132.

\bibitem[\protect\citename{{Kennicutt} and {Evans}, }2012]{KE12}
{Kennicutt}, R.~C., and {Evans}, N.~J. 2012.
\newblock {Star Formation in the Milky Way and Nearby Galaxies}.
\newblock {\em \araa}, {\bf 50}(Sept.), 531--608.

\bibitem[\protect\citename{{Kennicutt}, }1989]{Kennicutt89}
{Kennicutt}, Jr., R.~C. 1989.
\newblock {The star formation law in galactic disks}.
\newblock {\em \apj}, {\bf 344}(Sept.), 685--703.

\bibitem[\protect\citename{{Kennicutt}, }1998]{Kennicutt98}
{Kennicutt}, Jr., R.~C. 1998.
\newblock {The Global Schmidt Law in Star-forming Galaxies}.
\newblock {\em \apj}, {\bf 498}(May), 541--552.

\bibitem[\protect\citename{{Kirk} and {Myers}, }2011]{KM11}
{Kirk}, H., and {Myers}, P.~C. 2011.
\newblock {Young Stellar Groups and Their Most Massive Stars}.
\newblock {\em \apj}, {\bf 727}(Feb.), 64.

\bibitem[\protect\citename{{Kirk} and {Myers}, }2012]{KM12}
{Kirk}, H., and {Myers}, P.~C. 2012.
\newblock {Variations in the Mass Functions of Clustered and Isolated Young
  Stellar Objects}.
\newblock {\em \apj}, {\bf 745}(Feb.), 131.

\bibitem[\protect\citename{{Kirk} {et~al.}, }2014]{Kirk14}
{Kirk}, H., {Offner}, S.~S.~R., and {Redmond}, K.~J. 2014.
\newblock {The formation and evolution of small star clusters}.
\newblock {\em \mnras}, {\bf 439}(Feb.), 1765--1780.

\bibitem[\protect\citename{{Kirk} {et~al.}, }2016]{Kirk16}
{Kirk}, H., {Johnstone}, D., {Di Francesco}, J., {Lane}, J., {Buckle}, J.,
  {Berry}, D.~S., {Broekhoven-Fiene}, H., {Currie}, M.~J., {Fich}, M.,
  {Hatchell}, J., {Jenness}, T., {Mottram}, J.~C., {Nutter}, D., {Pattle}, K.,
  {Pineda}, J.~E., {Quinn}, C., {Salji}, C., {Tisi}, S., {Hogerheijde}, M.~R.,
  {Ward-Thompson}, D., and {The JCMT Gould Belt Survey Team}. 2016.
\newblock {The JCMT Gould Belt Survey: Dense Core Clusters in Orion B}.
\newblock {\em \apj}, {\bf 821}(Apr.), 98.

\bibitem[\protect\citename{{Koen}, }2006]{Koen06}
{Koen}, C. 2006.
\newblock {On the upper limit on stellar masses in the Large Magellanic Cloud
  cluster R136}.
\newblock {\em \mnras}, {\bf 365}(Jan.), 590--594.

\bibitem[\protect\citename{{K{\"o}nyves} {et~al.}, }2015]{Konyves+15}
{K{\"o}nyves}, V., {Andr{\'e}}, P., {Men'shchikov}, A., {Palmeirim}, P.,
  {Arzoumanian}, D., {Schneider}, N., {Roy}, A., {Didelon}, P., {Maury}, A.,
  {Shimajiri}, Y., {Di Francesco}, J., {Bontemps}, S., {Peretto}, N.,
  {Benedettini}, M., {Bernard}, J.-P., {Elia}, D., {Griffin}, M.~J., {Hill},
  T., {Kirk}, J., {Ladjelate}, B., {Marsh}, K., {Martin}, P.~G., {Motte}, F.,
  {Nguy{\^e}n Luong}, Q., {Pezzuto}, S., {Roussel}, H., {Rygl}, K.~L.~J.,
  {Sadavoy}, S.~I., {Schisano}, E., {Spinoglio}, L., {Ward-Thompson}, D., and
  {White}, G.~J. 2015.
\newblock {A census of dense cores in the Aquila cloud complex: SPIRE/PACS
  observations from the Herschel Gould Belt survey}.
\newblock {\em \aap}, {\bf 584}(Dec.), A91.

\bibitem[\protect\citename{{K{\"o}ppen} {et~al.}, }2007]{Koeppen07}
{K{\"o}ppen}, J., {Weidner}, C., and {Kroupa}, P. 2007.
\newblock {A possible origin of the mass-metallicity relation of galaxies}.
\newblock {\em \mnras}, {\bf 375}(Feb.), 673--684.

\bibitem[\protect\citename{{Kristensen} and {Dunham},
  }2018]{KristensenDunham18}
{Kristensen}, L.~E., and {Dunham}, M.~M. 2018.
\newblock {Protostellar half-life: new methodology and estimates}.
\newblock {\em \aap}, {\bf 618}(Oct.), A158.

\bibitem[\protect\citename{{Kroupa}, }1995a]{Kroupa95}
{Kroupa}, P. 1995a.
\newblock {Inverse dynamical population synthesis and star formation}.
\newblock {\em \mnras}, {\bf 277}(Dec.), 1491.

\bibitem[\protect\citename{{Kroupa}, }1995b]{Kroupa95a}
{Kroupa}, P. 1995b.
\newblock {The dynamical properties of stellar systems in the Galactic disc}.
\newblock {\em \mnras}, {\bf 277}(Dec.).

\bibitem[\protect\citename{{Kroupa}, }1995c]{Kroupa95MF}
{Kroupa}, P. 1995c.
\newblock {Unification of the nearby and photometric stellar luminosity
  functions}.
\newblock {\em \apj}, {\bf 453}(Nov.), 358.

\bibitem[\protect\citename{{Kroupa}, }2001]{Kroupa01}
{Kroupa}, P. 2001.
\newblock {On the variation of the initial mass function}.
\newblock {\em \mnras}, {\bf 322}(Apr.), 231--246.

\bibitem[\protect\citename{{Kroupa}, }2002a]{Kroupa02b}
{Kroupa}, P. 2002a.
\newblock {The Initial Mass Function of Stars: Evidence for Uniformity in
  Variable Systems}.
\newblock {\em Science}, {\bf 295}(Jan.), 82--91.

\bibitem[\protect\citename{{Kroupa}, }2002b]{Kroupa02}
{Kroupa}, P. 2002b.
\newblock {Thickening of galactic discs through clustered star formation}.
\newblock {\em \mnras}, {\bf 330}(Mar.), 707--718.

\bibitem[\protect\citename{{Kroupa}, }2005]{Kroupa05}
{Kroupa}, P. 2005 (Jan.).
\newblock {The Fundamental Building Blocks of Galaxies}.
\newblock {Page  629 of:} {Turon}, C., {O'Flaherty}, K.~S., and {Perryman},
  M.~A.~C. (eds), {\em The Three-Dimensional Universe with Gaia}.
\newblock ESA Special Publication, vol. 576.

\bibitem[\protect\citename{{Kroupa}, }2008]{Kroupa08}
{Kroupa}, P. 2008.
\newblock {Initial Conditions for Star Clusters}.
\newblock {Page  181 of:} {Aarseth}, S.~J., {Tout}, C.~A., and {Mardling},
  R.~A. (eds), {\em The Cambridge N-Body Lectures}.
\newblock Lecture Notes in Physics, Berlin Springer Verlag, vol. 760.

\bibitem[\protect\citename{{Kroupa} and {Bouvier}, }2003]{KB03}
{Kroupa}, P., and {Bouvier}, J. 2003.
\newblock {The dynamical evolution of Taurus-Auriga-type aggregates}.
\newblock {\em \mnras}, {\bf 346}(Dec.), 343--353.

\bibitem[\protect\citename{{Kroupa} and {Jerabkova}, }2018]{KJ18}
{Kroupa}, P., and {Jerabkova}, T. 2018.
\newblock {The Impact of Binaries on the Stellar Initial Mass Function}.
\newblock {\em arXiv:1806.10605}, June.

\bibitem[\protect\citename{{Kroupa} and {Weidner}, }2003]{KW03}
{Kroupa}, P., and {Weidner}, C. 2003.
\newblock {Galactic-Field Initial Mass Functions of Massive Stars}.
\newblock {\em \apj}, {\bf 598}(Dec.), 1076--1078.

\bibitem[\protect\citename{{Kroupa} {et~al.}, }1991]{KTG91}
{Kroupa}, P., {Gilmore}, G., and {Tout}, C.~A. 1991.
\newblock {The effects of unresolved binary stars on the determination of the
  stellar mass function}.
\newblock {\em \mnras}, {\bf 251}(July), 293--302.

\bibitem[\protect\citename{{Kroupa} {et~al.}, }1993]{KTG93}
{Kroupa}, P., {Tout}, C.~A., and {Gilmore}, G. 1993.
\newblock {The distribution of low-mass stars in the Galactic disc}.
\newblock {\em \mnras}, {\bf 262}(June), 545--587.

\bibitem[\protect\citename{{Kroupa} {et~al.}, }2013]{Kroupa13}
{Kroupa}, P., {Weidner}, C., {Pflamm-Altenburg}, J., {Thies}, I.,
  {Dabringhausen}, J., {Marks}, M., and {Maschberger}, T. 2013.
\newblock {\em {The stellar and sub-stellar initial mass function of simple and
  composite populations}}.
\newblock Page  115.

\bibitem[\protect\citename{{Kroupa} {et~al.}, }2018]{Kroupa18}
{Kroupa}, P., {Je{\v r}{\'a}bkov{\'a}}, T., {Dinnbier}, F., {Beccari}, G., and
  {Yan}, Z. 2018.
\newblock {Evidence for feedback and stellar-dynamically regulated bursty star
  cluster formation: the case of the Orion Nebula Cluster}.
\newblock {\em \aap}, {\bf 612}(Apr.), A74.

\bibitem[\protect\citename{{Krumholz}, }2014]{Krum14}
{Krumholz}, M.~R. 2014.
\newblock {The big problems in star formation: The star formation rate, stellar
  clustering, and the initial mass function}.
\newblock {\em \physrep}, {\bf 539}(June), 49--134.

\bibitem[\protect\citename{{Krumholz}, }2015]{Krum15}
{Krumholz}, M.~R. 2015.
\newblock {The Formation of Very Massive Stars}.
\newblock {Page ~43 of:} {Vink}, J.~S. (ed), {\em Very Massive Stars in the
  Local Universe}.
\newblock Astrophysics and Space Science Library, vol. 412.

\bibitem[\protect\citename{{Lada}, }2010]{Lada10}
{Lada}, C.~J. 2010.
\newblock {The physics and modes of star cluster formation: observations}.
\newblock {\em Philosophical Transactions of the Royal Society of London Series
  A}, {\bf 368}(Jan.), 713--731.

\bibitem[\protect\citename{{Lada} and {Lada}, }2003]{LL03}
{Lada}, C.~J., and {Lada}, E.~A. 2003.
\newblock {Embedded Clusters in Molecular Clouds}.
\newblock {\em \araa}, {\bf 41}, 57--115.

\bibitem[\protect\citename{{Lane} {et~al.}, }2016]{Lane+16}
{Lane}, J., {Kirk}, H., {Johnstone}, D., {Mairs}, S., {Di Francesco}, J.,
  {Sadavoy}, S., {Hatchell}, J., {Berry}, D.~S., {Jenness}, T., {Hogerheijde},
  M.~R., {Ward-Thompson}, D., and {The JCMT Gould Belt Survey Team}. 2016.
\newblock {The JCMT Gould Belt Survey: Dense Core Clusters in Orion A}.
\newblock {\em \apj}, {\bf 833}(Dec.), 44.

\bibitem[\protect\citename{{Larson}, }1998]{Larson98}
{Larson}, R.~B. 1998.
\newblock {Early star formation and the evolution of the stellar initial mass
  function in galaxies}.
\newblock {\em \mnras}, {\bf 301}(Dec.), 569--581.

\bibitem[\protect\citename{{Lee} {et~al.}, }2009]{Lee09}
{Lee}, J.~C., {Gil de Paz}, A., {Tremonti}, C., {Kennicutt}, Jr., R.~C.,
  {Salim}, S., {Bothwell}, M., {Calzetti}, D., {Dalcanton}, J., {Dale}, D.,
  {Engelbracht}, C., {Funes}, S.~J.~J.~G., {Johnson}, B., {Sakai}, S.,
  {Skillman}, E., {van Zee}, L., {Walter}, F., and {Weisz}, D. 2009.
\newblock {Comparison of H{$\alpha$} and UV Star Formation Rates in the Local
  Volume: Systematic Discrepancies for Dwarf Galaxies}.
\newblock {\em \apj}, {\bf 706}(Nov.), 599--613.

\bibitem[\protect\citename{{Lelli} {et~al.}, }2014]{Lelli+14}
{Lelli}, F., {Verheijen}, M., and {Fraternali}, F. 2014.
\newblock {Dynamics of starbursting dwarf galaxies. III. A H I study of 18
  nearby objects}.
\newblock {\em \aap}, {\bf 566}(June), A71.

\bibitem[\protect\citename{{Lelli} {et~al.}, }2017]{Lelli+17}
{Lelli}, F., {McGaugh}, S.~S., {Schombert}, J.~M., and {Pawlowski}, M.~S. 2017.
\newblock {One Law to Rule Them All: The Radial Acceleration Relation of
  Galaxies}.
\newblock {\em \apj}, {\bf 836}(Feb.), 152.

\bibitem[\protect\citename{{Lieberz} and {Kroupa}, }2017]{LK17}
{Lieberz}, P., and {Kroupa}, P. 2017.
\newblock {On the origin of the Schechter-like mass function of young star
  clusters in disc galaxies}.
\newblock {\em \mnras}, {\bf 465}(Mar.), 3775--3783.

\bibitem[\protect\citename{{Lim} {et~al.}, }2018]{Lim+18}
{Lim}, B., {Sung}, H., {Bessell}, M.~S., {Lee}, S., {Lee}, J.~J., {Oh}, H.,
  {Hwang}, N., {Park}, B.-G., {Hur}, H., {Hong}, K., and {Park}, S. 2018.
\newblock {Kinematic evidence for feedback-driven star formation in NGC 1893}.
\newblock {\em \mnras}, {\bf 477}(June), 1993--2003.

\bibitem[\protect\citename{{Lin} {et~al.}, }2019]{Lin+19}
{Lin}, Y., {Csengeri}, T., {Wyrowski}, F., {Urquhart}, J.~S., {Schuller}, F.,
  {Weiss}, A., and {Menten}, K.~M. 2019.
\newblock {Fragmentation and filaments at the onset of star and cluster
  formation. SABOCA 350 {\ensuremath{\mu}}m view of ATLASGAL-selected massive
  clumps}.
\newblock {\em \aap}, {\bf 631}(Nov.), A72.

\bibitem[\protect\citename{{Liptai} {et~al.}, }2017]{Liptai+17}
{Liptai}, D., {Price}, D.~J., {Wurster}, J., and {Bate}, M.~R. 2017.
\newblock {Does turbulence determine the initial mass function?}
\newblock {\em \mnras}, {\bf 465}(Feb.), 105--110.

\bibitem[\protect\citename{{Mahajan} {et~al.}, }2019]{Mahajan+19}
{Mahajan}, Smriti, {Ashby}, M.~L.~N., {Willner}, S.~P., {Barmby}, P., {Fazio},
  G.~G., {Maragkoudakis}, A., {Raychaudhury}, S., and {Zezas}, A. 2019.
\newblock {The Star Formation Reference Survey - III. A multiwavelength view of
  star formation in nearby galaxies}.
\newblock {\em \mnras}, {\bf 482}(Jan.), 560--577.

\bibitem[\protect\citename{{Ma{\'\i}z Apell{\'a}niz} {et~al.}, }2007]{Jesus07}
{Ma{\'\i}z Apell{\'a}niz}, J., {Walborn}, Nolan~R., {Morrell}, N.~I.,
  {Niemela}, V.~S., and {Nelan}, E.~P. 2007.
\newblock {Pismis 24-1: The Stellar Upper Mass Limit Preserved}.
\newblock {\em \apj}, {\bf 660}(2), 1480--1485.

\bibitem[\protect\citename{{Marks} and {Kroupa}, }2012]{MK12}
{Marks}, M., and {Kroupa}, P. 2012.
\newblock {Inverse dynamical population synthesis. Constraining the initial
  conditions of young stellar clusters by studying their binary populations}.
\newblock {\em \aap}, {\bf 543}(July), A8.

\bibitem[\protect\citename{{Marks} {et~al.}, }2011]{MKO11}
{Marks}, M., {Kroupa}, P., and {Oh}, S. 2011.
\newblock {An analytical description of the evolution of binary
  orbital-parameter distributions in N-body computations of star clusters}.
\newblock {\em \mnras}, {\bf 417}(Nov.), 1684--1701.

\bibitem[\protect\citename{{Marks} {et~al.}, }2012]{Marks12}
{Marks}, M., {Kroupa}, P., {Dabringhausen}, J., and {Pawlowski}, M.~S. 2012.
\newblock {Evidence for top-heavy stellar initial mass functions with
  increasing density and decreasing metallicity}.
\newblock {\em \mnras}, {\bf 422}(May), 2246--2254.

\bibitem[\protect\citename{{Mart{\'{\i}}n-Navarro}, }2016]{MartinNavarro16}
{Mart{\'{\i}}n-Navarro}, I. 2016.
\newblock {Revisiting the classics: is [Mg/Fe] a good proxy for galaxy
  formation time-scales?}
\newblock {\em \mnras}, {\bf 456}(Feb.), L104--L108.

\bibitem[\protect\citename{{Maschberger} and {Clarke}, }2008]{MC08}
{Maschberger}, T., and {Clarke}, C.~J. 2008.
\newblock {Maximum stellar mass versus cluster membership number revisited}.
\newblock {\em \mnras}, {\bf 391}(Dec.), 711--717.

\bibitem[\protect\citename{{Massey}, }2003]{Massey03}
{Massey}, P. 2003.
\newblock {MASSIVE STARS IN THE LOCAL GROUP: Implications for Stellar Evolution
  and Star Formation}.
\newblock {\em \araa}, {\bf 41}, 15--56.

\bibitem[\protect\citename{{Matteucci}, }1994]{Matteucci94}
{Matteucci}, F. 1994.
\newblock {Abundance ratios in ellipticals and galaxy formation}.
\newblock {\em \aap}, {\bf 288}(Aug.), 57--64.

\bibitem[\protect\citename{{Matzner} and {McKee}, }2000]{MatznerMcKee00}
{Matzner}, C.~D., and {McKee}, C.~F. 2000.
\newblock {Efficiencies of Low-Mass Star and Star Cluster Formation}.
\newblock {\em \apj}, {\bf 545}(Dec.), 364--378.

\bibitem[\protect\citename{{McGaugh}, }2004]{McGaugh04}
{McGaugh}, S.~S. 2004.
\newblock {The Mass Discrepancy-Acceleration Relation: Disk Mass and the Dark
  Matter Distribution}.
\newblock {\em \apj}, {\bf 609}(July), 652--666.

\bibitem[\protect\citename{{McGaugh}, }2005]{McGaugh05}
{McGaugh}, S.~S. 2005.
\newblock {The Baryonic Tully-Fisher Relation of Galaxies with Extended
  Rotation Curves and the Stellar Mass of Rotating Galaxies}.
\newblock {\em \apj}, {\bf 632}(Oct.), 859--871.

\bibitem[\protect\citename{{McGaugh}, }2012]{McGaugh12}
{McGaugh}, S.~S. 2012.
\newblock {The Baryonic Tully-Fisher Relation of Gas-rich Galaxies as a Test of
  {$\Lambda$}CDM and MOND}.
\newblock {\em \aj}, {\bf 143}(Feb.), 40.

\bibitem[\protect\citename{{McGaugh} {et~al.}, }2016]{McGaugh+2016}
{McGaugh}, S.~S., {Lelli}, F., and {Schombert}, J.~M. 2016.
\newblock {Radial Acceleration Relation in Rotationally Supported Galaxies}.
\newblock {\em Physical Review Letters}, {\bf 117}(20), 201101.

\bibitem[\protect\citename{{McQuinn} {et~al.}, }2015]{McQuinn+15}
{McQuinn}, K.~B.~W., {Skillman}, E.~D., {Dolphin}, A., {Cannon}, J.~M.,
  {Salzer}, J.~J., {Rhode}, K.~L., {Adams}, E.~A.~K., {Berg}, D., {Giovanelli},
  R., {Girardi}, L., and {Haynes}, M.~P. 2015.
\newblock {Leo P: An Unquenched Very Low-mass Galaxy}.
\newblock {\em \apj}, {\bf 812}(Oct.), 158.

\bibitem[\protect\citename{{McQuinn} {et~al.}, }2018]{McQuinn+18}
{McQuinn}, K.~B.~W., {Skillman}, E.~D., {Heilman}, T.~N., {Mitchell}, N.~P.,
  and {Kelley}, T. 2018.
\newblock {Galactic outflows, star formation histories, and time-scales in
  starburst dwarf galaxies from STARBIRDS}.
\newblock {\em \mnras}, {\bf 477}(July), 3164--3177.

\bibitem[\protect\citename{{Megeath} {et~al.}, }2012]{Megeath12}
{Megeath}, S.~T., {Gutermuth}, R., {Muzerolle}, J., {Kryukova}, E., {Flaherty},
  K., {Hora}, J.~L., {Allen}, L.~E., {Hartmann}, L., {Myers}, P.~C., {Pipher},
  J.~L., {Stauffer}, J., {Young}, E.~T., and {Fazio}, G.~G. 2012.
\newblock {The Spitzer Space Telescope Survey of the Orion A and B Molecular
  Clouds. I. A Census of Dusty Young Stellar Objects and a Study of Their
  Mid-infrared Variability}.
\newblock {\em \aj}, {\bf 144}(Dec.), 192.

\bibitem[\protect\citename{{Megeath} {et~al.}, }2016]{Megeath16}
{Megeath}, S.~T., {Gutermuth}, R., {Muzerolle}, J., {Kryukova}, E., {Hora},
  J.~L., {Allen}, L.~E., {Flaherty}, K., {Hartmann}, L., {Myers}, P.~C.,
  {Pipher}, J.~L., {Stauffer}, J., {Young}, E.~T., and {Fazio}, G.~G. 2016.
\newblock {The Spitzer Space Telescope Survey of the Orion A and B Molecular
  Clouds. II. The Spatial Distribution and Demographics of Dusty Young Stellar
  Objects}.
\newblock {\em \aj}, {\bf 151}(Jan.), 5.

\bibitem[\protect\citename{{Meurer} {et~al.}, }2009]{Meurer09}
{Meurer}, G.~R., {Wong}, O.~I., {Kim}, J.~H., {Hanish}, D.~J., {Heckman},
  T.~M., {Werk}, J., {Bland-Hawthorn}, J., {Dopita}, M.~A., {Zwaan}, M.~A.,
  {Koribalski}, B., {Seibert}, M., {Thilker}, D.~A., {Ferguson}, H.~C.,
  {Webster}, R.~L., {Putman}, M.~E., {Knezek}, P.~M., {Doyle}, M.~T.,
  {Drinkwater}, M.~J., {Hoopes}, C.~G., {Kilborn}, V.~A., {Meyer}, M.,
  {Ryan-Weber}, E.~V., {Smith}, R.~C., and {Staveley-Smith}, L. 2009.
\newblock {Evidence for a Nonuniform Initial Mass Function in the Local
  Universe}.
\newblock {\em \apj}, {\bf 695}(Apr.), 765--780.

\bibitem[\protect\citename{{Miller} and {Scalo}, }1979]{MS79}
{Miller}, G.~E., and {Scalo}, J.~M. 1979.
\newblock {The initial mass function and stellar birthrate in the solar
  neighborhood}.
\newblock {\em \apjs}, {\bf 41}(Nov.), 513--547.

\bibitem[\protect\citename{{Mor} {et~al.}, }2017]{Mor+17}
{Mor}, R., {Robin}, A.~C., {Figueras}, F., and {Lemasle}, B. 2017.
\newblock {Constraining the thin disc initial mass function using Galactic
  classical Cepheids}.
\newblock {\em \aap}, {\bf 599}(Mar.), A17.

\bibitem[\protect\citename{{Mor} {et~al.}, }2018]{Mor+18}
{Mor}, R., {Robin}, A.~C., {Figueras}, F., and {Antoja}, T. 2018.
\newblock {BGM FASt: Besan{\c{c}}on Galaxy Model for big data. Simultaneous
  inference of the IMF, SFH, and density in the solar neighbourhood}.
\newblock {\em \aap}, {\bf 620}(Dec.), A79.

\bibitem[\protect\citename{{Oey} and {Clarke}, }2005]{OC05}
{Oey}, M.~S., and {Clarke}, C.~J. 2005.
\newblock {Statistical Confirmation of a Stellar Upper Mass Limit}.
\newblock {\em \apjl}, {\bf 620}(Feb.), L43--L46.

\bibitem[\protect\citename{{Offner} {et~al.}, }2014]{Offner14}
{Offner}, S.~S.~R., {Clark}, P.~C., {Hennebelle}, P., {Bastian}, N., {Bate},
  M.~R., {Hopkins}, P.~F., {Moraux}, E., and {Whitworth}, A.~P. 2014.
\newblock {The Origin and Universality of the Stellar Initial Mass Function}.
\newblock {\em Protostars and Planets VI},  53--75.

\bibitem[\protect\citename{{Oh} and {Kroupa}, }2012]{OK12}
{Oh}, S., and {Kroupa}, P. 2012.
\newblock {The influence of stellar dynamical ejections and collisions on the
  relation between the maximum stellar and star cluster mass}.
\newblock {\em \mnras}, {\bf 424}(July), 65--79.

\bibitem[\protect\citename{{Oh} and {Kroupa}, }2016]{OK16}
{Oh}, S., and {Kroupa}, P. 2016.
\newblock {Dynamical ejections of massive stars from young star clusters under
  diverse initial conditions}.
\newblock {\em \aap}, {\bf 590}(May), A107.

\bibitem[\protect\citename{{Oh} and {Kroupa}, }2018]{OK18}
{Oh}, S., and {Kroupa}, P. 2018.
\newblock {Very massive stars in not so massive clusters}.
\newblock {\em \mnras}, {\bf 481}(Nov.), 153--163.

\bibitem[\protect\citename{{Oh} {et~al.}, }2015]{OKP15}
{Oh}, S., {Kroupa}, P., and {Pflamm-Altenburg}, J. 2015.
\newblock {Dependency of Dynamical Ejections of O Stars on the Masses of Very
  Young Star Clusters}.
\newblock {\em \apj}, {\bf 805}(June), 92.

\bibitem[\protect\citename{{Padoan} and {Nordlund}, }2002]{PN02}
{Padoan}, P., and {Nordlund}, {\AA}. 2002.
\newblock {The Stellar Initial Mass Function from Turbulent Fragmentation}.
\newblock {\em \apj}, {\bf 576}(Sept.), 870--879.

\bibitem[\protect\citename{{Papadopoulos} and {Thi}, }2013]{Papadopoulos13}
{Papadopoulos}, P.~P., and {Thi}, W.-F. 2013.
\newblock {The Initial Conditions of Star Formation: Cosmic Rays as the
  Fundamental Regulators}.
\newblock {Page ~41 of:} {Torres}, D.~F., and {Reimer}, O. (eds), {\em Cosmic
  Rays in Star-Forming Environments}.
\newblock Advances in Solid State Physics, vol. 34.

\bibitem[\protect\citename{{Parker} and {Goodwin}, }2007]{PG07}
{Parker}, R.~J., and {Goodwin}, S.~P. 2007.
\newblock {Do O-stars form in isolation?}
\newblock {\em \mnras}, {\bf 380}(Sept.), 1271--1275.

\bibitem[\protect\citename{{Pavl{\'\i}k} {et~al.}, }2019]{Pavlik+19}
{Pavl{\'\i}k}, V{\'a}clav, {Kroupa}, Pavel, and {{\v{S}}ubr}, Ladislav. 2019.
\newblock {Do star clusters form in a completely mass-segregated way?}
\newblock {\em \aap}, {\bf 626}(Jun), A79.

\bibitem[\protect\citename{{Pflamm-Altenburg} and {Kroupa}, }2006]{PAK06}
{Pflamm-Altenburg}, J., and {Kroupa}, P. 2006.
\newblock {A highly abnormal massive star mass function in the Orion Nebula
  cluster and the dynamical decay of trapezium systems}.
\newblock {\em \mnras}, {\bf 373}(Nov.), 295--304.

\bibitem[\protect\citename{{Pflamm-Altenburg} and {Kroupa}, }2008]{PAK08}
{Pflamm-Altenburg}, J., and {Kroupa}, P. 2008.
\newblock {Clustered star formation as a natural explanation for the
  H{$\alpha$} cut-off in disk galaxies}.
\newblock {\em \nat}, {\bf 455}(Oct.), 641--643.

\bibitem[\protect\citename{{Pflamm-Altenburg} and {Kroupa}, }2009]{Pflamm09}
{Pflamm-Altenburg}, J., and {Kroupa}, P. 2009.
\newblock {The Fundamental Gas Depletion and Stellar-Mass Buildup Times of
  Star-Forming Galaxies}.
\newblock {\em \apj}, {\bf 706}(Nov.), 516--524.

\bibitem[\protect\citename{{Pflamm-Altenburg} and {Kroupa}, }2010]{Pflamm10}
{Pflamm-Altenburg}, J., and {Kroupa}, P. 2010.
\newblock {The two-step ejection of massive stars and the issue of their
  formation in isolation}.
\newblock {\em \mnras}, {\bf 404}(May), 1564--1568.

\bibitem[\protect\citename{{Pflamm-Altenburg} {et~al.}, }2007]{Pflamm07}
{Pflamm-Altenburg}, J., {Weidner}, C., and {Kroupa}, P. 2007.
\newblock {Converting H{$\alpha$} Luminosities into Star Formation Rates}.
\newblock {\em \apj}, {\bf 671}(Dec.), 1550--1558.

\bibitem[\protect\citename{{Pflamm-Altenburg} {et~al.}, }2009]{Pflamm09b}
{Pflamm-Altenburg}, J., {Weidner}, C., and {Kroupa}, P. 2009.
\newblock {Diverging UV and H{$\alpha$} fluxes of star-forming galaxies
  predicted by the IGIMF theory}.
\newblock {\em \mnras}, {\bf 395}(May), 394--400.

\bibitem[\protect\citename{{Pflamm-Altenburg} {et~al.}, }2013]{Pflamm13}
{Pflamm-Altenburg}, J., {Gonz{\'a}lez-L{\'o}pezlira}, R.~A., and {Kroupa}, P.
  2013.
\newblock {The galactocentric radius dependent upper mass limit of young star
  clusters: stochastic star formation ruled out}.
\newblock {\em \mnras}, {\bf 435}(Nov.), 2604--2609.

\bibitem[\protect\citename{{Plummer}, }1911]{Plummer11}
{Plummer}, H.~C. 1911.
\newblock {On the problem of distribution in globular star clusters}.
\newblock {\em \mnras}, {\bf 71}(Mar.), 460--470.

\bibitem[\protect\citename{{Plunkett} {et~al.}, }2018]{Plunkett18}
{Plunkett}, A.~L., {Fern{\'a}ndez-L{\'o}pez}, M., {Arce}, H.~G., {Busquet}, G.,
  {Mardones}, D., and {Dunham}, M.~M. 2018.
\newblock {Distribution of Serpens South protostars revealed with ALMA}.
\newblock {\em \aap}, {\bf 615}(July), A9.

\bibitem[\protect\citename{{Portail} {et~al.}, }2017]{Portail+17}
{Portail}, M., {Wegg}, C., {Gerhard}, O., and {Ness}, M. 2017.
\newblock {Chemodynamical modelling of the galactic bulge and bar}.
\newblock {\em \mnras}, {\bf 470}(Sept.), 1233--1252.

\bibitem[\protect\citename{{Ram{\'{\i}}rez Alegr{\'{\i}}a} {et~al.},
  }2016]{Ramirez16}
{Ram{\'{\i}}rez Alegr{\'{\i}}a}, S., {Borissova}, J., {Chen{\'e}}, A.-N.,
  {Bonatto}, C., {Kurtev}, R., {Amigo}, P., {Kuhn}, M., {Gromadzki}, M., and
  {Carballo-Bello}, J.~A. 2016.
\newblock {Massive open star clusters using the VVV survey. V. Young clusters
  with an OB stellar population}.
\newblock {\em \aap}, {\bf 588}(Apr.), A40.

\bibitem[\protect\citename{{Randriamanakoto} {et~al.}, }2013]{Rand13}
{Randriamanakoto}, Z., {Escala}, A., {V{\"a}is{\"a}nen}, P., {Kankare}, E.,
  {Kotilainen}, J., {Mattila}, S., and {Ryder}, S. 2013.
\newblock {Near-infrared Adaptive Optics Imaging of Infrared Luminous Galaxies:
  The Brightest Cluster Magnitude-Star Formation Rate Relation}.
\newblock {\em \apjl}, {\bf 775}(Oct.), L38.

\bibitem[\protect\citename{{Randriamanakoto} {et~al.}, }2018]{Rand18}
{Randriamanakoto}, Z., {V{\"a}is{\"a}nen}, P., {Ryder}, S.~D., and
  {Ranaivomanana}, P. 2018.
\newblock {Young massive clusters in the interacting LIRG Arp 299: evidence for
  the dependence of star cluster formation and evolution on environment}.
\newblock {\em \mnras}, Oct.

\bibitem[\protect\citename{{Recchi} and {Kroupa}, }2015]{Recchi15}
{Recchi}, S., and {Kroupa}, P. 2015.
\newblock {The chemical evolution of galaxies with a variable integrated
  galactic initial mass function}.
\newblock {\em \mnras}, {\bf 446}(Feb.), 4168--4175.

\bibitem[\protect\citename{{Recchi} {et~al.}, }2009]{Recchi09}
{Recchi}, S., {Calura}, F., and {Kroupa}, P. 2009.
\newblock {The chemical evolution of galaxies within the IGIMF theory: the [
  {$\alpha$}/Fe] ratios and downsizing}.
\newblock {\em \aap}, {\bf 499}(June), 711--722.

\bibitem[\protect\citename{{Riedel} {et~al.}, }2018]{Riedel+18}
{Riedel}, A.~R., {Silverstein}, M.~L., {Henry}, T.~J., {Jao}, W.-C., {Winters},
  J.~G., {Subasavage}, J.~P., {Malo}, L., and {Hambly}, N.~C. 2018.
\newblock {The Solar Neighborhood. XLIII. Discovery of New Nearby Stars with
  mu<0.18/yr (TINYMO Sample)}.
\newblock {\em \aj}, {\bf 156}(Aug.), 49.

\bibitem[\protect\citename{{Rybizki} and {Just}, }2015]{RJ15}
{Rybizki}, J., and {Just}, A. 2015.
\newblock {Towards a fully consistent Milky Way disc model - III. Constraining
  the initial mass function}.
\newblock {\em \mnras}, {\bf 447}(Mar.), 3880--3891.

\bibitem[\protect\citename{{Salpeter}, }1955]{Salpeter55}
{Salpeter}, E.~E. 1955.
\newblock {The Luminosity Function and Stellar Evolution.}
\newblock {\em \apj}, {\bf 121}(Jan.), 161.

\bibitem[\protect\citename{{Sana} {et~al.}, }2012]{Sana+12}
{Sana}, H., {de Mink}, S.~E., {de Koter}, A., {Langer}, N., {Evans}, C.~J.,
  {Gieles}, M., {Gosset}, E., {Izzard}, R.~G., {Le Bouquin}, J.-B., and
  {Schneider}, F.~R.~N. 2012.
\newblock {Binary Interaction Dominates the Evolution of Massive Stars}.
\newblock {\em Science}, {\bf 337}(July), 444.

\bibitem[\protect\citename{{Scalo}, }1986]{Scalo86}
{Scalo}, J.~M. 1986.
\newblock {The stellar initial mass function}.
\newblock {\em Fundamentals of Cosmic Physics}, {\bf 11}(May), 1--278.

\bibitem[\protect\citename{{Schneider} {et~al.}, }2014]{Schneider+14}
{Schneider}, F.~R.~N., {Izzard}, R.~G., {de Mink}, S.~E., {Langer}, N.,
  {Stolte}, A., {de Koter}, A., {Gvaramadze}, V.~V., {Hu{\ss}mann}, B.,
  {Liermann}, A., and {Sana}, H. 2014.
\newblock {Ages of Young Star Clusters, Massive Blue Stragglers, and the Upper
  Mass Limit of Stars: Analyzing Age-dependent Stellar Mass Functions}.
\newblock {\em \apj}, {\bf 780}(Jan.), 117.

\bibitem[\protect\citename{{Schneider} {et~al.}, }2015]{Schneider+15}
{Schneider}, F.~R.~N., {Izzard}, R.~G., {Langer}, N., and {de Mink}, S.~E.
  2015.
\newblock {Evolution of Mass Functions of Coeval Stars through Wind Mass Loss
  and Binary Interactions}.
\newblock {\em \apj}, {\bf 805}(May), 20.

\bibitem[\protect\citename{{Schneider} {et~al.}, }2018]{Schneider18}
{Schneider}, F.~R.~N., {Sana}, H., {Evans}, C.~J., {Bestenlehner}, J.~M.,
  {Castro}, N., {Fossati}, L., {Gr{\"a}fener}, G., {Langer}, N.,
  {Ram{\'{\i}}rez-Agudelo}, O.~H., {Sab{\'{\i}}n-Sanjuli{\'a}n}, C.,
  {Sim{\'o}n-D{\'{\i}}az}, S., {Tramper}, F., {Crowther}, P.~A., {de Koter},
  A., {de Mink}, S.~E., {Dufton}, P.~L., {Garcia}, M., {Gieles}, M.,
  {H{\'e}nault-Brunet}, V., {Herrero}, A., {Izzard}, R.~G., {Kalari}, V.,
  {Lennon}, D.~J., {Ma{\'{\i}}z Apell{\'a}niz}, J., {Markova}, N., {Najarro},
  F., {Podsiadlowski}, P., {Puls}, J., {Taylor}, W.~D., {van Loon}, J.~T.,
  {Vink}, J.~S., and {Norman}, C. 2018.
\newblock {An excess of massive stars in the local 30 Doradus starburst}.
\newblock {\em Science}, {\bf 359}(Jan.), 69--71.

\bibitem[\protect\citename{{Schneider} {et~al.}, }2012]{Schneider+12}
{Schneider}, N., {Csengeri}, T., {Hennemann}, M., {Motte}, F., {Didelon}, P.,
  {Federrath}, C., {Bontemps}, S., {Di Francesco}, J., {Arzoumanian}, D.,
  {Minier}, V., {Andr{\'e}}, P., {Hill}, T., {Zavagno}, A., {Nguyen-Luong}, Q.,
  {Attard}, M., {Bernard}, J.-P., {Elia}, D., {Fallscheer}, C., {Griffin}, M.,
  {Kirk}, J., {Klessen}, R., {K{\"o}nyves}, V., {Martin}, P., {Men'shchikov},
  A., {Palmeirim}, P., {Peretto}, N., {Pestalozzi}, M., {Russeil}, D.,
  {Sadavoy}, S., {Sousbie}, T., {Testi}, L., {Tremblin}, P., {Ward-Thompson},
  D., and {White}, G. 2012.
\newblock {Cluster-formation in the Rosette molecular cloud at the junctions of
  filaments}.
\newblock {\em \aap}, {\bf 540}(Apr.), L11.

\bibitem[\protect\citename{{Schulz} {et~al.}, }2015]{Schulz+15}
{Schulz}, C., {Pflamm-Altenburg}, J., and {Kroupa}, P. 2015.
\newblock {Mass distributions of star clusters for different star formation
  histories in a galaxy cluster environment}.
\newblock {\em \aap}, {\bf 582}(Oct.), A93.

\bibitem[\protect\citename{{Shimajiri} {et~al.}, }2017]{Shimajiri+17}
{Shimajiri}, Y., {Andr{\'e}}, P., {Braine}, J., {K{\"o}nyves}, V., {Schneider},
  N., {Bontemps}, S., {Ladjelate}, B., {Roy}, A., {Gao}, Y., and {Chen}, H.
  2017.
\newblock {Testing the universality of the star-formation efficiency in dense
  molecular gas}.
\newblock {\em \aap}, {\bf 604}(Aug.), A74.

\bibitem[\protect\citename{{Smith} {et~al.}, }2005]{Smith+05}
{Smith}, N., {Stassun}, K.~G., and {Bally}, J. 2005.
\newblock {Opening the Treasure Chest: A Newborn Star Cluster Emerges from Its
  Dust Pillar in Carina}.
\newblock {\em \aj}, {\bf 129}(Feb.), 888--899.

\bibitem[\protect\citename{{Smith}, }2014]{Smith14}
{Smith}, R.~J. 2014.
\newblock {Variations in the initial mass function in early-type galaxies: a
  critical comparison between dynamical and spectroscopic results}.
\newblock {\em \mnras}, {\bf 443}(Sept.), L69--L73.

\bibitem[\protect\citename{{Smith} and {Lucey}, }2013]{SL13}
{Smith}, R.~J., and {Lucey}, J.~R. 2013.
\newblock {A giant elliptical galaxy with a lightweight initial mass function}.
\newblock {\em \mnras}, {\bf 434}(Sept.), 1964--1977.

\bibitem[\protect\citename{{Speagle} {et~al.}, }2014]{Speagle14}
{Speagle}, J.~S., {Steinhardt}, C.~L., {Capak}, P.~L., and {Silverman}, J.~D.
  2014.
\newblock {A Highly Consistent Framework for the Evolution of the Star-Forming
  ``Main Sequence'' from z \~{} 0-6}.
\newblock {\em \apjs}, {\bf 214}(Oct.), 15.

\bibitem[\protect\citename{{Stephens} {et~al.}, }2017]{Stephens+17}
{Stephens}, I.~W., {Gouliermis}, D., {Looney}, L.~W., {Gruendl}, R.~A., {Chu},
  Y.-H., {Weisz}, D.~R., {Seale}, J.~P., {Chen}, C.-H.~R., {Wong}, T.,
  {Hughes}, A., {Pineda}, J.~L., {Ott}, J., and {Muller}, E. 2017.
\newblock {Stellar Clusterings around ``Isolated'' Massive YSOs in the LMC}.
\newblock {\em \apj}, {\bf 834}(Jan.), 94.

\bibitem[\protect\citename{{Thies} {et~al.}, }2015]{Thies15}
{Thies}, I., {Pflamm-Altenburg}, J., {Kroupa}, P., and {Marks}, M. 2015.
\newblock {Characterizing the Brown Dwarf Formation Channels from the Initial
  Mass Function and Binary-star Dynamics}.
\newblock {\em \apj}, {\bf 800}(Feb.), 72.

\bibitem[\protect\citename{{Thomas} {et~al.}, }2005]{Thomas05}
{Thomas}, D., {Maraston}, C., {Bender}, R., and {Mendes de Oliveira}, C. 2005.
\newblock {The Epochs of Early-Type Galaxy Formation as a Function of
  Environment}.
\newblock {\em \apj}, {\bf 621}(Mar.), 673--694.

\bibitem[\protect\citename{{Tsujimoto}, }2011]{Tsujimoto11}
{Tsujimoto}, T. 2011.
\newblock {Chemical Signature Indicating a Lack of Massive Stars in Dwarf
  Galaxies}.
\newblock {\em \apj}, {\bf 736}(Aug.), 113.

\bibitem[\protect\citename{{Urquhart} {et~al.}, }2014]{Urquhart+14}
{Urquhart}, J.~S., {Figura}, C.~C., {Moore}, T.~J.~T., {Hoare}, M.~G.,
  {Lumsden}, S.~L., {Mottram}, J.~C., {Thompson}, M.~A., and {Oudmaijer}, R.~D.
  2014.
\newblock {The RMS survey: galactic distribution of massive star formation}.
\newblock {\em \mnras}, {\bf 437}(Jan.), 1791--1807.

\bibitem[\protect\citename{{{\v S}ubr} {et~al.}, }2008]{Subr+08}
{{\v S}ubr}, L., {Kroupa}, P., and {Baumgardt}, H. 2008.
\newblock {A new method to create initially mass segregated star clusters in
  virial equilibrium}.
\newblock {\em \mnras}, {\bf 385}(Apr.), 1673--1680.

\bibitem[\protect\citename{{van Dokkum} and {Conroy}, }2010]{vanDokkum10}
{van Dokkum}, P.~G., and {Conroy}, C. 2010.
\newblock {A substantial population of low-mass stars in luminous elliptical
  galaxies}.
\newblock {\em \nat}, {\bf 468}(Dec.), 940--942.

\bibitem[\protect\citename{{Vanbeveren}, }1982]{vanBeveren82}
{Vanbeveren}, D. 1982.
\newblock {On the difference between the initial mass function of single stars
  and of primaries of binaries}.
\newblock {\em \aap}, {\bf 115}(Nov.), 65--68.

\bibitem[\protect\citename{{Vazdekis} {et~al.}, }1997]{Vazdekis97}
{Vazdekis}, A., {Peletier}, R.~F., {Beckman}, J.~E., and {Casuso}, E. 1997.
\newblock {A New Chemo-evolutionary Population Synthesis Model for Early-Type
  Galaxies. II. Observations and Results}.
\newblock {\em \apjs}, {\bf 111}(July), 203--232.

\bibitem[\protect\citename{{Wang} {et~al.}, }2019]{WKJ18}
{Wang}, Long, {Kroupa}, Pavel, and {Jerabkova}, Tereza. 2019.
\newblock {Complete ejection of OB stars from very young star clusters and the
  formation of multiple populations}.
\newblock {\em \mnras}, {\bf 484}(2), 1843--1851.

\bibitem[\protect\citename{{Watts} {et~al.}, }2018]{Watts+18}
{Watts}, A.~B., {Meurer}, G.~R., {Lagos}, C.~D.~P., {Bruzzese}, S.~M.,
  {Kroupa}, P., and {Jerabkova}, T. 2018.
\newblock {Star formation in the outskirts of DDO 154: a top-light IMF in a
  nearly dormant disc}.
\newblock {\em \mnras}, {\bf 477}(July), 5554--5567.

\bibitem[\protect\citename{{Wegg} {et~al.}, }2017]{Wegg+17}
{Wegg}, C., {Gerhard}, O., and {Portail}, M. 2017.
\newblock {The Initial Mass Function of the Inner Galaxy Measured from OGLE-III
  Microlensing Timescales}.
\newblock {\em \apjl}, {\bf 843}(July), L5.

\bibitem[\protect\citename{{Weidner} and {Kroupa}, }2004]{WK04}
{Weidner}, C., and {Kroupa}, P. 2004.
\newblock {Evidence for a fundamental stellar upper mass limit from clustered
  star formation}.
\newblock {\em \mnras}, {\bf 348}(Feb.), 187--191.

\bibitem[\protect\citename{{Weidner} and {Kroupa}, }2005]{Weidner05}
{Weidner}, C., and {Kroupa}, P. 2005.
\newblock {The Variation of Integrated Star Initial Mass Functions among
  Galaxies}.
\newblock {\em \apj}, {\bf 625}(June), 754--762.

\bibitem[\protect\citename{{Weidner} and {Kroupa}, }2006]{Weidner06}
{Weidner}, C., and {Kroupa}, P. 2006.
\newblock {The maximum stellar mass, star-cluster formation and composite
  stellar populations}.
\newblock {\em \mnras}, {\bf 365}(Feb.), 1333--1347.

\bibitem[\protect\citename{{Weidner} {et~al.}, }2004]{Weidner04}
{Weidner}, C., {Kroupa}, P., and {Larsen}, S.~S. 2004.
\newblock {Implications for the formation of star clusters from extragalactic
  star formation rates}.
\newblock {\em \mnras}, {\bf 350}(June), 1503--1510.

\bibitem[\protect\citename{{Weidner} {et~al.}, }2009]{Weidner09}
{Weidner}, C., {Kroupa}, P., and {Maschberger}, T. 2009.
\newblock {The influence of multiple stars on the high-mass stellar initial
  mass function and age dating of young massive star clusters}.
\newblock {\em \mnras}, {\bf 393}(Feb.), 663--680.

\bibitem[\protect\citename{{Weidner} {et~al.}, }2010]{Weidner10}
{Weidner}, C., {Kroupa}, P., and {Bonnell}, I.~A.~D. 2010.
\newblock {The relation between the most-massive star and its parental star
  cluster mass}.
\newblock {\em \mnras}, {\bf 401}(Jan.), 275--293.

\bibitem[\protect\citename{{Weidner} {et~al.}, }2013a]{Weidner13c}
{Weidner}, C., {Ferreras}, I., {Vazdekis}, A., and {La Barbera}, F. 2013a.
\newblock {The (galaxy-wide) IMF in giant elliptical galaxies: from top to
  bottom}.
\newblock {\em \mnras}, {\bf 435}(Nov.), 2274--2280.

\bibitem[\protect\citename{{Weidner} {et~al.}, }2013b]{Weidner13b}
{Weidner}, C., {Kroupa}, P., {Pflamm-Altenburg}, J., and {Vazdekis}, A. 2013b.
\newblock {The galaxy-wide initial mass function of dwarf late-type to massive
  early-type galaxies}.
\newblock {\em \mnras}, {\bf 436}(Dec.), 3309--3320.

\bibitem[\protect\citename{{Weidner} {et~al.}, }2013c]{Weidner13}
{Weidner}, C., {Kroupa}, P., and {Pflamm-Altenburg}, J. 2013c.
\newblock {The m$_{max}$-M$_{ecl}$ relation, the IMF and IGIMF:
  probabilistically sampled functions}.
\newblock {\em \mnras}, {\bf 434}(Sept.), 84--101.

\bibitem[\protect\citename{{Weidner} {et~al.}, }2014]{Weidner14}
{Weidner}, Carsten, {Kroupa}, Pavel, and {Pflamm-Altenburg}, Jan. 2014.
\newblock {Sampling methods for stellar masses and the m$_{max}$-M$_{ecl}$
  relation in the starburst dwarf galaxy NGC 4214}.
\newblock {\em \mnras}, {\bf 441}(4), 3348--3358.

\bibitem[\protect\citename{{Whitmore} {et~al.}, }1999]{Whitmore+99}
{Whitmore}, B.~C., {Zhang}, Q., {Leitherer}, C., {Fall}, S.~M., {Schweizer},
  F., and {Miller}, B.~W. 1999.
\newblock {The Luminosity Function of Young Star Clusters in ``the Antennae''
  Galaxies (NGC 4038-4039)}.
\newblock {\em \aj}, {\bf 118}(Oct.), 1551--1576.

\bibitem[\protect\citename{{Wright} and {Mamajek}, }2018]{WM18}
{Wright}, N.~J., and {Mamajek}, E.~E. 2018.
\newblock {The kinematics of the Scorpius-Centaurus OB association from Gaia
  DR1}.
\newblock {\em \mnras}, {\bf 476}(May), 381--398.

\bibitem[\protect\citename{{Wuchterl} and {Tscharnuter}, }2003]{WT03}
{Wuchterl}, G., and {Tscharnuter}, W.~M. 2003.
\newblock {From clouds to stars. Protostellar collapse and the evolution to the
  pre-main sequence I. Equations and evolution in the Hertzsprung-Russell
  diagram}.
\newblock {\em \aap}, {\bf 398}(Feb.), 1081--1090.

\bibitem[\protect\citename{{Yan} {et~al.}, }2017]{Yan+17}
{Yan}, Z., {Jerabkova}, T., and {Kroupa}, P. 2017.
\newblock {The optimally sampled galaxy-wide stellar initial mass function.
  Observational tests and the publicly available GalIMF code}.
\newblock {\em \aap}, {\bf 607}(Nov.), A126.

\bibitem[\protect\citename{{Yan} {et~al.}, }2021]{Yan+21}
{Yan}, Zhiqiang, {Je{\v{r}}{\'a}bkov{\'a}}, Tereza, and {Kroupa}, Pavel. 2021.
\newblock {Downsizing revised: Star formation timescales for elliptical
  galaxies with an environment-dependent IMF and a number of SNIa}.
\newblock {\em \aap}, {\bf 655}(Nov.), A19.

\bibitem[\protect\citename{{Zhang} {et~al.}, }2001]{Zhang+01}
{Zhang}, Q., {Fall}, S.~M., and {Whitmore}, B.~C. 2001.
\newblock {A Multiwavelength Study of the Young Star Clusters and Interstellar
  Medium in the Antennae Galaxies}.
\newblock {\em \apj}, {\bf 561}(Nov.), 727--750.

\bibitem[\protect\citename{{Zhang} {et~al.}, }2018]{Zhang+18}
{Zhang}, Z.-Y., {Romano}, D., {Ivison}, R.~J., {Papadopoulos}, P.~P., and
  {Matteucci}, F. 2018.
\newblock {Stellar populations dominated by massive stars in dusty starburst
  galaxies across cosmic time}.
\newblock {\em \nat}, {\bf 558}(June), 260--263.

\bibitem[\protect\citename{{Zinnecker} and {Yorke}, }2007]{ZY07}
{Zinnecker}, H., and {Yorke}, H.~W. 2007.
\newblock {Toward Understanding Massive Star Formation}.
\newblock {\em \araa}, {\bf 45}(Sept.), 481--563.

\bibitem[\protect\citename{{Zonoozi} {et~al.}, }2016]{Zonoozi16}
{Zonoozi}, A.~H., {Haghi}, H., and {Kroupa}, P. 2016.
\newblock {A Possible Solution for the M/L-[Fe/H] Relation of Globular Clusters
  in M3. I. A Metallicity- and Density-dependent Top-heavy IMF}.
\newblock {\em \apj}, {\bf 826}(July), 89.

\bibitem[\protect\citename{{Zonoozi} {et~al.}, }2019]{Zonoozi+18}
{Zonoozi}, Akram~Hasani, {Mahani}, Hamidreza, and {Kroupa}, Pavel. 2019.
\newblock {Was the Milky Way a chain galaxy? Using the IGIMF theory to
  constrain the thin-disc star formation history and mass}.
\newblock {\em \mnras}, {\bf 483}(1), 46--56.

\end{thebibliography}
\label{refs}
  \bibliographystyle{cambridgeauthordate}

  \cleardoublepage





\end{document}